\documentclass[bibyear]{aa} % if the references are not structured 
\usepackage{graphicx,textcomp}
\usepackage{subcaption}
\usepackage[varg]{txfonts}
\usepackage{natbib} 
\usepackage{booktabs}

\usepackage{hyperref}
\hypersetup{colorlinks,allcolors=blue}

\begin{document} 

   \title{A catalogue of asteroseismically calibrated ages for APOGEE DR17}
   \subtitle{The predictions of a CatBoost machine learning model based on the [Mg/Ce] chemical clock and other stellar parameters.}

   %\subtitle{I. Overviewing the $\kappa$-mechanism}

    \author{Thibault Boulet\inst{\ref{IA-Porto},  \ref{UPorto}\thanks{\email{thibault.boulet@astro.up.pt}}}
        %\and Tiago Campante\inst{\ref{IA-Porto}, \ref{UPorto}}
        %\and Vardan Adibekyan \inst{\ref{IA-Porto},\ref{UPorto}}
        %\and Aldo Serenelli \inst{\ref{IEEC-CSIC}}
        %\and Andreas W. Neitzel \inst{\ref{IA-Porto},\ref{UPorto}}
    }

   %\author{ Thibault Boulet,
   %Tiago Campante, Vardan Adibekyan, Aldo Serenelli, Andreas Wilhelm Neitzel
          %\inst{1}
          %\and
          %\fnmsep\thanks{Just to show the usage
          %of the elements in the author field}
          %}

     \institute{Instituto de Astrof\'isica e Ci\^{e}ncias do Espa\c co, Universidade do Porto, CAUP, Rua das Estrelas, 4150-762           Porto, Portugal \label{IA-Porto}
            \and
            Departamento de Física e Astronomia, Faculdade de Ciências, Universidade do Porto, Rua do Campo Alegre, 4169-007 Porto, Portugal. \label{UPorto}
            %\and
            %Institute of Space Sciences (IEEC-CSIC), Campus UAB, E-08193 Bellaterra, Spain %\label{IEEC-CSIC}
    }

   %\institute{Astrophysics center of Porto University,
              %\email{Thibault.boulet@astro.up.pt}
    %     \and
    %         Astrophysics center of Barcelona University ,\\
             %\email{}
             %\thanks{}
    %         } 

   \date{Received 21 September 2023 / Accepted 24 December 2023}
 
  \abstract
   {The formation history and evolution of the Milky Way through cosmological time is a complex field of research requiring the sampling of highly accurate stellar ages for all Galaxy components. Such highly reliable ages are starting to become available thanks to the synergy of asteroseismology, spectroscopy, stellar modelling, and machine learning analysis in the era of all-sky astronomical surveys.}
   {Our goal is to provide an accurate list of ages for the Main Red Star Sample of the APOGEE DR17 catalogue. In order to reach this goal, ages obtained under asteroseismic constraints are used to train a machine learning model.}
   {As our main objective is to obtain reliable age predictions without the need for asteroseismic parameters, the optimal choice of stellar non-asteroseismic parameters was investigated to obtain the best performances on the test set. The stellar parameters T$_{\texttt{eff}}$ and L, the abundances of [CI/N],[Mg/Ce], and [$\alpha$/Fe], the U(LSR) velocity, and the vertical height from the Galactic plane `Z' were used to predict ages with a categorical gradient boost decision trees model. The model was trained on two merged samples of the TESS Southern Continuous Viewing Zone and the Second APOKASC catalogue to avoid a data shift and to improve the reliability of the predictions. Finally, the model was tested on an independent data set of the K2 Galactic Archaeology Program.}
   {A model with a median fractional age error of 20.8\% is obtained. Its prediction variance between the validation and the training set is 4.77\%. For stars older than 3 Gyr, the median fractional error in age ranges from 7\% to 23\%. For stars with ages ranging from 1 to 3 Gyr, the median fractional error in age ranges from 26\% to 28\%. For stars younger than 1 Gyr, the median fractional error is 43\%. The optimised model applies to 125,445 stars from the Main Red Star Sample of the APOGEE DR17 catalogue. Our analysis of the ages confirms previous findings regarding the flaring of the young Galactic disc towards its outer regions. Additionally, we find an age gradient among the youngest stars within the Galactic plane. Finally, we identify two groups of a few metal-poor ([Fe/H] < -1 dex) young stars (Age < 2 Gyr) with similar peculiar chemical abundances and halo kinematics. These are likely the outcomes of the predicted third and latest episode of gas infall in the solar vicinity ($\sim$ 2.7 Gyr ago).
 }
   {We make a catalogue of asteroseismically calibrated ages for 125,445 red giants from the APOGEE DR17 catalogue available to the community. The analysis of the associated stellar parameters corroborates the predictions of different literature models.
   }

   \keywords{Galaxy: formation -- Galaxy: evolution -- Galaxy: abundances -- catalogs --  asteroseismology}

   \maketitle
%-------------------------------------------------------------------

\section{Introduction}\label{sect:Introduction}

Galactic archaeology is the study of the formation and evolution of the Milky Way \citep{Miglio-Plato-2017}, with stellar age precision being crucial for progress \citep{Hekker2018}. The isochrone placement method \citep{Soderblom-2010} has yielded precise results \citep{Nissen-2015, DelgadoHarpsGTO2019, Casali-2020}, although these are limited to the stellar cluster population, turn-off stars, and solar analogues \citep{Morel-2021}.

Solar-like red giants are the major target of interest for studying the Galaxy because of their intrinsic brightness, seismic constraints, and observational ubiquity \citep{Hayden-2015, Hekker2018}. Red giants reveal Galactic chemical evolution information, except for certain elements affected by stellar diffusion. These elements are atomic carbon (CI), nitrogen \citep{Masseron-2015, Martig-2016,Ness-2016, Hasselquist-2019}, and lithium \citep{Deal-2021}, but also include sodium and aluminium for stellar masses above 1.8 $\text{M}_{\sun}$ \citep{Smiljanic-2016}.

As red giants are the preferred targets for studying the Galaxy, the APOGEE survey \citep{Majewski-APOGEE-2017} stands out as the most suited mission, having probed the vastest number of them across a large fraction of the celestial sphere in both the Northern and Southern Hemispheres in the infrared H band (1.51$\mu$m-1.70$\mu$m). The latest public release \citep[APOGEE DR17;][]{APOGEE-DR17} contains data on 657,000 unique stars.

In general, for individual field stars, age precision is limited to about 40\% \citep{Lebreton-2009} at any evolutionary stage. Asteroseismic constraints improve the precision to 10-20\% \citep{Soderblom-2010, Lebreton-2014,Second_APOKASC_Catalog:2018}.\\ However, there is a limitation to using asteroseismic constraints for dating methods. The majority of stars observed by all-sky spectroscopic surveys do not benefit from asteroseismic data. The disparity between the availability of asteroseismic data and chemical abundance data has motivated the search for age-abundance relations, also known as `chemical clocks' \citep{daSilva-2012, Nissen-2015}.\\ Chemical clock modelling improved dating precision within the solar neighbourhood \citep{Feuillet-2018,DelgadoHarpsGTO2019, Sharma-Hayden-2020,Hayden-2021,Sharma-2021,MOYA-2022}. Nevertheless, applying chemical clocks to vast regions of the Galaxy beyond the solar neighbourhood is inefficient because of the significant scatter in abundance. This was observed, for example, in red clump stars by \cite{Casamiquela-2021} beyond 1 kpc from the Sun.

Recent advancements relying on a data-driven approach have enabled the expansion of dating capabilities from the solar neighbourhood to the entire Milky Way Galactic disc. This approach is based on the spectroscopic determination of age for red giants, and relies on a function to model the flux of reference stars at each wavelength.\\
The Cannon \citep{The_Cannon,Ness-2018} was a pioneering method to implement this data-driven approach using APOGEE DR12 \citep{APOGEE-DR12} stellar spectra from stars sampled from the Second APOKASC  asteroseismic catalogue \citep[APOKASC-2;][]{Second_APOKASC_Catalog:2018}. Following the release of The Cannon, two other methods based on the same principle were developed. "ASTRO-NN" \citep{ASTRO-NN} relies on a neural network to deal with high-resolution spectra (R$\sim$22 500) from APOGEE DR14 \citep{APOGEE-DR13-DR14}. On the other hand, "DD-Payne" \citep{DD-Payne} uses the same training scheme as "The Cannon"  but combines it with a flexible and efficient tool for the simultaneous determination of several stellar parameters with full spectral fitting called "Payne"\citep{The-Payne}. DD-Payne was used to predict stellar parameter values for 6 million stars from $\sim$ 8 million low-resolution (R$\sim$1800) spectra from LAMOST DR5 \citep{LAMOST-2012}. These three methods have achieved a maximum age precision of 30\% because of the inherent limitations in extracting information from the subtle differences in red giant stellar spectra.

In order to avoid the limitations of these data-driven methods, a promising approach was developed by \cite{Anders-XGBoost}. Instead of relying on stellar spectra, it directly utilises the stellar parameters from the APOGEE-\textit{Kepler} catalogue \citep{Miglio-2021} as features for training a machine learning model, specifically an \texttt{XGBoostRegressor} \citep{XGBoost-2016}.

The work presented in this article differs from that of \cite{Anders-XGBoost} in three major respects. Firstly, the model presented in the present article is trained with a \texttt{CatBoostRegressor} \citep{Catboost-2017} instead of an \texttt{XGBoostRegressor}.
Secondly, a different set of stellar parameters are used to train the model, including the [Mg/Ce] chemical clock. Finally, the training set is not only made of APOGEE red giants from the \textit{Kepler} \citep{KEPLER2010} field but also incorporates red giants observed with the Transiting Exoplanet Survey Satellite \citep[TESS;][]{tess:2014} in its Southern Continuous Viewing Zone (TESS SCVZ, hereafter). TESS, a recent asteroseismological mission, overcomes the limitations of the previous missions COROT \citep{CORTOT-2007}, \textit{Kepler,} and K2 \citep{Rendle-K2-2019}, offering advantages for studying the vertical and radial structure of the Milky Way.

The goals of the present work are to compile a catalogue of asteroseismically calibrated ages for stars within the Main Red Sample of the APOGEE DR17 catalogue and to subsequently analyse the distribution of stellar parameters associated with the obtained ages.\\
The research sample studied here is described in Section \ref{sect:Sample description}. Section \ref{sect:Selection of a machine learning model} provides basic concepts in machine learning and justifies the choice of the selected model. Section \ref{sec:Feature Selection} deals with the choice of features for the model. Section \ref{sec:Model optimisation} details all the various optimisation processes employed to improve the accuracy of the predictive model.
In Section \ref{sect:Results}, the optimal performances of the model and the associated results are presented, and in Section \ref{sect:Discussion}, the results are discussed. The conclusions
of this article are outlined in Section \ref{sec:Conclusion}.

%%%%%%%%%%%%%%%%%%%%%%%%%%%%%%%%%%%%%%%%%%%%%%%%%%%%%%%%%%%%%%%%%%%%%%%%%%%%

\begin{figure*}[ht]
     \centering
     \begin{subfigure}[b]{0.32\textwidth}
         \centering
         \includegraphics[width=\textwidth]{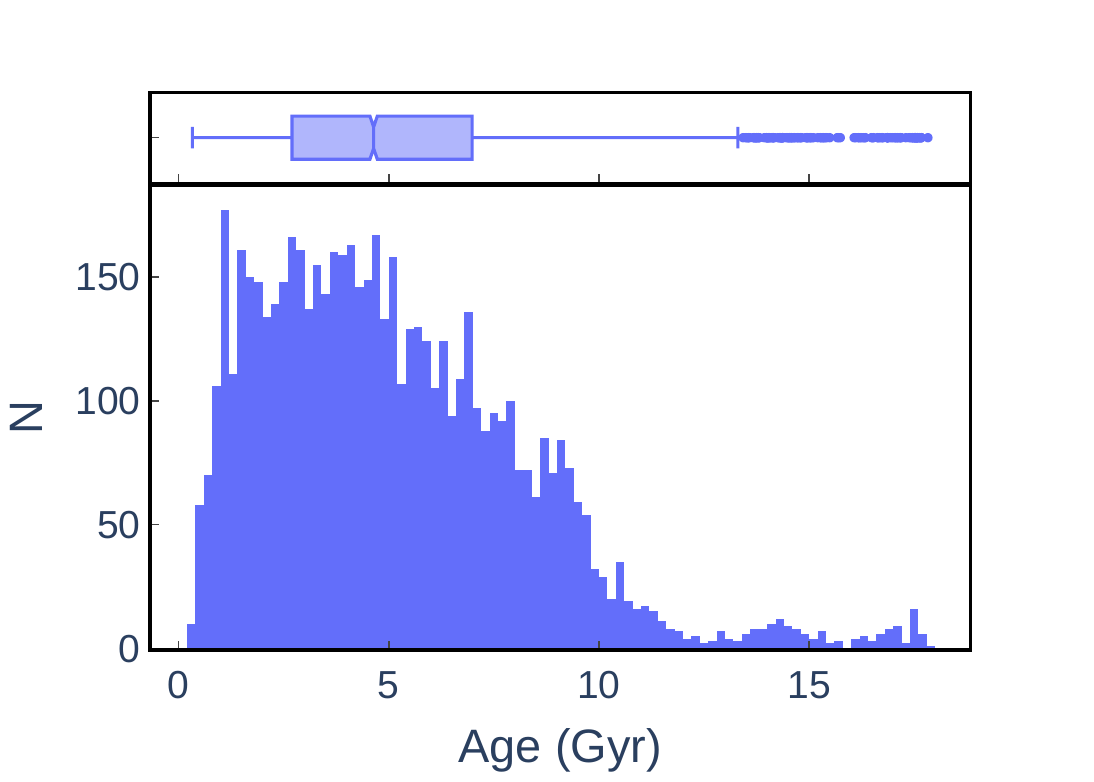}
         \caption{}
         \label{fig:Age-Histogram-APOKASC}
     \end{subfigure}
     \hfill
     \begin{subfigure}[b]{0.32\textwidth}
         \centering
         \includegraphics[width=\textwidth]{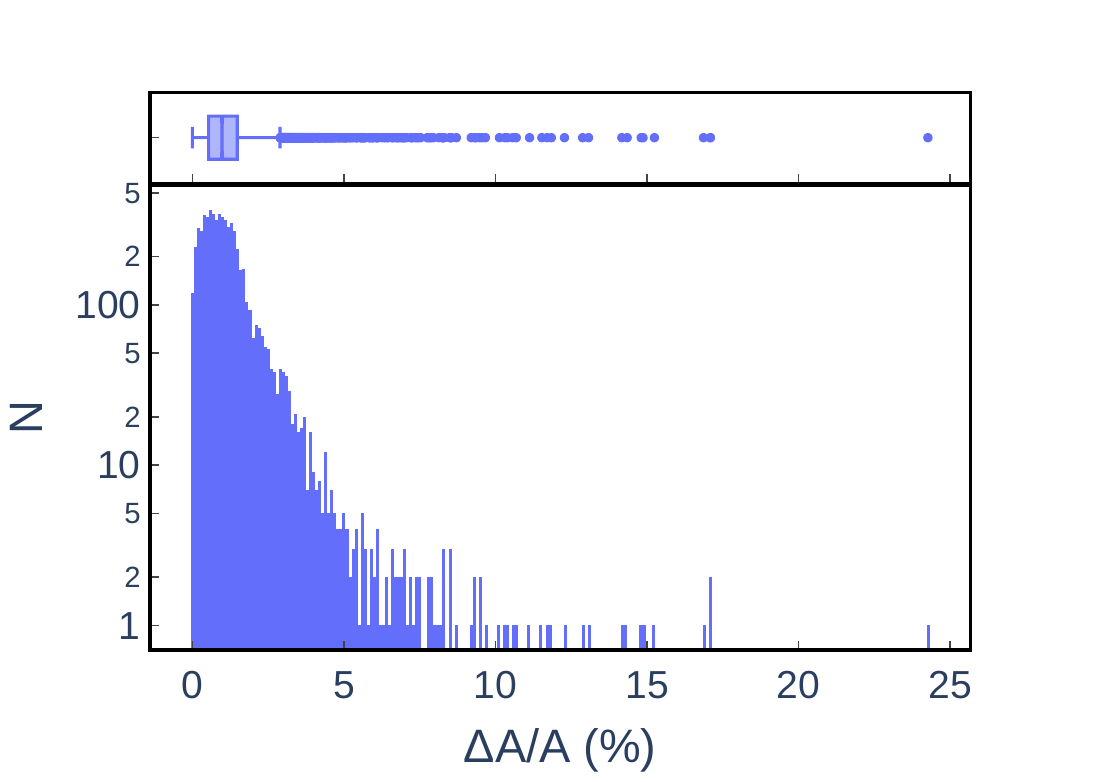}
         \caption{}
         \label{fig:Age-Uncertainty-Histogram-APOKASC}
     \end{subfigure}
     \hfill
     \begin{subfigure}[b]{0.32\textwidth}
         \centering
         \includegraphics[width=\textwidth]{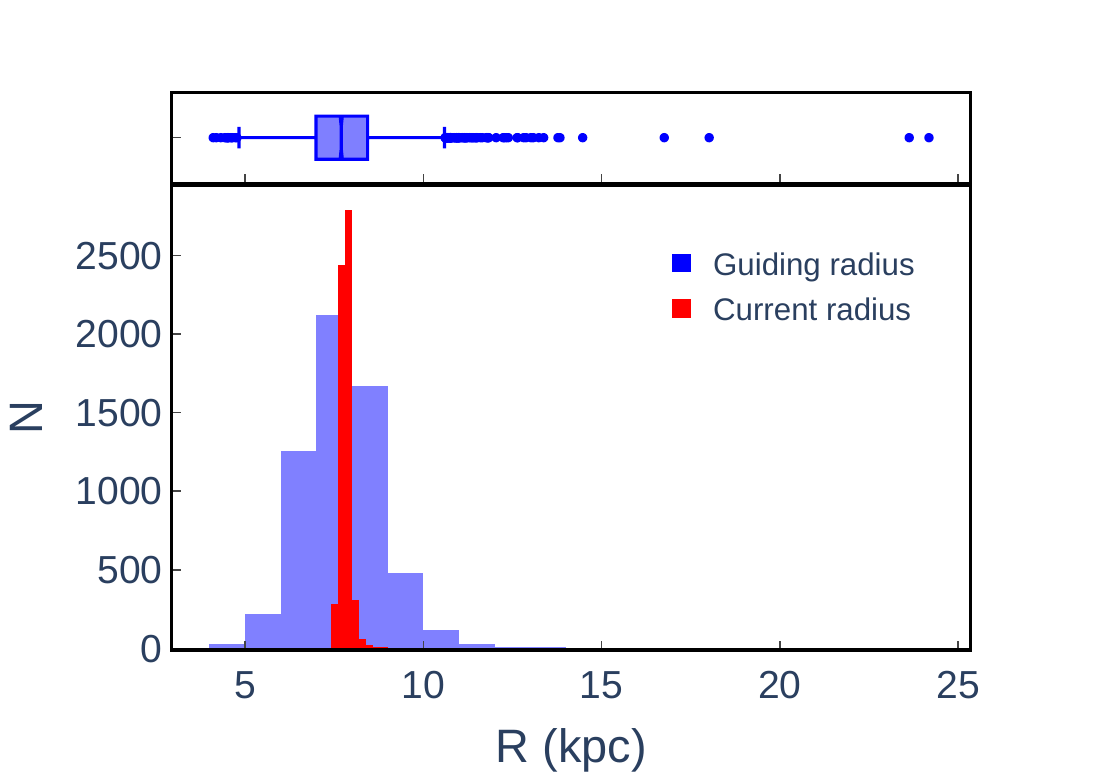}
         \caption{}
         \label{fig:Age-Guiding-Radius-APOKASC}
     \end{subfigure}
     \hfill
     \begin{subfigure}[b]{0.32\textwidth}
         \centering
         \includegraphics[width=\textwidth]{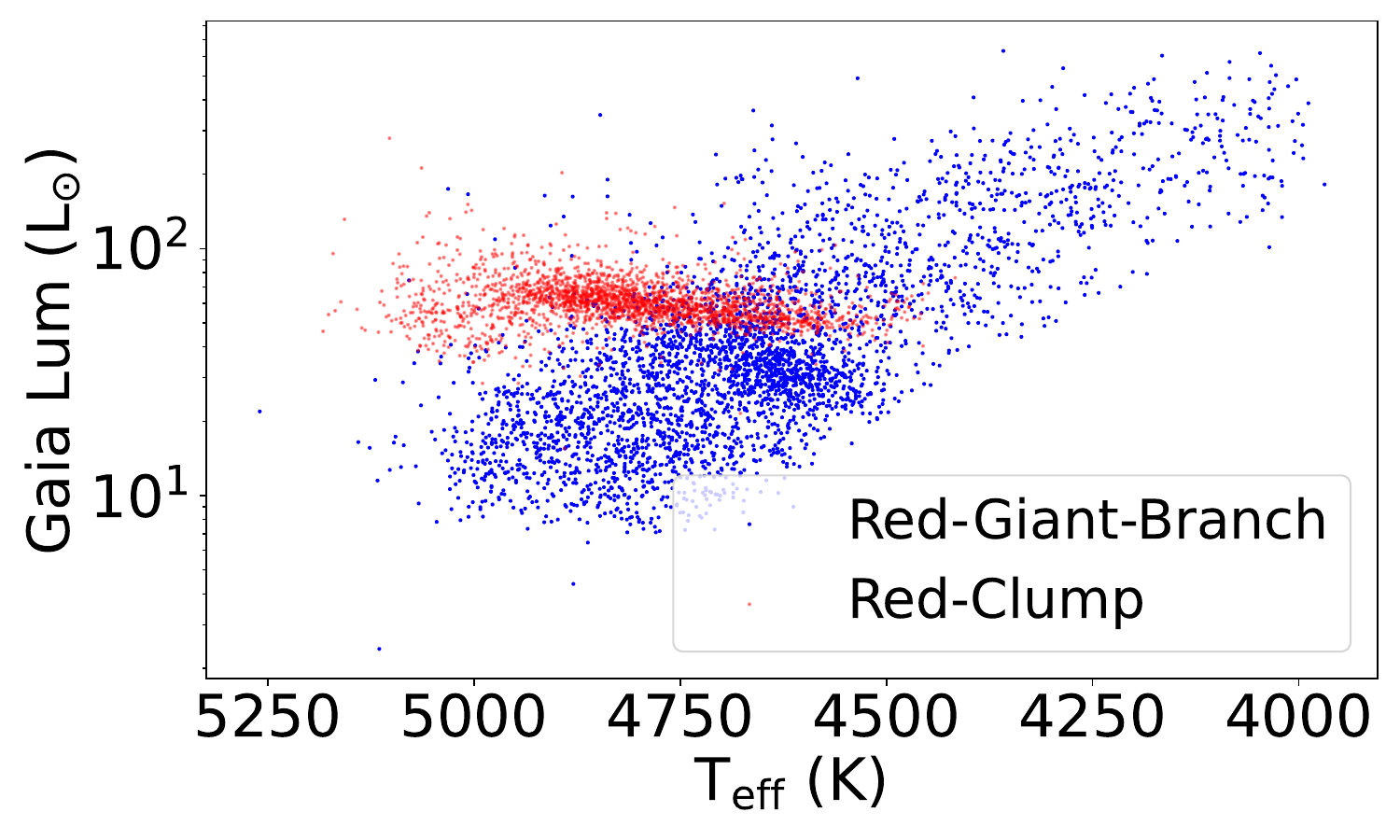}
         \caption{}
         \label{fig:HR-Diagram-APOKASC}
     \end{subfigure}
     \begin{subfigure}[b]{0.32\textwidth}
         \centering
         \includegraphics[width=\textwidth]{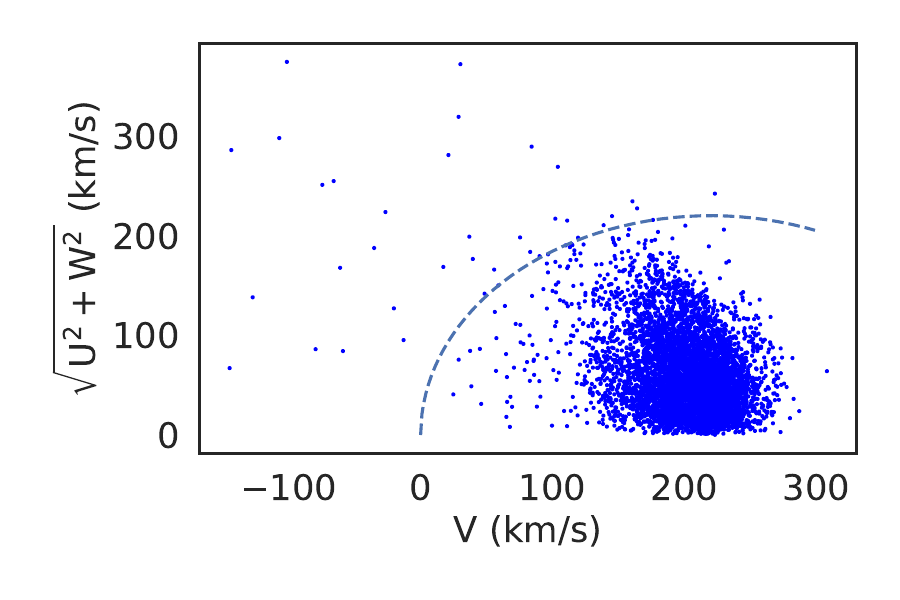}
         \caption{}
         \label{fig:Toomre-APOKASC}
     \end{subfigure}
     \begin{subfigure}[b]{0.32\textwidth}
         \centering
         \includegraphics[width=\textwidth]{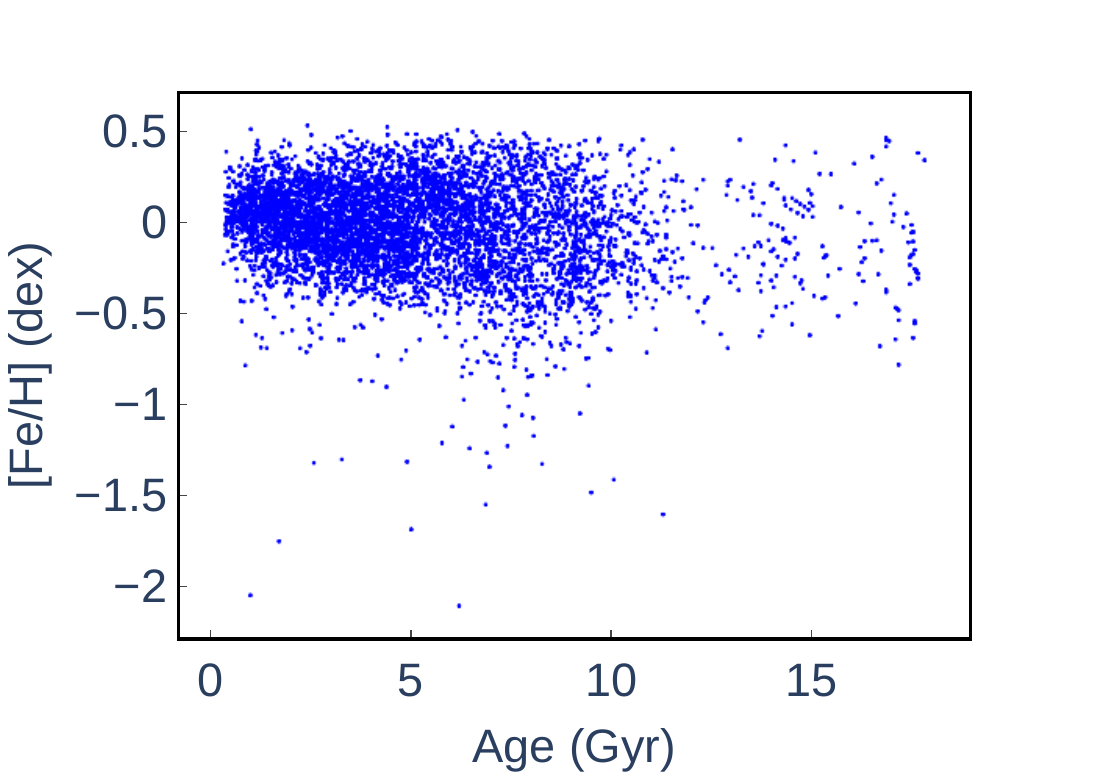}
         \caption{}
         \label{fig:Metalicity-Gradient-APOKASC}
     \end{subfigure}
     \begin{subfigure}[b]{0.4\textwidth}
         \centering
         \includegraphics[width=\textwidth]{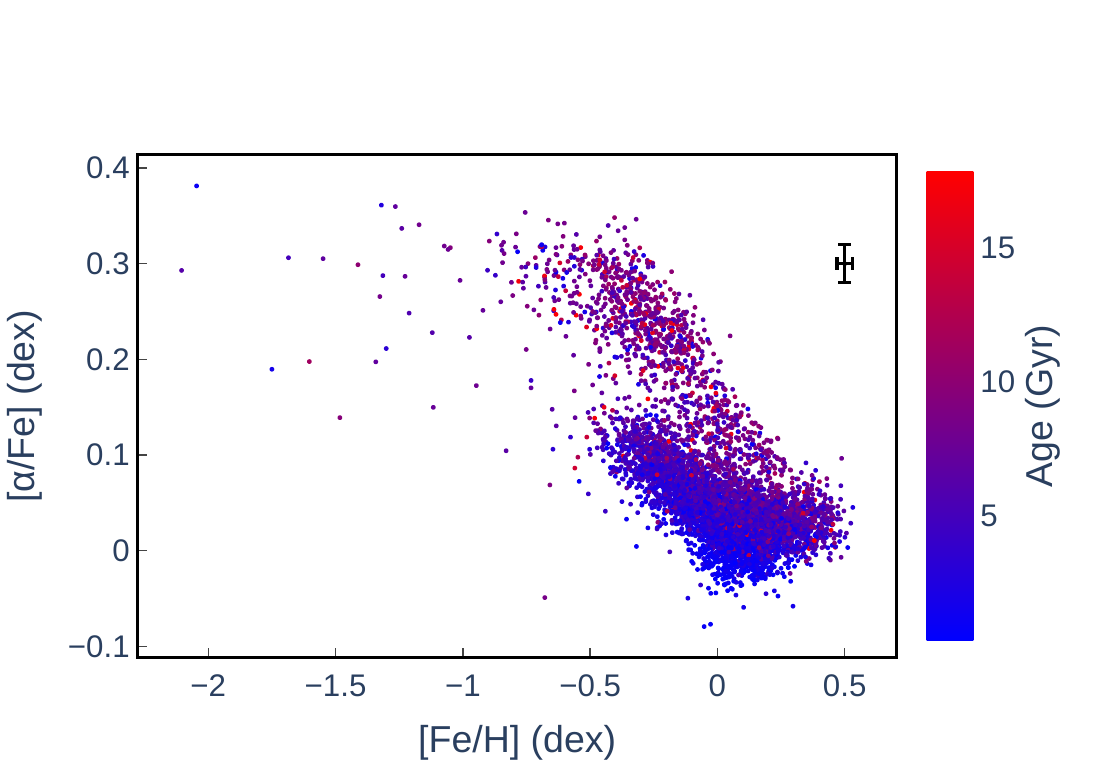}
         \caption{}
         \label{fig:Alpha-Dichotomy-APOKASC}
     \end{subfigure}
        \caption{Ensemble of plots summarising the information on the APOKASC-2 component of the research sample. Panel \ref{fig:Age-Histogram-APOKASC}: Age histogram. A box plot is systematically added to the histograms. Panel \ref{fig:Age-Uncertainty-Histogram-APOKASC}: Histogram of the random fractional uncertainties on age. Panel \ref{fig:Age-Guiding-Radius-APOKASC}: Histogram comparison of the current galactocentric distances and the guiding radii. Panel \ref{fig:HR-Diagram-APOKASC}: HR diagram of the sample. Panel \ref{fig:Toomre-APOKASC}: Toomre diagram of the velocities in the galactocentric referential. Panel \ref{fig:Metalicity-Gradient-APOKASC}: Age vs metallicity.
        Panel \ref{fig:Alpha-Dichotomy-APOKASC}: [$\alpha$/Fe] vs [Fe/H] plane. The black point with error bars depicts the mean uncertainty for both parameters.}
        \label{fig:Summarising-Plots-APOKASC}
\end{figure*}

\begin{figure*}[ht]
     \centering
     \begin{subfigure}[b]{0.32\textwidth}
         \centering
         \includegraphics[width=\textwidth]{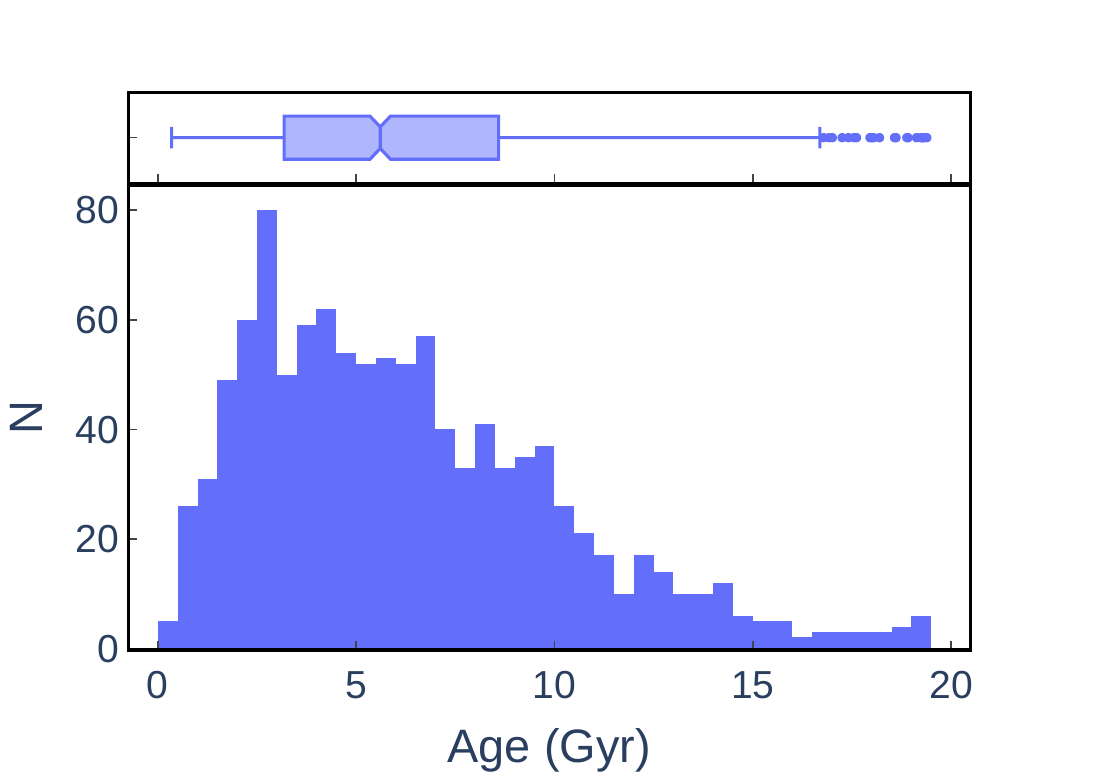}
         \caption{}
         \label{fig:Age-Histogram}
     \end{subfigure}
     \hfill
     \begin{subfigure}[b]{0.32\textwidth}
         \centering
         \includegraphics[width=\textwidth]{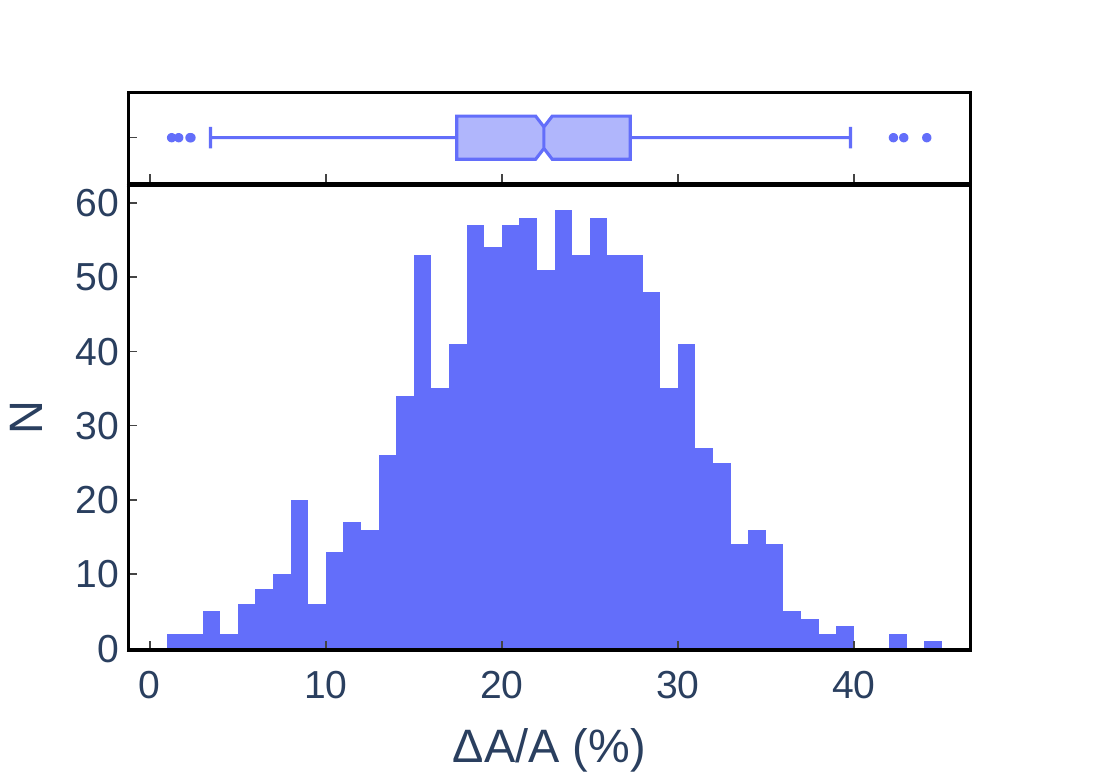}
         \caption{}
         \label{fig:Age-Uncertainty-Histogram}
     \end{subfigure}
     \hfill
     \begin{subfigure}[b]{0.32\textwidth}
         \centering
         \includegraphics[width=\textwidth]{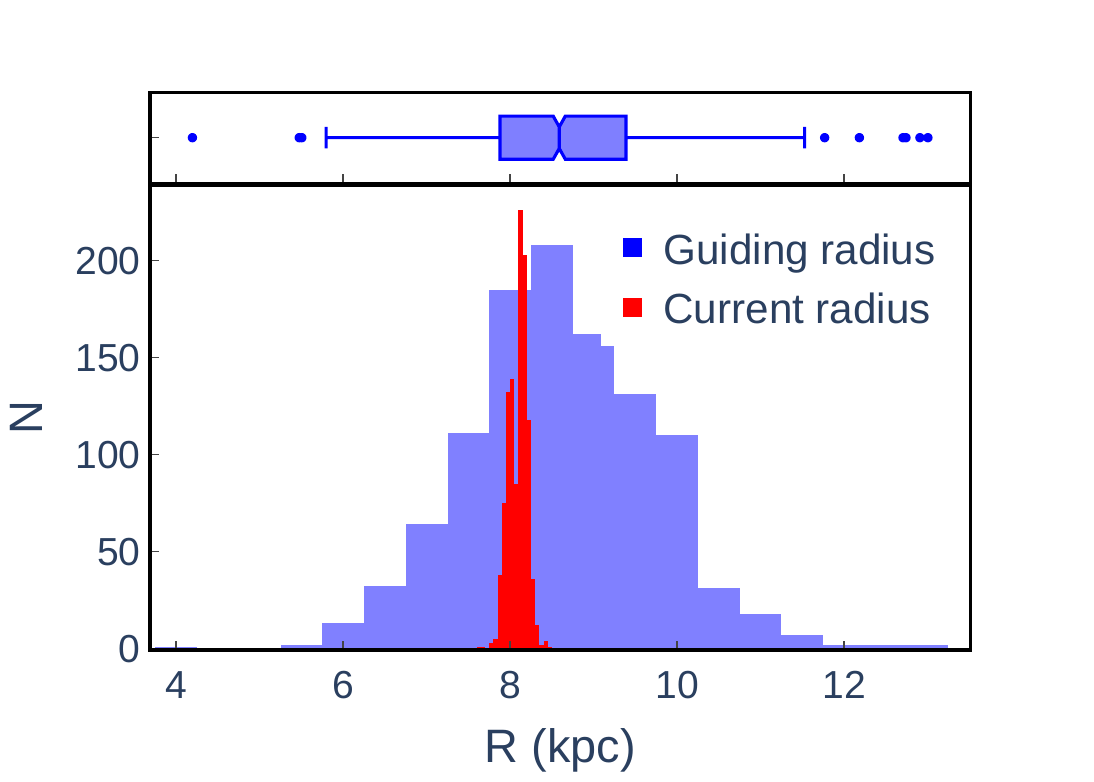}
         \caption{}
         \label{fig:Age-Guiding-Radius}
     \end{subfigure}
     \hfill
     \begin{subfigure}[b]{0.32\textwidth}
         \centering
         \includegraphics[width=\textwidth]{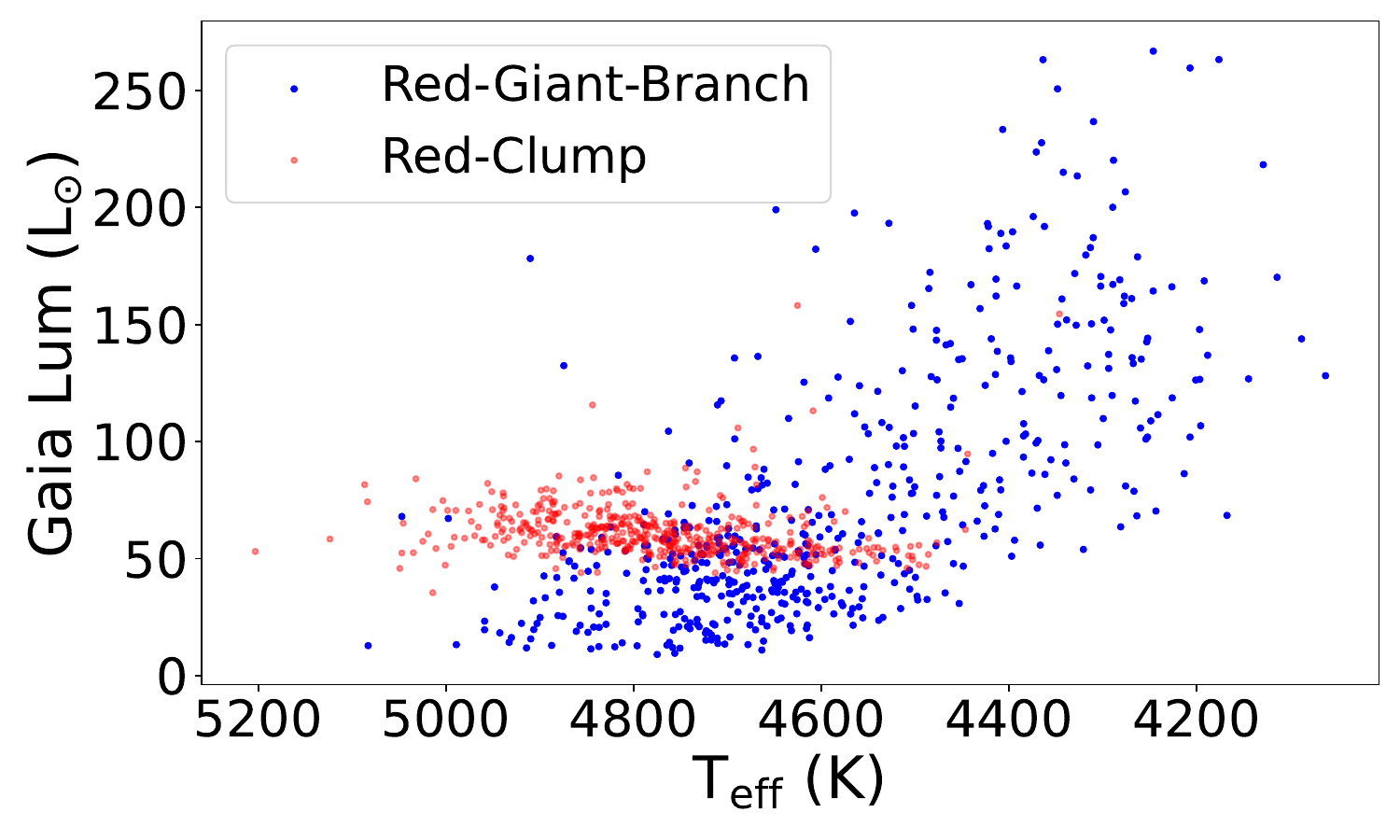}
         \caption{}
         \label{fig:HR-Diagram}
     \end{subfigure}
     \begin{subfigure}[b]{0.32\textwidth}
         \centering
         \includegraphics[width=\textwidth]{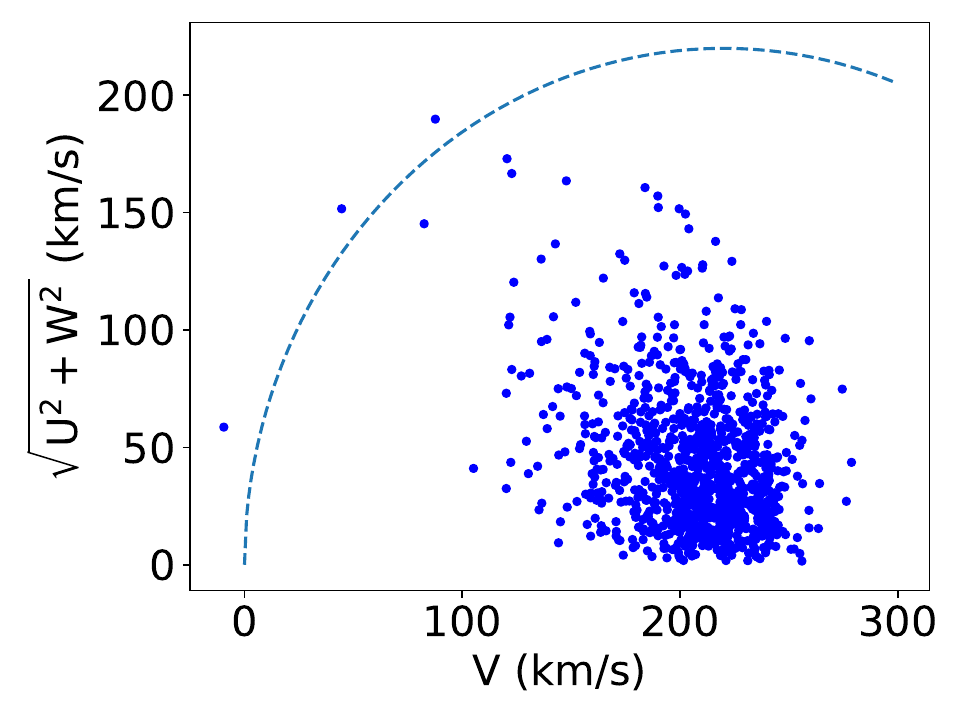}
         \caption{}
         \label{fig:Toomre}
     \end{subfigure}
     \begin{subfigure}[b]{0.32\textwidth}
         \centering
         \includegraphics[width=\textwidth]{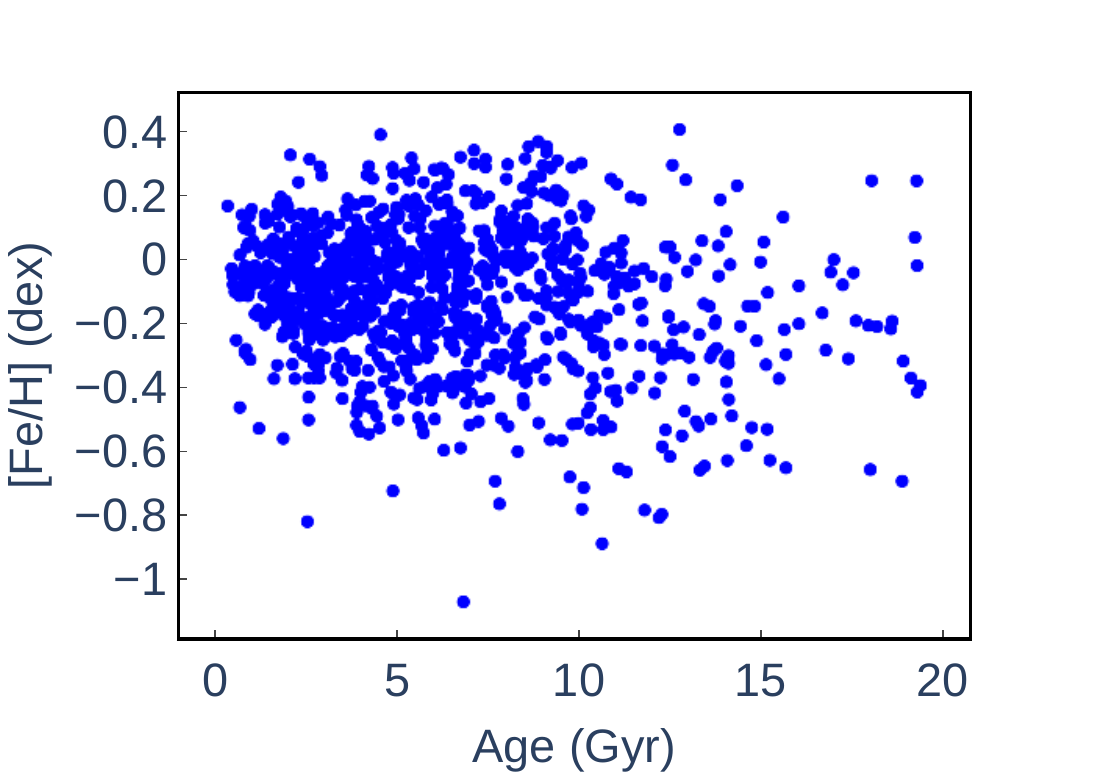}
         \caption{}
         \label{fig:Metalicity-Gradient}
     \end{subfigure}
     \begin{subfigure}[b]{0.4\textwidth}
         \centering
         \includegraphics[width=\textwidth]{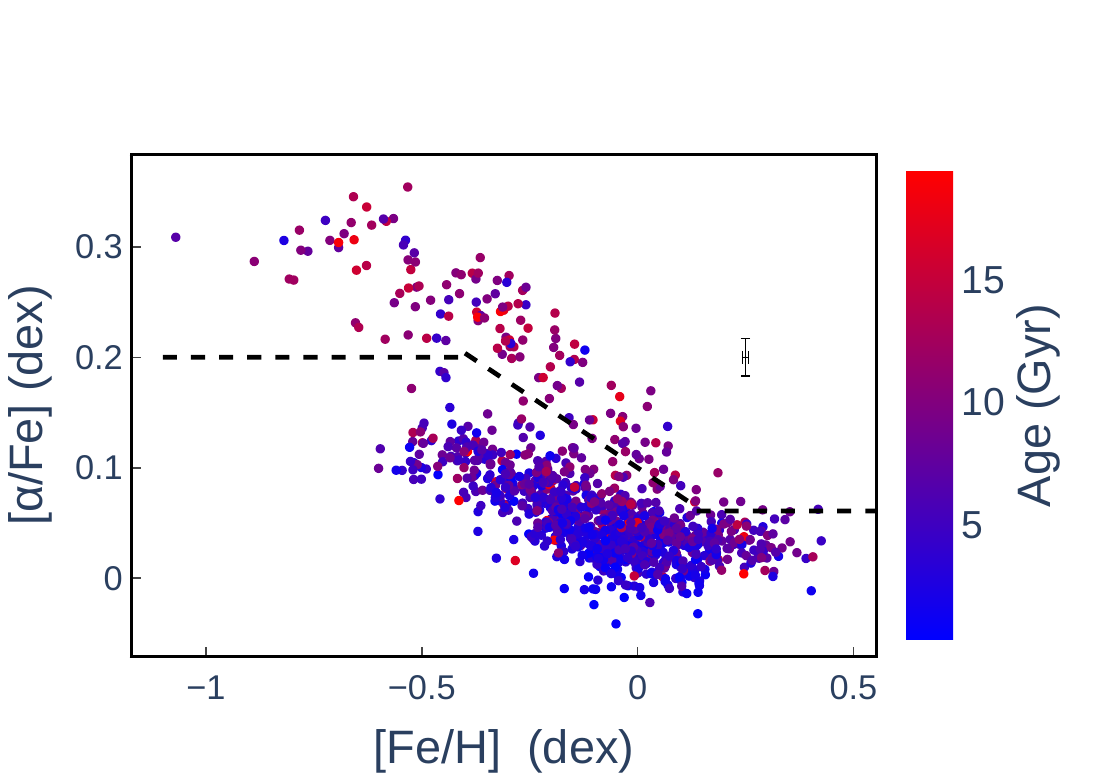}
         \caption{}
         \label{fig:Alpha-Dichotomy}
     \end{subfigure}
        \caption{Ensemble of plots summarising the information on the MCK component of the research sample. We refer to Figure \ref{fig:Summarising-Plots-APOKASC} for the description of the panels.}
        \label{fig:Summarising-Plots}
\end{figure*}

\section{Sample description}\label{sect:Sample description}

A sequential approach was adopted to achieve the highest accuracy in predicting ages for APOGEE DR17. Initially, one training set was used that is made of stars from the APOKASC-2 catalogue. However, upon observing a decline in model performance when tested on a TESS SCVZ sample, the decision was made to merge the two datasets, creating what is referred to as the MCK-APOKASC sample. The rationale for this combination is discussed in Section \ref{subsect:Identification of a data shift}. The resulting merged dataset yields more robust predictions in both regions, contributing to an overall improvement in the prediction accuracy of the model.

The selection of APOKASC-2 as the primary dataset was driven by three key factors. Firstly, this catalogue contains the highest number of red giants in the \textit{Kepler} field, that is, 6676 evolved stars. Secondly, it offers high-quality stellar ages due to asteroseismic parameters obtained using five independent techniques from continuous monitoring by \textit{Kepler} over a four-year period. Notably, the resulting asteroseismic constraints allowed the authors to reach fractional age uncertainties of mainly between 0.6\% and 5\%, as illustrated in Figure \ref{fig:Age-Uncertainty-Histogram-APOKASC}. However, it is important to note that these uncertainties are of random origin and do not reflect systematic errors in inputs or theoretical age inferences \citep{Second_APOKASC_Catalog:2018}. Thirdly, APOKASC-2 provides a dynamically sampled representation of a large portion of the Galactic disc, as depicted in Figure \ref{fig:Age-Guiding-Radius-APOKASC}.\\
To ensure the highest accuracy in stellar age data, the APOKASC-2 sample was refined. Only stars with evolutionary states determined through the asteroseismology method described in \cite{Elsworth-2017} were retained, excluding those identified spectroscopically. The resulting APOKASC-2 sample spans a galactocentric radius exceeding 5 kpc, as calculated using the \textit{astropy} Python package \citep{astropy:2013,astropy:2018,astropy:2022}. The closest stars to the Sun are located just beyond the Local Bubble \citep{Local_Bubble}; that is, with a heliocentric spherical radius (R$_{Helio}$) surpassing 300 pc.\\
Figure \ref{fig:Summarising-Plots-APOKASC} summarises the key characteristics of the sample. The age histogram (refer to Figure \ref{fig:Age-Histogram-APOKASC}) reveals some stars with ages exceeding that of the Universe \citep[13.77 Gyr;][]{Planck-Collaboration-2020}. The guiding radius computed using the \texttt{Galpy} code \citep{GALPY} spans a wider range than current stellar positions (refer to Figure \ref{fig:Age-Guiding-Radius-APOKASC}). Evolutionary states illustrated in the HR diagram (refer to Figure \ref{fig:HR-Diagram-APOKASC}) clearly identify red clump stars. The velocities were computed using the method outlined in \cite{Galactic-Velocities-CalculusJ}, incorporating the latest data from \cite{Bensby-et-al-2003}, \cite{Reference-1-Galactic-Vel} and \cite{Bland-Hawthorn-2016}. Additionally, \texttt{Galpy} was used to confirm the consistency between the two techniques used to compute velocity. The resulting stellar velocities are illustrated in a Toomre diagram, as presented in Figure \ref{fig:Toomre-APOKASC}, revealing the identification of 32 kinematically distinct halo stars. The expected blurred metallicity gradient with age \citep{Nissen-2020} is displayed in Figure \ref{fig:Metalicity-Gradient-APOKASC}, and the chemical dichotomy in the sample is illustrated in Figure \ref{fig:Alpha-Dichotomy-APOKASC}.\\
In order to decide which $\alpha$-elements to retain in the computation of the [$\alpha$/Fe] ratio, several [X/Fe] versus [Fe/H] plots were compared. The goal was to find the sharpest separation between the two $\alpha$-populations. Eventually, the analysis led to retaining the mean of [Si/Fe] and [Mg/Fe] as the [$\alpha$/Fe]. No flagged abundances were found in the sample for these two $\alpha$-elements.

The additional sample is the result of the cross-match between the age-asteroseismic catalogue from \citep[MCK;][]{MacKereth-2021} and APOGEE DR17. This sample is made of 1025 stars and spans a spatial extension of approximately 2 kpc. As in the APOKASC-2 sample, its closest stars to the Sun are located slightly beyond the Local Bubble. Our motivations for its selection were the same as for APOKASC-2; one is the reliability of the ages, as they were derived using stellar modelling with asteroseismic constraints. The ages in the MCK catalogue display a mean fractional uncertainty of close to 22\%.  Also, its stars reveal that MCK managed to sample a substantial part of the Galactic disc dynamically.
The main characteristics of the sample are summarised in Figure \ref{fig:Summarising-Plots}.\\ Overall, the MCK sample displays similar properties to the APOKASC-2 sample, but there are a few differences. The guiding radius spans a narrower range of distances. There are fewer halo stars, which is expected given that the MCK sample is smaller. Finally, the fractional age uncertainty distribution is wider. \\
As MCK relied on the SkyMapper effective temperatures, we conducted a straightforward ordinary square regression. The aim of this analysis, performed with the built-in Python \textit{stats-model} package, was to verify the agreement between the two sets of effective temperatures. The results of the regression, available in Appendix \ref{Appendix:Temperature-Comparisons}, indicate that the two T$_{\texttt{eff}}$ scales are compatible when considering uncertainties. Consequently, we made the decision to utilise the APOGEE DR17 effective temperature for the MCK sample.

%%%%%%%%%%%%%%%%%%%%%%%%%%%%%%%%%%%%%%%%%%%%
\section{Selection of a machine learning model}\label{sect:Selection of a machine learning model}
%%%%%%%%%%%%%%%%%%%%%%%%%%%%%%%%%%%%%%%%%%%%

\subsection{The training scheme}\label{sect:The training scheme}

In supervised machine learning, data are divided into a training set used to train the model and a testing set used to assess the model's performance on unseen data, ensuring an evaluation of its generalisation ability.
To attain the best complexity for the model, we carried out three optimisation steps. Initially, emphasis was placed on the selection of a set of features. Subsequently, we fine-tuned the hyperparameters using the \textit{GridSearchCV} method. Finally, we determined the optimal set of random seeds.

The \textit{GridSearchCV} method, a scikit-learn \citep{scikit-learn} class, was employed to construct a grid of models with all combinations of selected hyperparameters. This process allows the identification of the model with the best hyperparameter values through cross-validation, employing a 10 K-fold of the training set.\\
Cross-validation involves dividing the training set into multiple subsets known as folds. The model is trained on several folds and is validated on a separate fold (the validation set) not used during training. This process is repeated, and the performance metrics are averaged across folds to provide a more robust assessment of the model's generalisation performance. The validation set helps tune hyperparameters and prevents overfitting by simulating how the model might perform on the testing set or any unseen data.

The random seed is used to initialise the random number generator used by machine learning algorithms. The random number generator is used in many different ways during the training process; for example, it is used to initialise the weights of the model and to select samples for each batch. The reason for testing different sets of values for the random seed is that even small differences in the sequence of random numbers can have a large impact on the final accuracy of the model.

The optimisation process was initiated by the selection of a set of features for a default grid of hyperparameter values. After computation of the predicted target values, a graphical check was conducted to ensure that the spread in predictions was well reproduced. Subsequently, an assessment for overfitting and underfitting was made using the root mean squared error (RMSE) metric.\\ If the RMSE on the validation set exceeded that of the training set, it was concluded that the model was overfitting. Overfitting was avoided when the relative difference in RMSE between the training set and the validation set known as the variance of the model was small.\\
In this study, the variance threshold was set to 5\%, an arbitrary but motivated choice, similar to typical p-value thresholds in statistical tests.\\ Once overfitting was ruled out, a check for underfitting was performed by ensuring that the RMSE on the test set was lower than the baseline RMSE. The baseline RMSE is obtained on the test set and is achieved with a model trained on a single feature, namely the one with the highest feature importance (refer to Section \ref{sec:Feature importance}).

\subsection{Regression trees in machine learning}

The choice was made to employ a machine learning technique based on tree-based models for the research objective, specifically using decision trees for regression. The primary advantage of opting for tree-based models lies in their ability to effectively capture non-linear relationships between features (the parameters of a model) and labels (the variables the model predicts).\\
However,  according to \cite{Hastie2009} (we refer to
their Table 10.1 for more details), this type of learner confers several comparative advantages over other machine learning methods. These advantages concern their performances in terms of the natural handling of mixed data types, the handling of missing values, the robustness to outliers in the input space, the insensitivity to the monotone transformation of features, the computable scalability, and the ability to deal with irrelevant inputs. Nevertheless, the use of a single tree leads to weak predictive power, which is the reason for the creation of the ensemble learning approach. This approach is designed to train different trees on the same dataset and let each model make its predictions. In the end, a meta-model aggregates predictions of the individual models. The final predictions are therefore more robust and less prone to errors. In the case of boosting-ensemble, when the base learner is a regression tree, the most suited ensemble approach is gradient boosting.
In gradient boosting, each tree is trained using the residual errors of its predecessor as labels. The first tree is initially trained on the input dataset (X, y), where X represents the feature values matrix and y is the column vector of label values. The predictions ($\hat{y_{1}}$) from the first tree are then used to calculate the residual errors ($r_{1}$ = $y$ - $\hat{y_{1}}$).\\
Next, the second tree is trained using the feature matrix X and the residual errors ($r_{1}$) of the first tree as labels. The predicted residuals ($\hat{r_{1}}$) from the second tree are then used to calculate the residuals of residuals, labelled as $r_{2}$ = $r_{1}$ - $\hat{r_{1}}$.\\
An important factor in training gradient-boosted trees to enhance performance is shrinkage. In this context, shrinkage involves multiplying each residual error by the learning rate ($\epsilon$). Notably, it is crucial to be aware of the trade-off between the learning rate and the number of trees in the model's final performances.\\
This process is iteratively repeated until all N trees in the ensemble are trained. Once all the trees are trained, introducing an unknown instance of data prompts each tree to make predictions, and the final predicted label value ($y_{pred}$) is determined from Equation \ref{eq:Gradient-Boosting}:

\begin{equation}\label{eq:Gradient-Boosting}
    y_{\text{pred}} = y_{1} + \epsilon r_{1} + \epsilon r_{2} + \ldots + \epsilon r_{N}
.\end{equation}

Gradient-boosting algorithms, such as XGBoost \citep{XGBoost-2016} and CatBoost \citep{Catboost-2017,Catboost-2018}, share common characteristics, including efficient handling of large datasets and support for parallel processing. Notably, they are recognised as state-of-the-art performers. Comparing their performances on the MCK-APOKASC sample using \texttt{XGBoostRegressor} and \texttt{CatBoostRegressor} models reveals better results with CatBoost in terms of the variance between the validation and training sets. Specifically, the variance with CatBoost (see Table \ref{tab:Perf-Summary-Table}) is, on average, two times smaller than with XGBoost. Consequently, we decided to continue our analysis with  \texttt{CatBoostRegressor} as our machine learning model. 

The performance of CatBoost can be attributed to its robust decision-tree algorithm, leveraging Oblivious Trees \citep{Oblivious-Tree-2016} for outlier handling. Its L2 regularisation approach, applied to both leaves and nodes, is more effective at preventing overfitting compared to other algorithms. Additionally, CatBoost benefits from a distinct hyperparameter \textit{random\_strength}, controlling randomness in the tree construction process to prevent overfitting and improve generalisation. For details on the tuned hyperparameters, we refer to Table \ref{tab:Hyparameters-CatBoost} in Appendix \ref{Appendix:CatBoost_Hyperparam}.

The final optimised results derived from the MCK-APOKASC sample are summarised in Table \ref{tab:Perf-Summary-Table}. These results match the case displaying the highest accuracy in predictions among 1000 different random configurations of the three different random seeds. These configurations imply the random splitting of the training--test sets (90\%-10\%), and the random instantiation of a \texttt{CatBoostRegressor} and the \texttt{RandomSampler} method (refer to Section \ref{sect:Oversampling imbalanced data}). The variance is sufficiently small to confidently consider that the model is not overfitting the data set. Also, the baseline test reveals that the model is not underfitting the data set.\\

%%%%%%%%%%%%%%%%%%%%%%%%%%%%%%%%%%%%%%%%%%%%%%%%%%%%%%%%%%%%%%%%%%%%%%%%%%%%%%%%%%%%%%%%%%

\subsection{Feature importance metric}
\label{sec:Feature importance}

The importance of the features in the model predictions was evaluated using the Shapley value technique. A Shapley value quantifies the average impact of a  feature on a model output magnitude. This technique satisfies a set of axioms that make it more reliable than other feature-importance calculation techniques \citep{Shapley-Values-Young-1985}.\\
Tree-based models in scikit-learn have a built-in feature-importance calculation method based on the Gini impurity index. However, Gini impurity-based feature importances may lead to inaccurate results, as a large number of distinct values tend to lead the associated feature to a higher importance score with Gini impurity, even if it might not be as informative as suggested.\\In our analysis, we frequently noticed that the rank of the most important features was permuted between the results based on the Gini index and the Shapley values. Therefore, the importance of the features is based on the Shapley values. The Shapley values were computed thanks to the SHAP Python package \citep{SHAP}. The plot of the Shapley features importance obtained on the MCK-APOKASC test set is displayed in Section \ref{sect:Final performance test set} (refer to Figure \ref{fig:Shap-values}).

\section{Feature selection}\label{sec:Feature Selection}

In this section, the rationale for selecting each age-correlated feature is described. The model was trained with a progressively expanding set of features, each known or expected to be age dependent. The addition of each new feature depended on its capacity to improve the capture of age dispersion, depict overall trends, reduce the variance of the model, and improve accuracy in age determination. There is a separate and dedicated section (refer to Section \ref{subsect:Additional variable}) dealing with an extra feature with no direct correlation to age. Notably, [Fe/H] was not a selected feature in the model for two reasons. First, there is an expected blurring of the age--metallicity relationship for large data sets with wide dispersion in age \citep{Nissen-2020}, rendering the correlation between [Fe/H] and age statistically insignificant. Second, there is a negligible Shapley value for [Fe/H] compared to other features when included in the model. Eventually, the impact of [Fe/H] was accounted for through the [$\alpha$/Fe] ratio, which serves as a proxy for metallicity.

\subsection{[Mg/Ce]}
\label{subsec:Feature [Mg/Ce]}

The Galactic chemical clock, chosen as the primary feature, showed the best  performance in age prediction. To identify this clock, abundance correlations with age were computed using the 20 calibrated abundances from APOGEE DR17. The APOGEE calibrated abundances, denoted as [X/Fe], are obtained by aligning solar-metallicity stars to [X/M]=0. These abundances exclude stars with suspect or known incorrect values based on various criteria set by the APOGEE consortium.

The Spearman coefficient ($\rho$) was used for correlation calculations because it does not require the assumption of a linear relationship. Stars with masses of greater than 1.8 $\text{M}_{\sun}$ were excluded for [Na/Fe] and [Al/Fe] (refer to Section \ref{sect:Introduction}). The APOGEE DR17 did not provide [S/Fe] abundances for the stars in the training sample because of the unreliability of the associated spectra. Titanium (Ti) was excluded from the study due to persistent discrepancies between the APOGEE DR16 [Ti/Fe]--[Fe/H] trend and the optical trend as identified and discussed in \cite{APOGEE-DR16}, which continue to be observed in APOGEE DR17. [CI/Fe] and [N/Fe] were not considered as Galactic chemical clocks, as they track stellar evolution rather than chemical Galactic evolution \citep{Hasselquist-2019}.\\
The Spearman coefficients are detailed in Table \ref{tab:Spearman-Table}, calculated using the \textit{scipy.stats} package \citep{Scipy}. Chromium ([Cr/Fe]) showed no statistical evidence of correlation with age and was excluded. [X/Ce] abundances, particularly [O/Ce] and [Mg/Ce], displayed the strongest correlation with age (refer to Figure \ref{fig:Spearman-Matrix-Ce}, in Appendix \ref{Appendix:Spearman}). These robust correlations align with the findings of \cite{Casali-2020} and \cite{Casamiquela-2021}, who demonstrate that combinations of $\alpha$ and s-process elements make the most effective chemical clocks.

[Mg/Ce] was chosen as the Galactic chemical clock rather than [O/Ce] because it displays the smallest intrinsic dispersion ($\sigma_{\text{[Mg/Ce]}}=0.22$) in the data. Extensive studies on the non-local thermodynamic equilibrium (NLTE) effects of magnesium in the H-band \citep{Osorio-2020, APOGEE-DR17}, on its reliability in chemical enrichment studies \citep{magnesium-rationale-paper-1-2011, Kobayashi-2020}, and on the recommendation of its use as a reference element \citep{Weinberg-2019,Weinberg-2022} further justify the choice of magnesium over oxygen to obtain the chemical clock with the most reliable performance.

%%%%%%%%%%%%%%%%%%%%%%%%%%%%%%%%%%%%%%%%%%%%%%%%%%%%%%

\begin{table}[tbp]
\centering
\begin{tabular}{ccc}
\hline
%\rowcolor[HTML]{C0C0C0} 
{[}X/Fe{]}  & $\rho$ & p-value \\ \hline
\textbf{{[}Mg/Fe{]}} & 0.61           & 0.0         \\ \hline
\textbf{{[}O/Fe{]}}  & 0.60           & 0.0         \\ \hline
\textbf{{[}Ni/Fe{]}} & 0.43           & 3.14e-256         \\ \hline
\textbf{{[}K/Fe{]}}  & 0.42           & 8.21e-253         \\ \hline
\textbf{{[}Si/Fe{]}} & 0.40           & 2.03e-221         \\ \hline
\textbf{{[}Co/Fe{]}} & 0.34           & 9.73e-162         \\ \hline
\textbf{{[}Al/Fe{]}} & 0.34           & 5.51e-158         \\ \hline
\textbf{{[}S/Fe{]}}  & 0.32           & 1.94e-139         \\ \hline
\textbf{{[}Ca/Fe{]}} & 0.25           & 6.24e-83          \\ \hline
\textbf{{[}Cr/Fe{]}} & 0.04          & 5.66e-3            \\ \hline
\textbf{{[}Na/Fe{]}} & -0.06          & 2.74e-6            \\ \hline
\textbf{{[}Mn/Fe{]}} & -0.21          & 2.30e-57         \\ \hline
\textbf{{[}V/Fe{]}}  & -0.22          & 2.51e-67          \\ \hline
\textbf{{[}Ce/Fe{]}} & -0.34          & 8.86e-155         \\ \hline
\end{tabular}
\caption{Spearman correlation coefficients with their p-values for the relations involving [X/Fe] vs. age. The null hypothesis assumes no correlation between a given [X/Fe] and age.}
\label{tab:Spearman-Table}
\end{table}

\subsection{\texorpdfstring{\texttt{[}$\alpha$\texttt{/Fe]}}{[alpha/Fe]}}\label{subsec:Feature [alpha/Ce]}

The second selected feature was the $\alpha$-dichotomy ratio used to isolate the Galactic disc in two distinct chemical populations \citep{Adibekyan-2012}. When using regression trees, as long as [$\alpha$/Fe] is a parameter of the model, there is no need to separate the analysis into the $\alpha$-rich and $\alpha$-poor components of the Galactic disc, as done in \cite{DelgadoHarpsGTO2019}. Moreover, limiting the training set to the $\alpha$-poor disc diminishes model performance due to fewer training data.

[$\alpha$/Fe] has proved to be efficient in separating each of the Galactic components in the [$\alpha$/Fe] versus [Fe/H] diagram \citep{Spitoni-2016,Rojas-Arriagada-2017,Hawkins-Wyse-2018}.  This chemical tagging property is exploitable by regression trees as they can identify distinct hidden trends in the data, matching regions with different chemical-enrichment histories.

Stellar age modelling codes, specifically BeSPP \citep{BeSPP2013} and PARAM \citep{PARAM}, employed for calculating ages in APOKASC-2 and MCK, respectively, do not take into account the stellar Galactic population membership (bulge, disc, halo). Consequently, this information is not incorporated into the final CatBoost model. In other words, the model assigns similar age confidence to bulge or halo stars if their parameters match the training set, regardless of population.

The final age catalogue (refer to Section \ref{sect:Discussion of the APOGEE age map}) reveals that 2\% of the stars are in the halo, with the rest in the disc. Probabilities of the Galactic population membership of each star were calculated using the method described in \cite{Bensby-et-al-2003,Bensby-2005}, updated with priors from \cite{Anguiano-2020}.

\subsection{[CI/N]}\label{subsec:Feature [CI/N]}

The third feature selected was the carbon-to-nitrogen ratio. It is important to know that the APOGEE catalogue comprises two types of carbon abundance: [C/Fe], derived from carbon molecule lines, and [CI/Fe], computed from neutral carbon lines. To emphasise the use of atomic carbon abundance, the carbon-to-nitrogen ratio is depicted as [CI/N].

Studies on [CI/N] as an age indicator in red giants have confirmed the correlation with age, albeit with some dispersion (\cite{Hasselquist-2019} and references therein). According to \cite{Karakas-2010}, given the mass range (0.64 < M(M$_{\odot}$) < 3.48) of the MCK-APOKASC sample, stars likely underwent only the first dredge-up, as their mass remains below the critical threshold of 5 M$_{\odot}$. This suggests a significant role for [CI/N] in indicating the evolutionary state within MCK-APOKASC, particularly impacting the surface composition of low- to intermediate-mass stars (0.8 < M(M$_{\odot}$) < 8) \citep{Karakas-2010}.

The scatter in the [CI/N] versus age relationship arises from various factors, including mixing processes, nucleosynthesis, and chemical evolution. In the model, the influence of chemical evolution was included by incorporating the Galactic chemical clock [Mg/Ce] as well as the [$\alpha$/Fe] ratio. While [CI/N] is valuable for age predictions in red giants, it alone has limitations in providing accurate ages for individual stars \citep{Salaris-2005}. Additional factors, such as metallicity and effective temperature, influence stellar evolution and must be considered for reliable age estimates. Therefore, for robust age estimates, it is essential to combine the [CI/N] ratio with other stellar parameters.

\begin{figure*}[htb]
       \centering
       \includegraphics[width=1.0\linewidth]{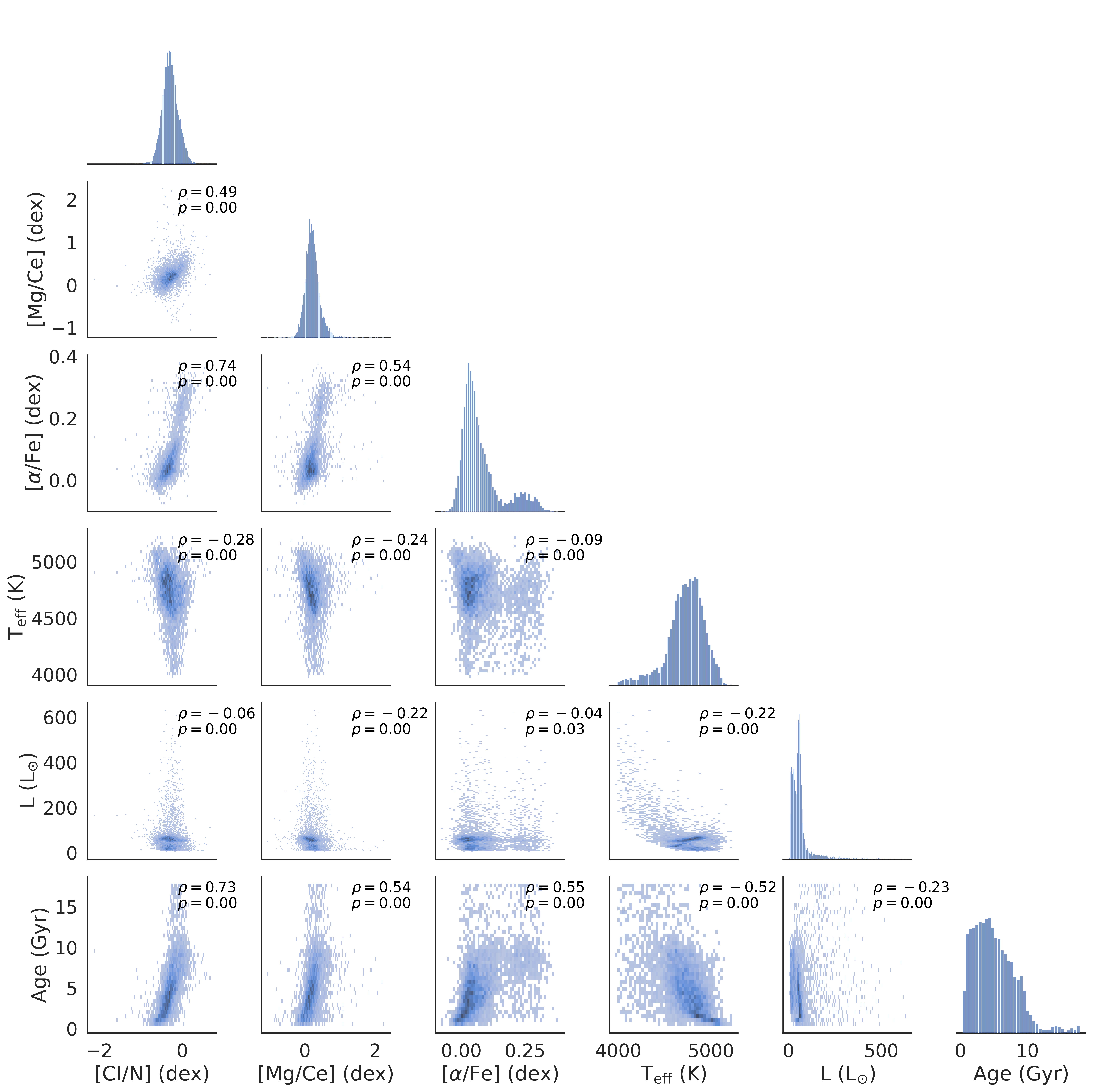}
       \caption{Corner plot of the retained stellar features, with a significant correlation with age, in the full training sample (MCK-APOKASCK). The diagonal depicts the histogram associated with each feature and the triangular bulk displays the correlation trends between each feature. Each plot displays the associated Spearman correlation coefficient ($\rho$) with the p-value of the test.}\label{fig:APOKASC_2_Corner_Plot_Cont_Features}
\end{figure*}

\subsection{\texorpdfstring{T\textsubscript{\texttt{eff}}, Z, and L}{Teff, Z, and L}}\label{Sec:TZL}

The remaining features selected were the effective temperature, the vertical distance from the disc, and the luminosity. Given that the total evolutionary lifetime of a star on the main sequence scales with its mass \citep{serenelli:2017}, incorporating features related to stellar mass significantly improves model performance. This improvement is particularly evident in better fitting age dispersion. As the effective temperature is linked to stellar mass through asteroseismic scaling relations for red giants \citep{RG-Scaling-Relations}, it was selected as a relevant feature.
The vertical distance from the Galactic disc (\textit{Z}) is crucial for model accuracy, considering the known vertical gradient in the stellar-mass distribution across the Galactic disc \citep{Vertical-Mass-Gradient-Miglio-2012,Vertical-Mass-Gradient-Casagrande-2015,Hon-2021}. \textit{Z} ranked among the most impactful features, leading to an improvement in  accuracy (refer to Figure \ref{fig:Shap-values}).

To address the generation of fractional residuals in age exceeding values of 100\%, luminosity (L) was chosen over log(g). Log(g) results in outliers reaching up to 150\%, while L successfully mitigates this issue. Including both L and log(g) does not improve accuracy but increases the variance of the model, likely because of the lower Shapley feature importance of log(g) compared to luminosity (SHAP = 0.053 vs. SHAP = 0.063).

As luminosities are not provided in the second APOKASC catalogue, we computed them using the same method as in \cite{MacKereth-2021} for consistency reasons. This method relies on bolometric corrections and the use of a 3D dust map library. The bolometric corrections in the $K_{s}$ band were computed using the \texttt{bolometric-corrections code} \citep{Bolometric-corretion-code-2014,Bolometric-corretion-code-2018-b,Bolometric-corretion-code-2018-a}, with a preference for the $K_{s}$ band given its lower sensitivity to extinction.\\
As bolometric corrections depend on reliable distances, stars with negative parallaxes were removed, and a fractional parallax uncertainty criterion ($f_{pu} = \sigma_{p}/p$, where $\sigma_{p}$ is uncertainty on parallax and p is parallax) was applied to filter out stars with $f_{pu}$ > 0.2 \citep{Bailer-Jones-2021}. After applying these filters, 6466 APOKASC-2 stars remain. Finally, the reddening ( E(B-V) ) was computed using the \texttt{MWDust} code \citep{Bovy-2016-c} and the 3D dust map library from \cite{Green-2019-3D-dust-map}.

\subsection{Summary of features}\label{Sect:Features summary}

The first feature configuration set obtained is ([CI/N],[Mg/Ce], [$\alpha$/Fe], T$_{\texttt{eff}}$, L, Z). \\
Except for \textit{Z}, which shows no significant Spearman correlation with age ($\rho$=-0.05, p=4.7×10$^{-6}$), all features exhibit visible correlations with age (refer to Figure \ref{fig:APOKASC_2_Corner_Plot_Cont_Features}). 
%Notably, [Mg/Ce], [$\alpha$/Fe], and T$_{\texttt{eff}}$ shared similar correlation strengths with age, while [CI/N] and L showed the highest and weakest correlations with age, respectively.
However, it is important to note that the correlation with age for [$\alpha$/Fe] is significant only for the $\alpha$-poor component, aligning with the findings of \cite{DelgadoHarpsGTO2019}. Additionally, T$_{\texttt{eff}}$ exhibits noticeable scatter, increasing as the temperature diminishes towards 4000 K.

%%%%%%%%%%%%%%%%%%%%%%%%%%%%%%%%%%%%%%%%%%%%%%%%%%%%%%%%%%%%%%%%%%%%%%%%%%%%%%%%%%%%%%%

\begin{table*}[t]
\centering
\begin{tabular}{@{}c c c c @{}}
\toprule
                       & Variance & Median Error & Baseline Test \\ \midrule
CatBoost       & 4.77\%            & 20.8\%                  & 0.173 $<$ 0.236  \\ \bottomrule
\end{tabular}
\caption{Summary of the best performances obtained on the MCK-APOKASCK sample. The median residual error is the mean difference between the ages of reference and those predicted by the model, divided by the reference ages. The two other metrics are defined in Section \ref{sect:The training scheme}.}
\label{tab:Perf-Summary-Table}
\end{table*}

%%%%%%%%%%%%%%%%%%%%%%%%%%%%%%%%%%%%%%%%%%%%%%%%%%%%%%%%%%%%%%%%%%%%%%%%%%%%%%%%%%%%%%%%%%%%%%%%%%%%

\section{Model optimisation}\label{sec:Model optimisation}

\subsection{Rescaling the age distribution}\label{sec:Rescaling}

Regression trees, which are designed to minimise mean squared error (MSE), may exhibit bias in the presence of skewed target variables, such as age (refer to Figure \ref{fig:APOKASC_2_Corner_Plot_Cont_Features}). Higher values disproportionately influence the optimisation process, leading to imbalanced splits in a tree's nodes and potential isolation of data tails (\cite{Hastie2009}, refer to their section 10.7).\\
To address this, log-transforming the data reduces the impact of high values, improving accuracy for the majority. Tests with age as the target variable demonstrate a 33\% mean error in residuals and a fractional residual maximum outlier of 359\%. Logarithmic transformation reduces these values to 28\% and 281\%, respectively, improving prediction accuracy. It is important to note here that residuals were computed by converting ages back to the linear scale.

Ages older than the age of the Universe were not included in the model training (13.77 < Age(Gyr) < 20). There are several reasons for this decision. Firstly, the model has been observed to predict such ages to a noticeable extent. This is expected because these ages were included as inputs for the model. Secondly, these ages are known to be mainly due to unconstrained systematic errors in the stellar modelling \citep{Second_APOKASC_Catalog:2018}. Thirdly, setting these ages arbitrarily to the age of the Universe would result in the generation of fabricated data with age-stellar parameter inconsistencies. This last point underscores a fundamental limitation of machine learning models in general; namely their lack of built-in methods to explicitly incorporate uncertainties in the input parameters. Models assume that the input data are accurate and the training data are representative of the underlying distribution. Therefore, to mitigate the generation of potential machine learning artefacts, the machine learning model was eventually applied solely to APOGEE stars with a fractional luminosity uncertainty inferior to 30\% (refer to Section \ref{sect:Discussion of the APOGEE age map}).

\subsection{Oversampling imbalanced data} \label{sect:Oversampling imbalanced data}

A noticeable data imbalance exists between ages older and younger than 10 Gyr, where `imbalance' in machine learning refers to a skewed or unequal distribution of data classes. This imbalance was suspected to contribute to increased mean fractional residuals at the oldest ages. To address this issue, the `oversampling' technique was applied using the Imbalance-Learn package \citep{Imbalanced-Learn}. Random oversampling was chosen as it involves duplicating existing data without the need to synthesise any. Importantly, oversampling was applied only to the training set. The approach sets a threshold at 10 Gyr, classifying data beyond this age as the minority class and everything below as the majority class. While experimenting with different thresholds, the one at 10 Gyr yields the best age accuracy performances. Overall, this oversampling significantly improved accuracy performances.

\subsection{Identification of a data shift}\label{subsect:Identification of a data shift}

As mentioned in Section \ref{sect:Sample description}, the model optimisation followed a sequential process. Initially, the model was trained exclusively on the APOKASC-2 sample. Subsequently, its predictive performance was assessed on the MCK sample to validate its generalisation capabilities. The relevance of the MCK sample lies in the distinct derivation of its asteroseismic constraints and ages compared to the second APOKASC catalogue. Specifically, the asteroseismic parameters in the MCK sample were derived from TESS light curves.
 The optimised model, trained on the APOKASC-2 sample, led to several key outcomes:
\begin{itemize}
  \item The maximum fractional error in predicting age is 131\%, indicating instances where the model predictions deviate significantly.
  \item The overall median fractional age error is 21\%, decreasing as the stellar age increases, except for a specific range between 11 and 13.77 Gyr.
  \item The standard deviation of the fractional error in a given age bin ($\sigma$) varies between $\sigma$=10 and $\sigma$=25, showing moderate fluctuations in model accuracy across different age ranges.
\end{itemize}

However, when the model trained on the APOKASC-2 sample is tested on the MCK sample, notable changes in its characteristics are observed:
\begin{itemize}
  \item Numerous fractional age errors increase significantly, reaching values close to 600\%.
  \item The median of the fractional age error distribution shows a global increase, reaching 25\%. The most substantial increments occur for age intervals of $0 < \text{Age(Gyr)} < 1$ and $1 < \text{Age(Gyr)} < 3$, with corresponding values of 80\% and 40\%, respectively.
  \item The standard deviation associated with the fractional age error exhibits a continuously increasing trend from the youngest to the oldest age bins, ranging from $\sigma$=5 to $\sigma$=100.
\end{itemize}

Consequently, these findings indicate that the model, trained solely on the APOKASC-2 sample, exhibits a serious deterioration in its prediction accuracy when applied to the MCK sample.
Kolmogorov-Smirnov (KS) tests  indicate the possible presence of a data shift between the two data sets (refer to Table \ref{tab:KSmirnov-Table}). These tests were conducted by considering variables related to stellar dynamics as well as chemical ratios. The test results indicate that all variables tested, except W(LSR), exhibit a significant difference in the origin of their distribution.\\
The most pronounced differences are observed in U(LSR) and the guiding radius. The difference in [$\alpha$/Fe] is not surprising, as previous research by \cite{Queiroz-2020} and \cite{Hayden-2015} demonstrated its distribution shapeshift, with APOGEE DR16 and APOGEE DR12 data, respectively, at different distances from the Galactic centre and heights above the Galactic plane.

\begin{table}[h]
\centering
\begin{tabular}{lcc}
\hline
\multicolumn{3}{c}{\textbf{Kolmogorov-Smirnov Test}}                                             \\ \hline
\multicolumn{1}{c}{Variable} & \multicolumn{1}{l}{Test Statistic} & p-value                      \\ \hline
\multicolumn{3}{c}{\textit{\textbf{Statistically Significant}}}                                  \\ \hline
U(LSR)                       & 0.569                              & 1.00e-267                    \\ \hline
Guiding Radius              & 0.317                              & 9.22e-79                     \\ \hline
zmax                         & 0.175                              & 4.21e-24                     \\ \hline
V(LSR)                       & 0.126                              & 1.24e-12                     \\ \hline
{[}$\alpha$/Fe{]}            & 0.107                              & \multicolumn{1}{l}{2.53e-09} \\ \hline
ecc                            & 0.077                              & 5.53e-05                     \\ \hline
{[}Mg/Ce{]}                  & 0.068                              & \multicolumn{1}{l}{5.52e-04} \\ \hline
\multicolumn{3}{c}{\textit{\textbf{Statistically Insignificant}}}                                \\ \hline
W(LSR)                       & 0.025                              & 6.39e-01                     \\ \hline
\end{tabular}
\caption{KS test results between APOKASC-2 and MCK samples, sorted by the test statistic. Velocity components U(LSR), V(LSR), and W(LSR) are in the Local Standard of Rest. Other variables, including guiding radius, maximum vertical height (zmax), and eccentricity (ecc), are from \texttt{Galpy}. Abundance ratios are from APOGEE DR17.}
\label{tab:KSmirnov-Table}
\end{table}

\subsection{Additional feature}\label{subsect:Additional variable}

To address the deterioration of the performance of our model, the APOKASC-2 sample was merged with the MCK sample, and an additional feature was added to the model. This approach effectively resolves the issue. Indeed, it results in improved reliability, robustness, and accuracy of the model predictions for both APOKASC-2 and MCK data.

The most pronounced difference between the two samples is the sign shift in radial velocity U(LSR). In the merged sample, stars below the Galactic plane mostly display negative speeds and stars above the Galactic plane mostly display positive speeds. The associated mean and standard deviations in U(LSR) are ($\mu$ = - 19.5, $\sigma$ = 42.5) for the MCK sample and ($\mu$ = + 47.5, $\sigma$ = 51.3) for the APOKASC-2 sample.\\ This asymmetry in U(LSR) in the region (8 < R(kpc) < 9, -1 < Z(kpc) < 1) is expected from the robust measurements of the three-dimensional velocity moments presented in the detailed Galactic disc kinematics study with LAMOST K giants in \cite{Ping-Jie-2021} (refer to their figure 6). Symmetries and asymmetries in the (U, Z) and (W, Z) planes are considered indicators of breathing and bending velocity motions in the Milky Way \citep{Ping-Jie-2021} but also in other disc galaxies \citep{Kumar-2022}.\\
\cite{Ping-Jie-2021} showed that the local asymmetry discussed earlier does not have a consistent shape and extent throughout the entire Galactic disc. In fact, it does not exist in some regions of the Galactic disc.
Therefore, the U(LSR) feature was added to the model in order for it to be able to generalise its predictions across the disc.\\
Finally, although U(LSR) shows no significant Spearman correlation with age ($\rho$ = 0.059, p = 8.99$\cdot$10$^{-7}$), incorporating U(LSR) further improves the performance of the model by reducing the fractional error, particularly for the youngest and oldest age groups.\\

\begin{figure*}[ht]
     \centering
     \begin{subfigure}[b]{0.45\textwidth}
         \centering
         \includegraphics[width=\textwidth]{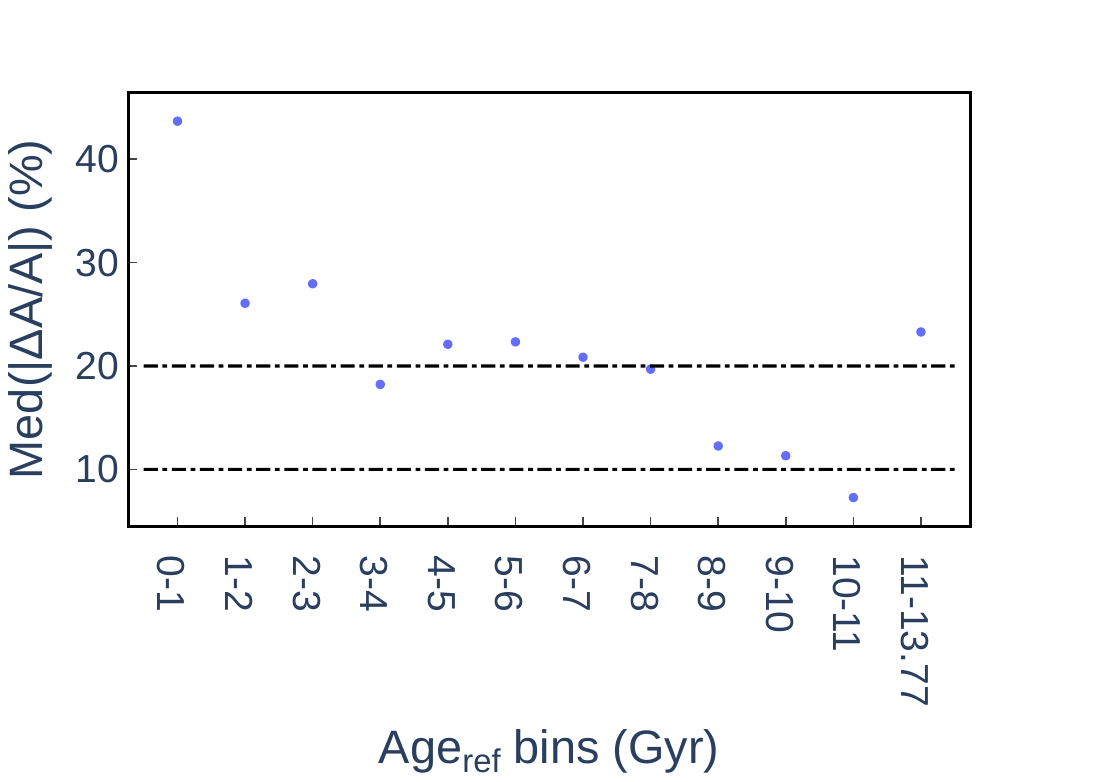}
         \caption{}
         \label{fig:Median_Frac_Residual_per_bin}
     \end{subfigure}
     \hfill
     \begin{subfigure}[b]{0.45\textwidth}
         \centering
         \includegraphics[width=\textwidth]{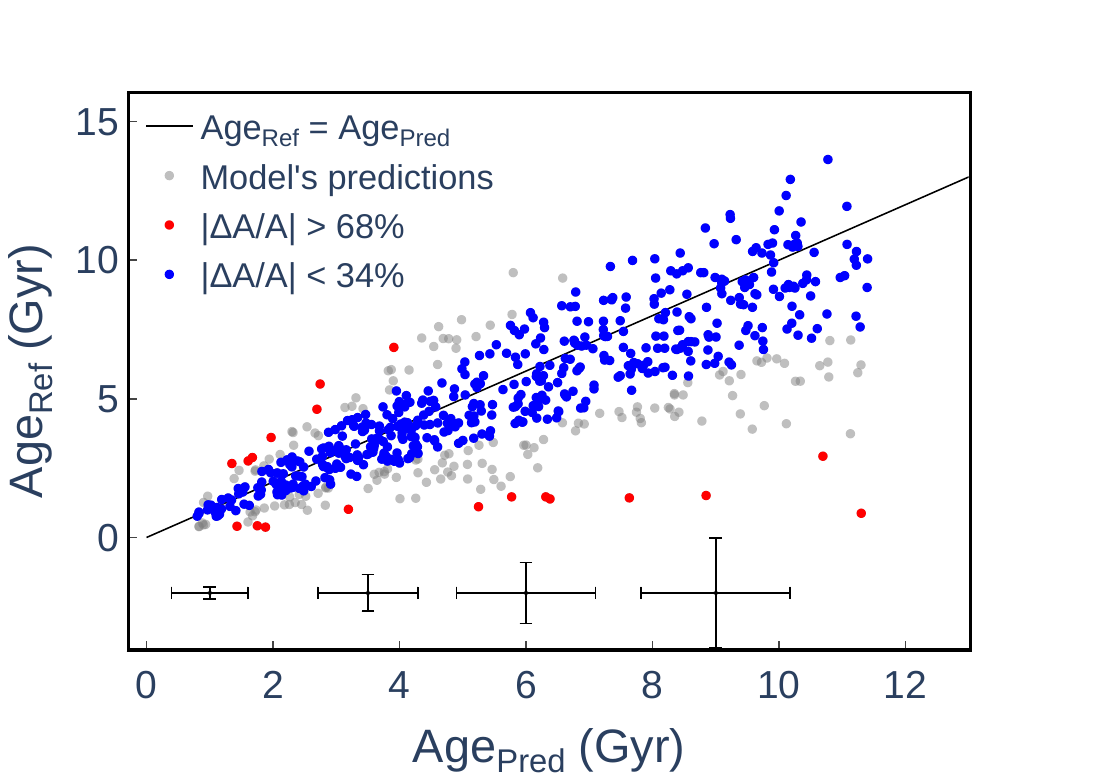}
         \caption{}
         \label{fig:Map_Relative_Residual-Outliers}
     \end{subfigure}
     \hfill
     \begin{subfigure}[b]{0.45\textwidth}
         \centering
         \includegraphics[width=\textwidth]{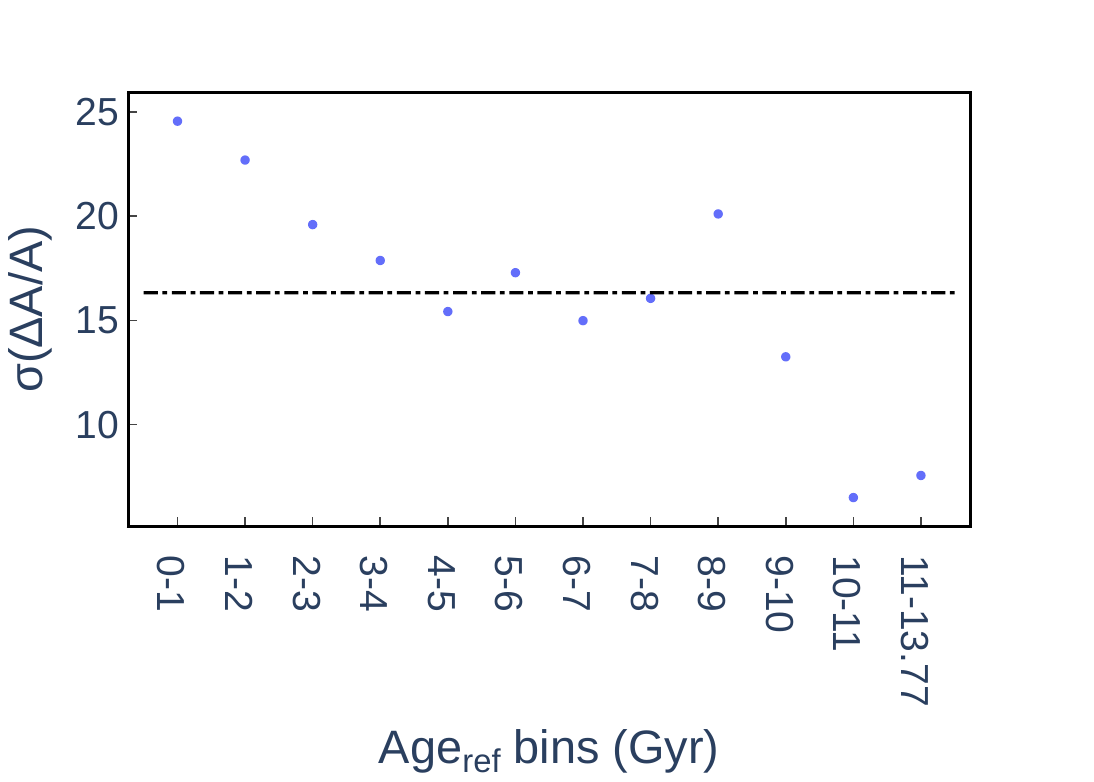}
         \caption{}\label{fig:Dispersion_in_residuals_per_bin}
     \end{subfigure}
     \hfill
     \begin{subfigure}[b]{0.45\textwidth}
         \centering
         \includegraphics[width=\textwidth]{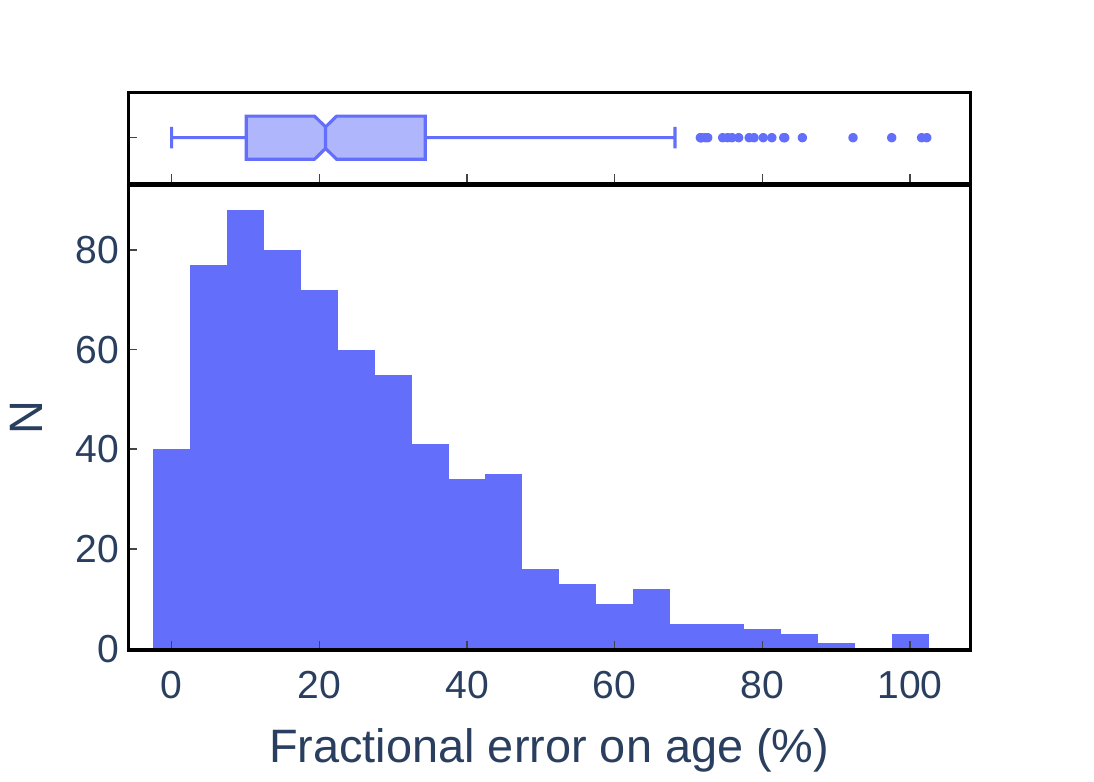}
         \caption{}\label{fig:Histogram_Relative_Residuals}
     \end{subfigure}
     \hfill
     \begin{subfigure}[b]{0.45\textwidth}
         \centering
         \includegraphics[width=\textwidth]{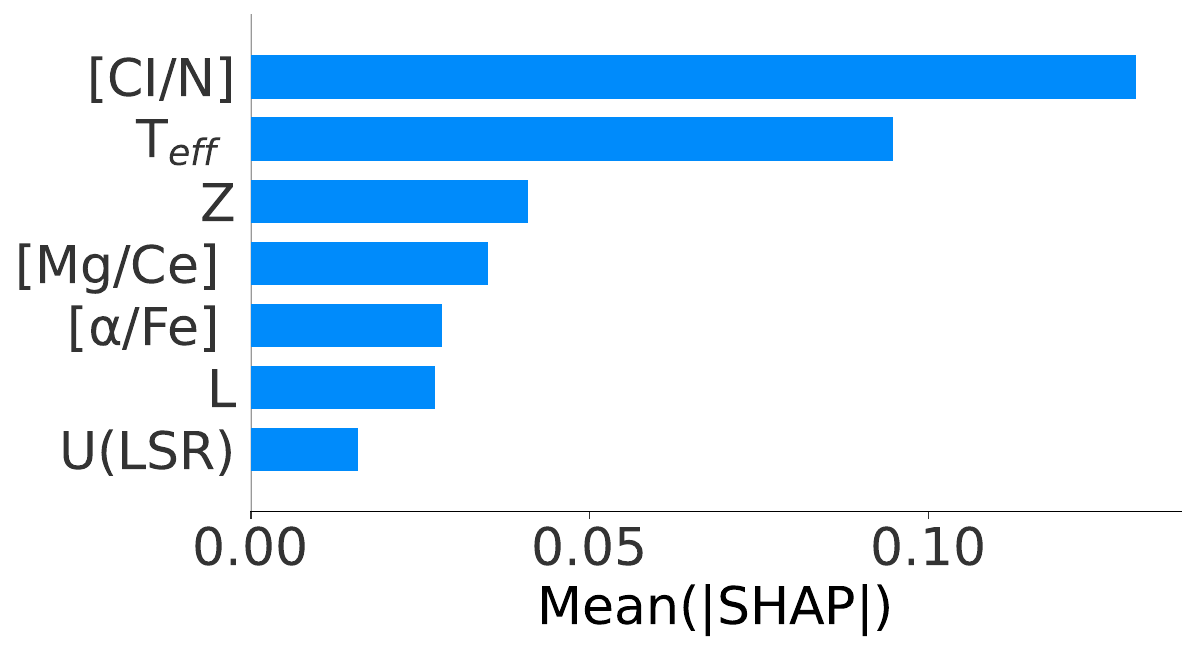}
         \caption{}
         \label{fig:Shap-values}
     \end{subfigure}
     \hfill
     \begin{subfigure}[b]{0.42\textwidth}
         \centering
         \includegraphics[width=\textwidth]{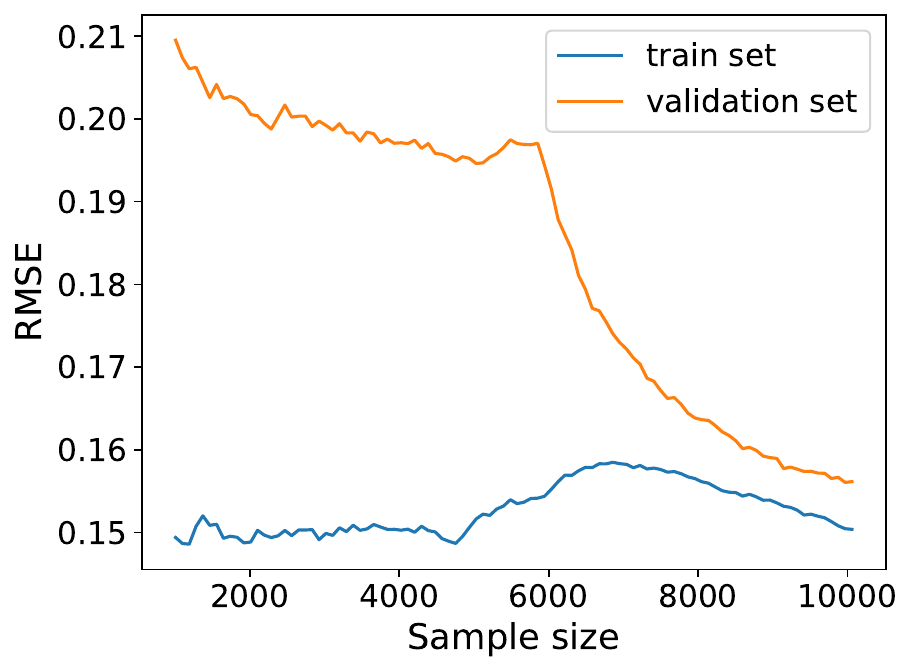}
         \caption{}
         \label{fig:Learning_Curve}
     \end{subfigure}
     \hfill
        \caption{Ensemble of plots summarising the final performances on the MCK-APOKASC training-test sample. Panel \ref{fig:Median_Frac_Residual_per_bin}: Evolution of the median of the absolute value for the fractional error on age per bin. Panel \ref{fig:Map_Relative_Residual-Outliers}: Comparison of the ages of reference and those predicted by the model. The black line is the identity function. |$\Delta$A/A| is the absolute fractional error on age. The black error bars represent the means of the errors in age. Panel \ref{fig:Dispersion_in_residuals_per_bin}: Evolution of the standard deviation of the fractional error on age per bin. The vertical-horizontal bar depicts the mean value for the whole age range. Panel \ref{fig:Histogram_Relative_Residuals}: Histogram of the absolute fractional error on age. Panel \ref{fig:Shap-values}: Bar plot of the feature importance for all the retained features. Panel \ref{fig:Learning_Curve}: Plot of the learning curves.}
        \label{fig:Summarising_Model_Performances}
\end{figure*}

\begin{figure*}[ht]
     \centering
     \begin{subfigure}[b]{0.46\textwidth}
         \centering
         \includegraphics[width=\textwidth]{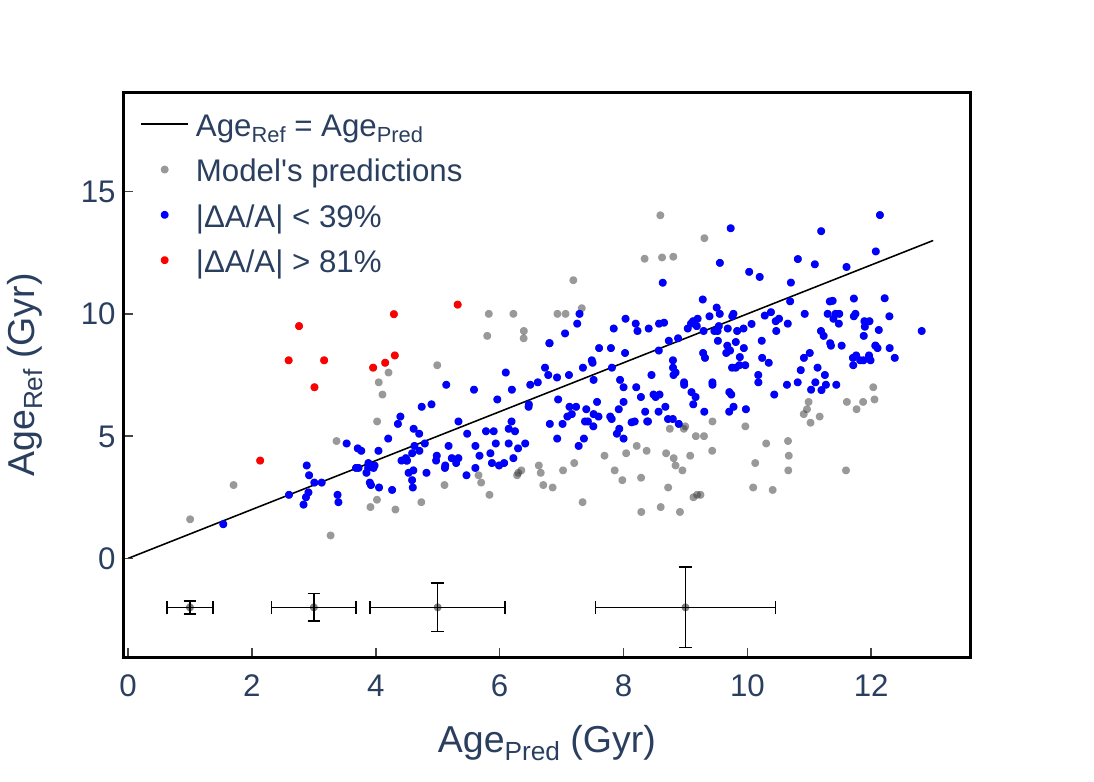}
         \caption{}\label{fig:Map-Residuals-K2-GALAH}
     \end{subfigure}
     \hfill
     \begin{subfigure}[b]{0.46\textwidth}
         \centering
         \includegraphics[width=\textwidth]{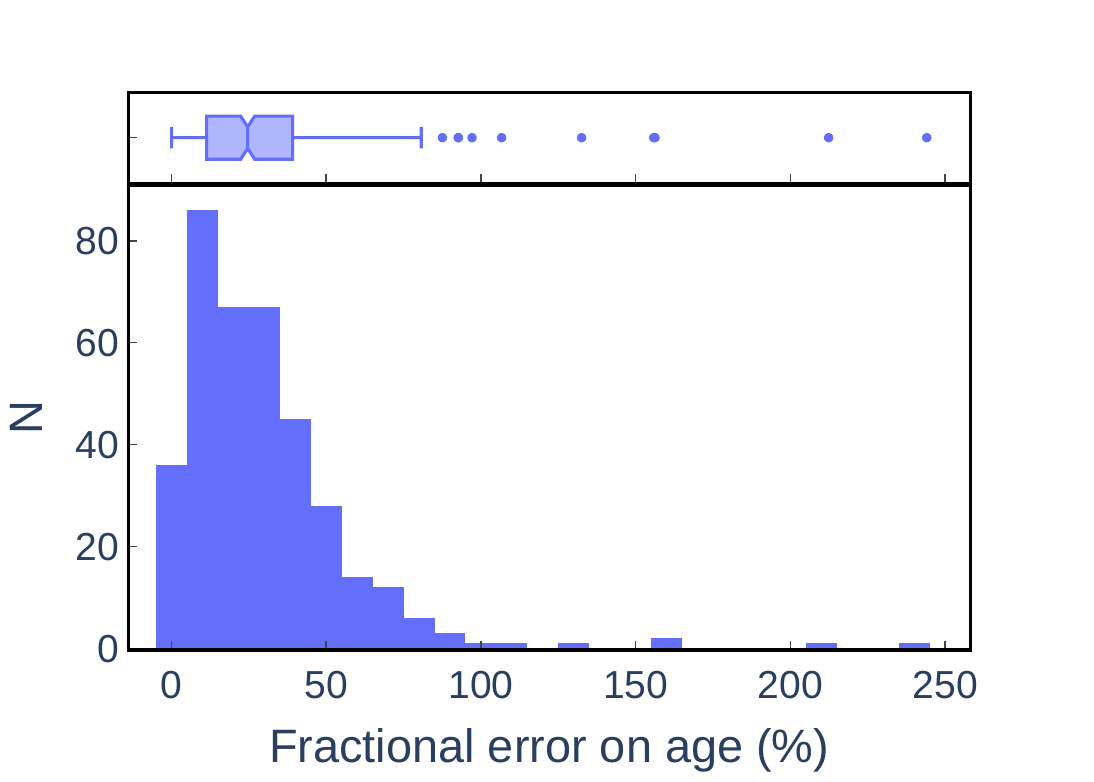}
         \caption{}
         \label{fig:Histogram-Residuals-K2-values}
     \end{subfigure}
     \hfill
        \caption{Ensemble of plots summarising the performance of the model on the K2-GALAH sample. Panel \ref{fig:Map-Residuals-K2-GALAH}: Plot of the scatter between the predicted ages and reference ages. The black line serves to visualise the residuals. Panel \ref{fig:Histogram-Residuals-K2-values}: Histogram of the absolute residuals between the predicted ages and the reference ages in absolute value.}
        \label{fig:Summarising_Model_Performances_Unseen_Data}
\end{figure*}

\subsection{Unreliable [Ce/Fe] abundances}\label{subsect:Unreliable [Ce/Fe] abundances}

The research by \cite{Casali-2023} using APOGEE DR17-TESS-\textit{Kepler}-K2 data suggests that combining cerium with $\alpha$-elements is a promising proxy for understanding star formation. However, it also suggests that uncertainties, especially in [Ce/Fe], are likely underestimated. A comparative study by \cite{Contursi-2023} with \textit{Gaia} DR3 \citep{GAIA-DR3-Citation-1,GAIA-DR3-Citation-2}, Forsberg's catalogue \citep{Forsberg-2019}, GALAH DR3 \citep{Buder-2021}, and APOGEE DR17 reveals improved agreement by rejecting low [Ce/Fe] values.

Based on these studies, criteria were implemented: excluding [Ce/Fe] with uncertainties greater than 0.2 dex, removing flagged abundances, and establishing a threshold at [Ce/Fe] = - 0.46 dex based on the median minus 1.5 times the interquartile range ($M - 1.5 \cdot IQR$). The unreliable values of [Ce/Fe] account for 2.6\% of the dataset.\\
No threshold was applied to the highest values of cerium (0.5 $<$ [Ce/Fe] (dex) $<$ 1.4) for two reasons. Firstly, they are expected from the study of \cite{Contursi-2023}, as these values have been observed in Baryum stars, which are known to have higher levels of barium, cerium, zirconium, ytterbium, and lanthanum \citep{Ba-Stars-2016,Ba-Stars-2019}. Secondly, excluding these values from the training sample leads to a model with poorer accuracy performances. Rejecting unreliable cerium abundances improves the performance of the model, particularly reducing standard deviation in fractional age error in the range 11-13.77 Gyr.

%%%%%%%%%%%%%%%%%%%%%%%%%%%%%%%%%%%%%%%%%%%%%%%%%%%%%%%%%%%%%%%%%%%%%%%%%%%%%%%%%%%%%%%%%%%%%%%%%%%%%%%%%%%%

\section{Results}\label{sect:Results}

\subsection{Final performances on the test set}\label{sect:Final performance test set}

As a result of the full optimisation, the performance of the model reveals an overall decline in the median fractional error on age per age bin up to 11 Gyr (refer to Figure \ref{fig:Median_Frac_Residual_per_bin}). The median fractional error is approximately 20\% in the age range (3 < Age(Gyr) < 8), while it is around 10\% in the range (8 < Age(Gyr) < 10). For the range (10 < Age(Gyr) < 11), the median fractional error decreases further to approximately 7\%. Furthermore, the oldest stars exhibit a median fractional error of 23\%.\\ Regarding the youngest stars, the age range (1 < Age(Gyr) < 3) corresponds to a fractional error of approximately 27\%, while the range (0 < Age(Gyr) < 1) exhibits a fractional error of approximately 43\%. Despite their higher fractional error, the predictions for these very young cases are considered as accurate as the older cases due to their lower age values.\\ There is a consistent decline in the standard deviation of the fractional error per age bin (refer to Figure \ref{fig:Dispersion_in_residuals_per_bin}). This indicates that the predictions become increasingly robust as one deals with increasing age values. Among all the stars analysed, there are only two instances where the fractional errors slightly surpass 100\%, as depicted in Figure \ref{fig:Histogram_Relative_Residuals}. In Figure \ref{fig:Map_Relative_Residual-Outliers}, the blue points represent cases with an absolute fractional age error of lower than the third quartile of the distribution. Conversely, the red points lie outside the range of the box plot of Figure \ref{fig:Histogram_Relative_Residuals} and are consequently categorised as statistical outliers. It is important to bear in mind that the fractional error for a given age serves as a measure of the proportion by which the predicted age should be adjusted (either increased or decreased) to align with the reference age.

An increased amount of data generally improves model performance until a ceiling is reached; this threshold is determined by the quality of available information. Learning curves, comparing training and validation performance, help identify this ceiling. In this study, using RMSE as the scoring metric, the small gap observed between the learning curves depicts a relative difference of 4.77\%, also known as the variance of the model (Figure \ref{fig:Learning_Curve}). This suggests high performance on the MCK-APOKASC sample. Consequently, the \texttt{CatBoostRegressor} model exhibits sufficient quality for a reliable application to the APOGEE Main Red Star Sample (refer to Section \ref{sect:Discussion of the APOGEE age map}).

\begin{figure*}[htb]
       \centering
       \includegraphics[width=1\linewidth]{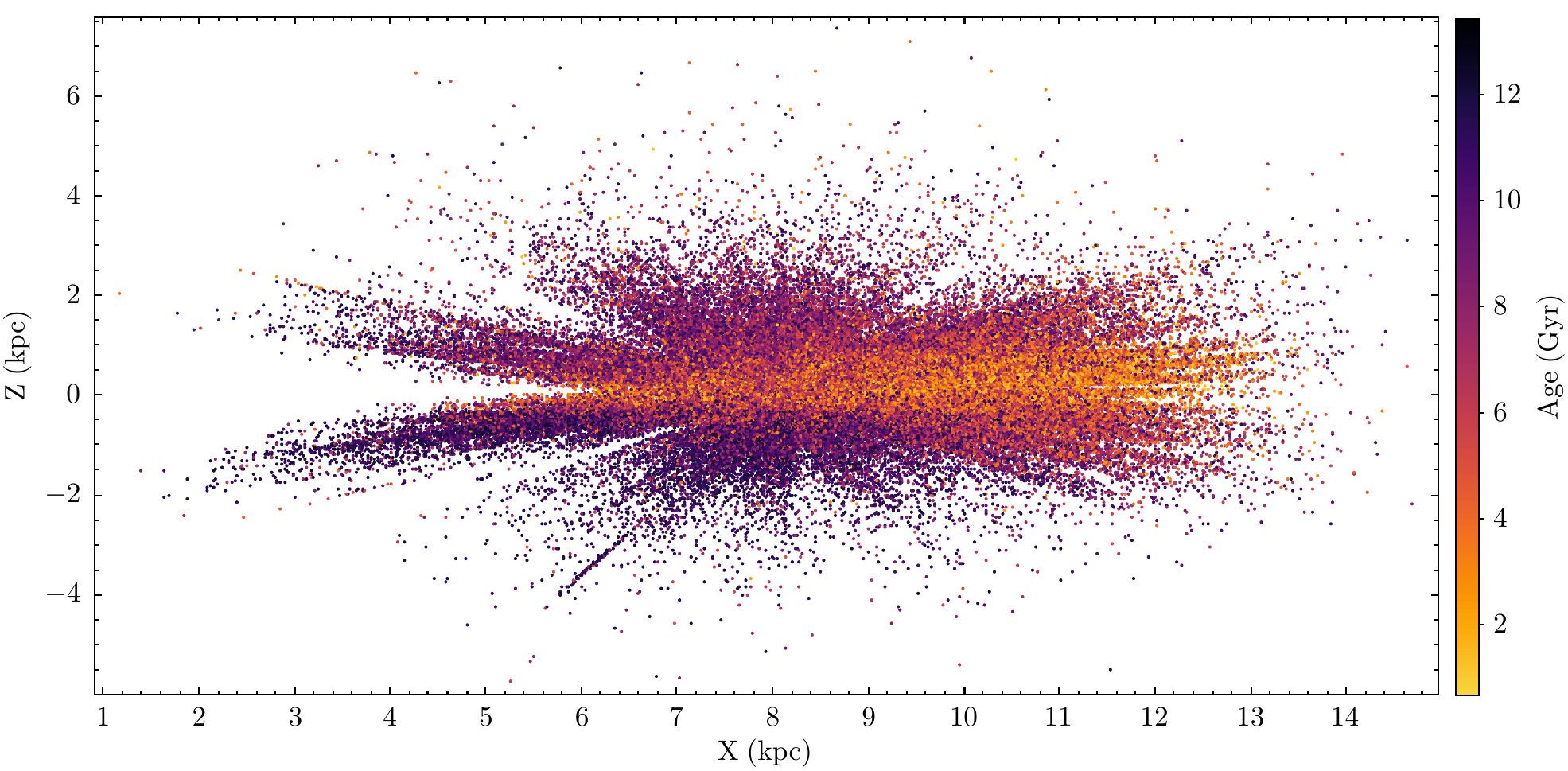}
       \caption{APOGEE age map for the sample of red giants computed with CatBoost.}
    \label{fig:APOGEE-MAP-Low-Resolution}
\end{figure*}

\subsection{Performance on an independent set}\label{sect: Additional test set}

Given the inclusion of the MCK sample in the training set to address the data shift (refer to Section \ref{subsect:Identification of a data shift}), we performed an extra evaluation to gauge the model's capability to generalise to new, independent data testing on a stellar age sample from \cite{Zinn-2022} (K2-GALAH, hereafter). This catalogue is made of red giants enriched with asteroseismic parameters derived from various sources, including anterior K2 Galactic Archaeology Program data \citep{Stello-2017,Zinn-2019} and APOGEE DR16 \citep{APOGEE-DR16} spectroscopic data for calibration. Their ages were computed using asteroseismic masses, GALAH DR3 temperatures \citep{Buder-2021}, and stellar modelling with the code BSTEP \citep{BSTEP-2018}. GALAH data were used for age-abundance analysis.

To ensure a fair evaluation of the model on the K2-GALAH catalogue, only stars with fractional age uncertainties of lower than or equal to 30\% were sampled. Also, stars with fractional luminosity uncertainties of greater than 30\% were excluded. Consequently, the K2-GALAH testing sample comprises 371 stars (refer to Figure \ref{fig:Location-Merged-Sample}).

Figure \ref{fig:Summarising_Model_Performances_Unseen_Data} displays residual distribution plots for the K2-GALAH test set. Lower and upper thresholds of absolute fractional residuals correspond to the third quartile and the upper edge in the box plot of Figure \ref{fig:Histogram-Residuals-K2-values}. Error bars represent mean errors in ages for each age bin, except the last one, where bars represent mean errors for stars with predicted ages of between 6 and 13 Gyr.%\LEt{}

Given the smaller size of the K2-GALAH test set (371 stars vs. 653 stars in MCK-APOKASC), the Wilcoxon-Mann-Whitney (WMW) test was chosen over the KS test. The reason behind this choice is that the KS test is recognised as being potentially unreliable when sample sizes are significantly different, as it relies upon the comparison of the empirical cumulative distribution function of the two samples. The WMW test yielded the following test statistic and p-value: ($\rho$ = 68820.5, p = 1.000). Therefore, insufficient evidence exists to conclude differences in the distribution of fractional age errors between the two samples. Consequently, contrary to results in Section \ref{subsect:Identification of a data shift}, no significant difference in model performance is observed when applied to independent data. This consistency can be explained by the training data effectively capturing key stellar parameters of the distribution in the reference Main Red Star sample from APOGEE DR17 (for more details refer to Section \ref{sect:Discussion}).

%%%%%%%%%%%%%%%%%%%%%%%%%%%%%%%%%%%%%%%%%%%%%%%%%%%%%%%%%%%%%%%%%5%%%%%%%%%%%%%%%%%%%%%%%%%%%%%%%%%%%%%%%%%%%%%%%%%%%%%%%%%%%%%%%%%%%%%%%%%

\subsection{The age map}\label{sect:Discussion of the APOGEE age map}

The APOGEE Main Red Star Sample (MRS) comprises 372,000 stars randomly selected from the Two Micron All Sky Survey (2MASS) photometric catalogue \citep{2MASS}. The MRS was designed to select red giants based on colour-magnitude criteria made to provide a clear set of rules for a robust selection function reconstruction.

To ensure the relevance of the MRS in Galactic archaeology, kinematic data are crucial. Obtaining such information involves a sequential process: first, cross-referencing the MRS with the \textit{Gaia} DR3 catalogue, followed by excluding stars with negative parallax or fractional parallax error of greater than 20\%, as described in Section \ref{Sec:TZL}. The refined sample, named MRS-\textit{Gaia}, contains 283,196 stars, ensuring reliable astrometric information for deriving kinematic parameters.

Given the potential contamination from undesired targets due to colour--magnitude criteria, we inspected the MRS log(g) histogram (refer to Fig \ref{fig:Histogram-log(g)-MRS-Gaia}, in Appendix \ref{Appendix:Extra statiscal plots}). The analysis reveals a bimodal distribution, indicating contamination with main sequence stars, which make up  42\%.

To handle this issue, the stars with log(g) values surpassing 3.7 dex are excluded. This threshold is selected because it marks the shift from the declining trend of the initial component to the rising trend of the second component. Considering that a log(g) of around 3.5 dex is the theoretical upper boundary for red giants, the selection of the 3.7 dex threshold matches the usual 0.1 dex uncertainty associated with log(g) determination using spectroscopic methods. As a result of this refinement, the MRS-\textit{Gaia} sample size is subsequently reduced to 176,516 stars.

When applying a \texttt{CatBoostRegressor}, the reliability of the model is generally higher when used on data with values similar to those in the initial training set. Therefore, the MRS-\textit{Gaia} sample is restricted to values seen during the training phase, except for the variable `Z', which captures the vertical age gradient of the Galactic disc. Without the need for restrictions on Z, the model has effectively captured the Z trend with age previously found in \cite{Ness-2018} and \cite{Anders-XGBoost}, as demonstrated in Appendix \ref{Appendix:Age-map extra plots}.

Table \ref{tab:Restriction-Conditions-Age-Dating} illustrates each range of values for which a restriction in the application of the model was necessary. Given that CatBoost assumes the accuracy of input data and the representativeness of training data for the underlying distribution, age calculations were confined to stars with a luminosity uncertainty of lower than 30\%  to mitigate the potential introduction of machine-learning artefacts. Additionally, stars displaying flagged abundances in the features [CI/N], [Mg/Ce], and [$\alpha$/Fe] were excluded from the age-determination process. Moreover, in accordance with the observations detailed in Section \ref{subsect:Unreliable [Ce/Fe] abundances}, only stars meeting the criteria of [Ce/Fe] surpassing -0.46 dex and having errors below 0.2 dex were considered for age determination.

Ultimately, out of the 176,516 stars in the MRS-\textit{Gaia} sample, 125,445 stars were selected for age determination. In summary, 51,071 could not be dated mainly because of their unreliable abundances and luminosities. Additionally, but to a lesser extent, this is also due to their parameter values not being encountered during the training phase.
The age map associated with these data is displayed in Figure \ref{fig:APOGEE-MAP-Low-Resolution}.

\begin{table}[ht]
\centering
\begin{tabular}{@{}cccc@{}}
\toprule
                               & Min   & Max   & Units      \\ \midrule
{[}CI/N{]} & -2.13 & 0.68  & dex        \\ \midrule
{[}Mg/Ce{]}                    & -0.50  & 0.88  & dex        \\ \midrule
{[}$\alpha$/Fe{]} & -0.10 & 0.36  & dex        \\ \midrule
T$_{\texttt{eff}}$                           & 3968  & 5324  & K          \\ \midrule

L                    & 2.39   & 631 & $\text{L}_{\sun}$ \\ \bottomrule
\end{tabular}
\caption{Range of values used for age computation within the MRS-\textit{Gaia} sample.}
\label{tab:Restriction-Conditions-Age-Dating}
\end{table}

%%%%%%%%%%%%%%%%%%%%%%%%%%%%%%%%%%%%%%%%%%%%%%%%%%%%%%%%%%%%%%%%%%%%%%%%%%%%%%%%%%%
\section{Discussion}\label{sect:Discussion}

\subsection{Completeness of the training set}\label{Sect:Completeness of the training set}

In section \ref{sect: Additional test set}, we demonstrate that the CatBoost model trained on MCK-APOKASC is able to make predictions that extend effectively to stars with reliable ages in the K2 Galactic program, without a decrease in performance. Nevertheless, to assess the potential limitation of the model when predicting stellar ages in wider and future surveys, it is crucial to discuss whether the APOGEE stars in the training set fairly sample the underlying distribution of targets in the MRS-\textit{Gaia} sample. Moreover, as MRS-\textit{Gaia} is a refinement of the MRS sample, it is also crucial to discuss whether MRS accurately reflects the broader population of red giants in the Galaxy.

Given the disparity in the size of the MCK-APOKASC and MRS-\textit{Gaia} samples, the non-normally distributed nature of the samples, and the fact they are not independent of each other, the application of semi-parametric (KS-test), non-parametric (MWM test), or parametric statistical tests to compare them is precluded. In situations where statistical tests cannot be employed, a conventional strategy involves employing box plots to scrutinise the distributional properties and draw comparisons between the two datasets.

These tests were conducted for the main stellar parameters ([Fe/H], T$_{\texttt{eff}}$ , and log(g)), the chemical dichotomy feature [$\alpha$/Fe], and the guiding radius; this latter is a proxy for the stellar birth radius. In Appendix \ref{Appendix:Extra statiscal plots}, the comparison of the box plots indicates that the medians of the two distributions are nearly identical, or that they differ by an amount smaller than the typical uncertainty associated with each respective feature. As an example, the temperature difference is less than 100 K and the differences in log(g) and abundance are less than 0.1 dex.\\
As anticipated, the interquartile ranges (IQRs) are smaller for almost all the tested features in the MCK-APOKASC sample, except for [Fe/H], which displays an IQR that is almost identical to that of the MRS-\textit{Gaia} sample. Notably, the upper fences of each box plot consistently encompass the IQR of the larger MRS-\textit{Gaia} sample. Consequently, one can confidently conclude that the training sample fairly underlines the MRS-\textit{Gaia} sample.

In order to determine whether or not the Main Red Star sample effectively represents the population of red giants in the Galactic disc, we conducted a review of the APOGEE literature. Within the APOGEE framework, the accurate computation of selection biases is feasible for samples chosen in a genuinely random manner. This is due to the fact that only under such conditions can the sample selection function be reconstructed \citep{Bovy-2016-b}. The stars satisfying this criterion are referred to as the Main Red Star sample. Notably for previous APOGEE data releases, \cite{Nandakumar-2017} showed that there is a negligible selection function effect on the metallicity distribution function (MDF) and the vertical metallicity gradients for APOGEE, RAVE \citep{RAVE-2006}, and LAMOST \citep{LAMOST-2012} using two stellar population synthesis models. This outcome suggests that it is feasible to combine data from different surveys when studying the MDF in common fields.\\
The selection function and completeness of the APOGEE DR17 Main Red Star sample are discussed in an article in preparation by members of the SDSS-IV collaboration (J. Imig et al. in prep). Consequently, one cannot currently address the potential limitations of the Main Red Star sample in fairly underlying the population of red giants in the Galactic disc.\\
Nevertheless, the APOGEE documentation online has already provided an analysis of previous selection functions relying on the Python code \textit{apogee}, which are described in \cite{Bovy-2016} and in full detail in the associated documentation online \footnote{https://github.com/jobovy/apogee}. These analyses revealed that APOGEE has covered an increasingly large portion of the sky, with a far higher selection fraction in many fields of the Main Red Star sample between DR12 and DR16. Notably, it is already known from APOGEE documentation that the number of stars passed from 357,167 in DR16 to 372,458 in DR17. Finally, APOGEE has probed the vastest number of red giants for a great fraction of the sky in both the Northern and Southern Hemisphere \citep{APOGEE-DR17}.

%%%%%%%%%%%%%%%%%%%%%%%%%%%%%%%%%%%%%%%%%%%%%%%%%%%%%%%%%%%%%%%%%%%%%%%%%%%%%%%%%%%%%%%%%%%%%%%%%%%%%%%%%%%%%%%%%%%%%%%%%%%%%%%%
\subsection{Comparison with other age maps}\label{Sect:Comparison with other age maps}

In the pioneering work of \cite{Ness-2018}, age labels were provided for 73,180 red giant stars. The associated mean fractional error on age was reported to be 40\%. The age map obtained with the CatBoost model is presented similarly to Figure 14 in \cite{Ness-2018}. Hereafter, we refer to this age map as the Ness map.

Similarly, \cite{Anders-XGBoost} contributed a catalogue of 178,825 red giants from APOGEE DR17. Their training sample consists of stars exclusively from the \textit{Kepler} field (3,060 stars) with ages sourced from \cite{Miglio-2021}. Achieving a median statistical uncertainty of 17\% with an \texttt{XGBoostRegressor}, \cite{Anders-XGBoost} conducted validation plots that reproduced expected trends in chemistry, position, and kinematics with age. However, evaluating potential overfitting and underfitting, or assessing the bias and the variance of the model, is challenging due to the absence of learning curves in their study.

As discussed in Section \ref{sect:Selection of a machine learning model}, using the APOKASC-MCK training dataset for a \texttt{CatBoostRegressor} model results in more stable predictions (i.e. lower variance of the model) compared to training with an \texttt{XGBoostRegressor} model. Additionally, it is important to highlight that the dataset used in this study offers relevant advantages over the datasets used in the studies by \cite{Ness-2018} and \cite{Anders-XGBoost}. For example, it includes more stars from the \textit{Kepler} field (APOKASC-2 and APOGEE-\textit{Kepler}) and also incorporates stars from the MCK catalogue. The inclusion of MCK stars is crucial, as explained in Section \ref{subsect:Identification of a data shift}, because relying solely on data from the APOKASC catalogue for training leads to a significant drop in model performance when applied to stars from the MCK catalogue.

The enhancement in machine learning model performance is inherently linked to the amount of available data. This improvement can be attributed to several factors: a larger dataset provides more examples for learning, reducing the risk of overfitting. It better represents the diversity of cases, mitigating potential bias and allowing for the use of more complex models without overfitting concerns. Additionally, a more extensive dataset reduces variance, resulting in more stable predictions while minimising statistical uncertainty. Consequently, the model can adjust its parameters more robustly and reliably.

\begin{figure*}[ht]
     \centering
     \includegraphics[width=1.0\linewidth]{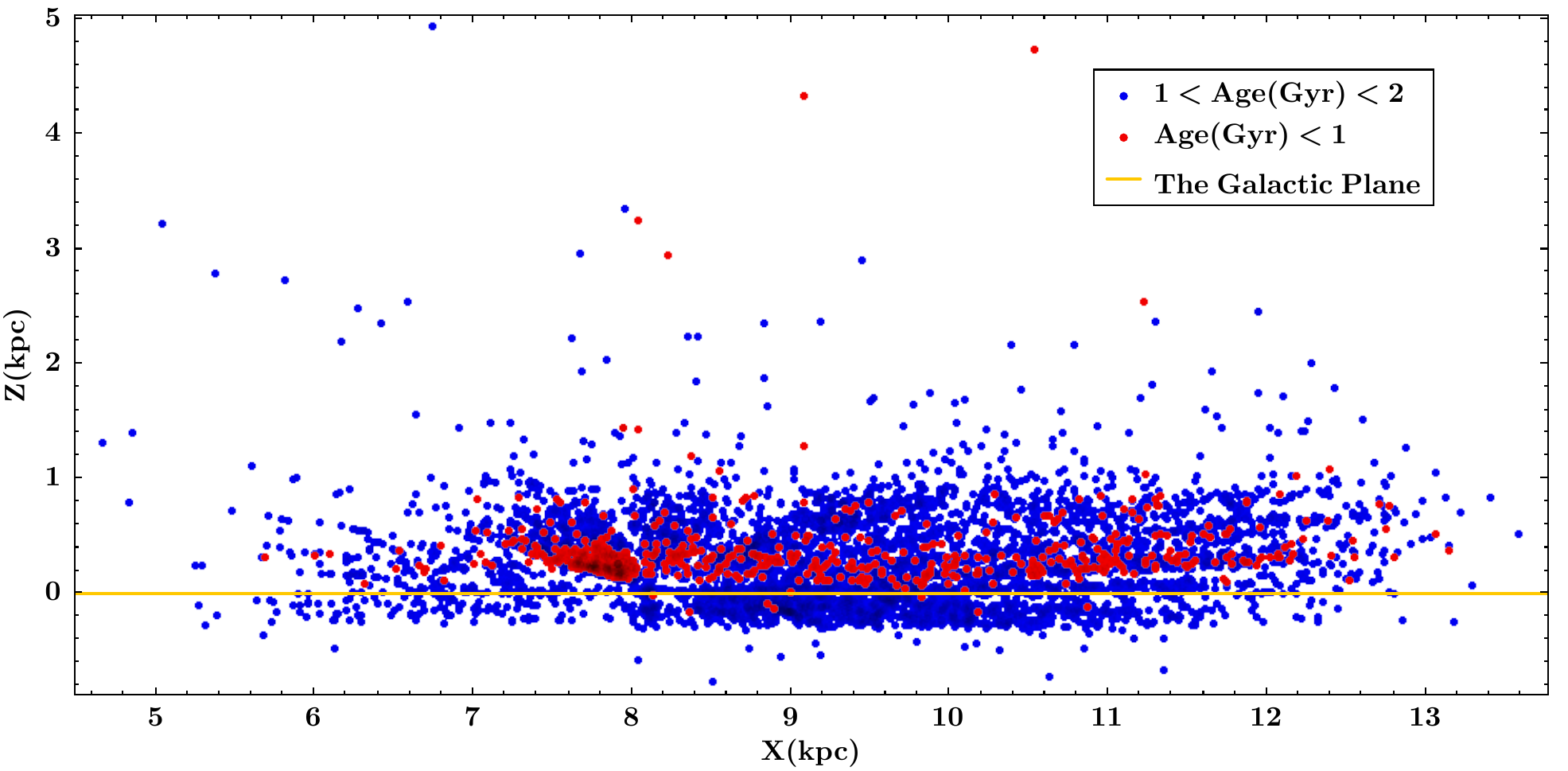}
     \caption{Plot of the youngest stars in the (X, Z) plane within the APOGEE map.}
     \label{fig:Young-Age-Gradient}
\end{figure*}

Comparing the age map presented in this article (refer to Figure \ref{fig:APOGEE-MAP-Low-Resolution}) with those of Ness and Anders reveals several differences. 
\begin{itemize}
    \item  The map presented here spans a more comprehensive age range than the Ness map but a similar age range to the Anders map. 
    \item The stars in our map reach a greater vertical extension (-6 < Z(kpc) < 7) than in the maps of Ness and Anders, but a smaller extension along the Sun--Galactic centre axis (1.2 < X(kpc) < 14.7).
    \item Our age map fills the gap of data found in the Ness map in the region (Z < -3 kpc, 0 < X(kpc) < 6) and that in the region  ( Z $\approx$ - 2 kpc, 4 < X(kpc) < 6) for the map of Anders.
\end{itemize}
However, some similarities are also apparent.
\begin{itemize}
\item Our map shows the flaring of the young Galactic disc (Age < 6 Gyr), as already outlined in the maps of Ness and Anders (refer to Figure \ref{fig:APOGEE-MAP-Low-Resolution}). 
\item The youngest stars (Age $\le$ 2 Gyr) are mostly found close to the Galactic plane (Z = 0 kpc) (refer to Figure \ref{fig:Young-Age-Gradient}), as also revealed by the Ness and Anders maps. 
\item   The expected gap of data within the Galactic plane towards the bright Galactic centre for X < 6 kpc is also seen in all three maps. This gap prevents us from gaining insight into the age distribution close to the Galactic plane for this inner part of the Galaxy.\\
\end{itemize} 

\begin{figure*}[ht]
     \centering
         \includegraphics[width=1\linewidth]{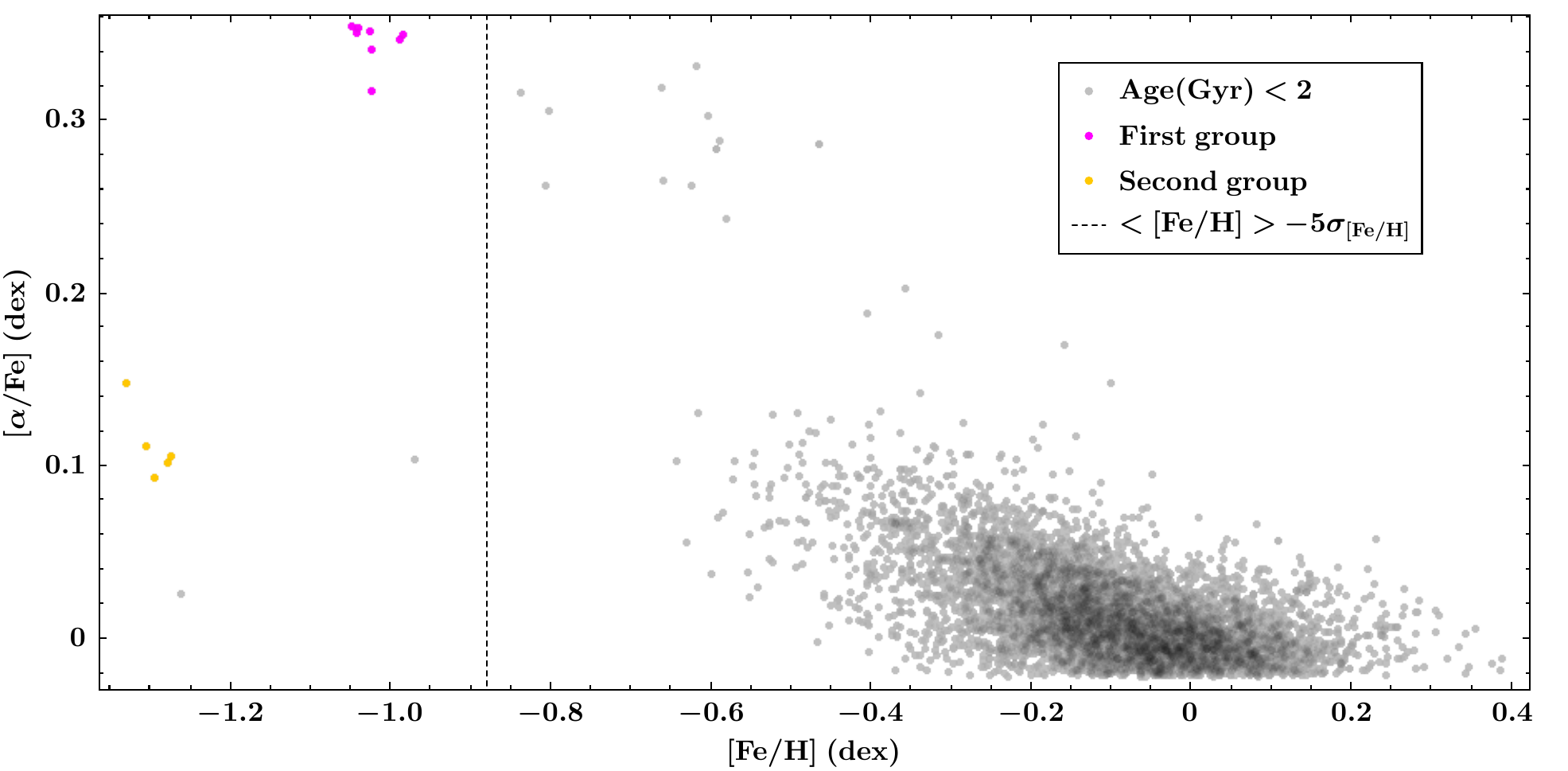}
         \caption{Plot of [$\alpha$/Fe] against [Fe/H] for the stars younger than 2 Gyr.}
         \label{fig:APOGEE-Young-Alpha-Fe}
\end{figure*}
\begin{figure*}[ht]
     \centering
        \includegraphics[width=0.9\linewidth]{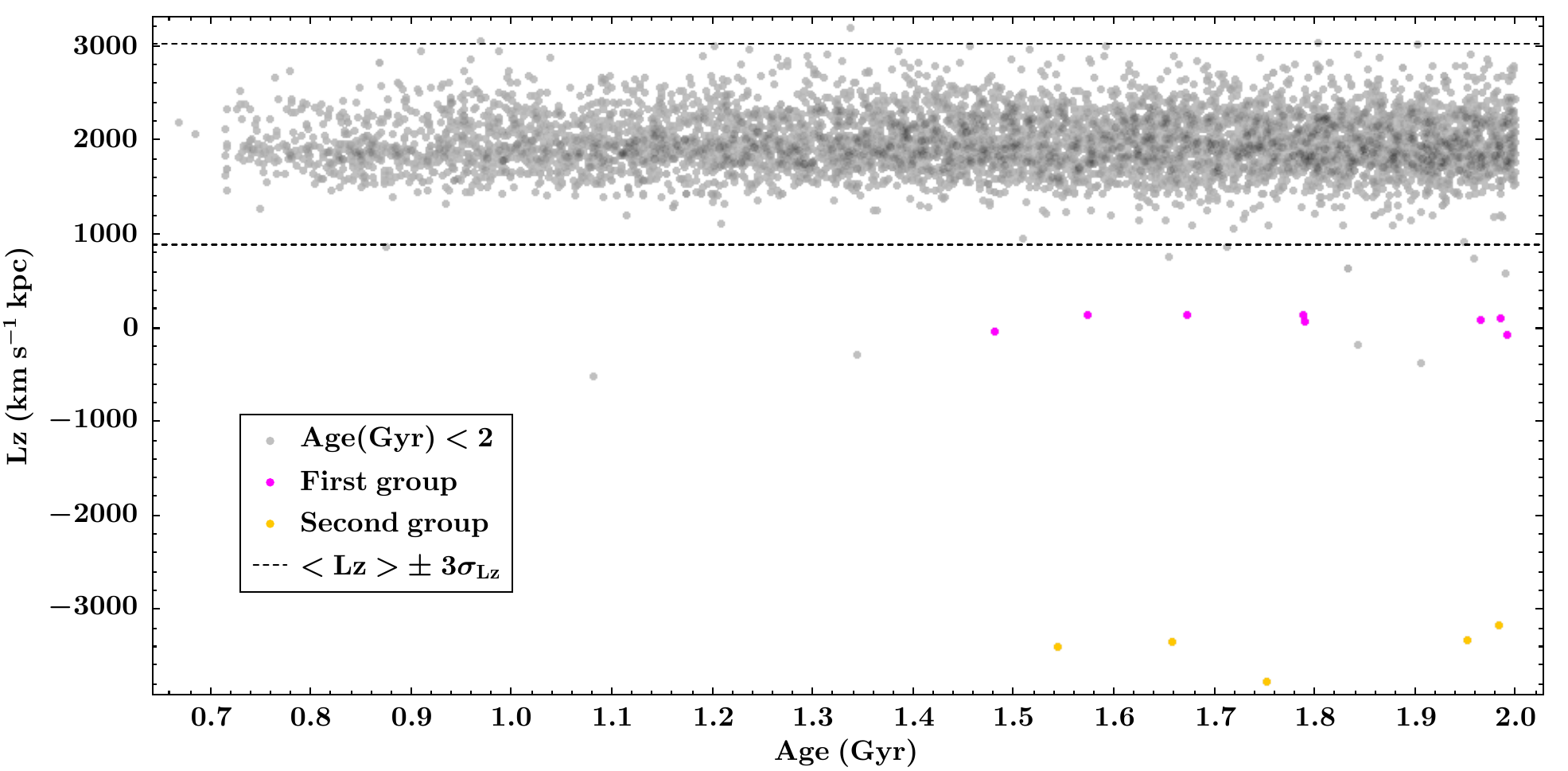}
        \caption{Plot of the vertical angular momentum against the stellar age for the stars younger than 2 Gyr.}
        \label{fig: L$_{z}$-Age-Structure}
\end{figure*}

\subsection{Unprecedented features in APOGEE age maps}\label{Sect:New Features Unveiled in APOGEE Age Map Analysis}

The analysis of the APOGEE age map reveals new features not previously seen in other maps.
The level of age resolution contributes to revealing the theoretically expected age gradient among the youngest stars within the Galactic plane (refer to Figure \ref{fig:Young-Age-Gradient}). In order to unveil this gradient,  the youngest stars were divided into two groups. The first group is made of stars younger than one billion years and the second contains stars of between one billion and two billion years old. As an example, while both groups share approximately the same median vertical height above the Galactic plane (m $\approx$ 0.30 kpc), the IQR of the youngest group (IQR $\approx$ 0.17) is significantly lower than that of the older group (IQR $\approx$ 0.47). This matches the findings of \cite{Mackereth-2019} and the simulation results from \cite{Martig-2014}, where in periods of no gas accretion, new stars are born within the Galactic plane and are later kinematically heated. The kinematical heating is thought to be mainly due to disc growth with a combination of spiral arms and bars coupled with overdensities in the disc and vertical bending waves \citep{Martig-2014,Aumer-2016}. 

Additionally, our results reveal that the newborn stars (Age < 1 Gyr) in the solar neighbourhood (7.5 $<$ X(kpc) $<$ 8.5) are clustered along a plane shifted approximately 300 pc from the Galactic plane. In contrast, beyond the solar neighbourhood, these stars are more scattered and there is a non-negligible number of them closer to the Galactic plane. Also, one notices a few newborn stars on the south side of the Galactic plane. In the solar neighbourhood, the resulting distribution of these newborn stars is expected from the analysis of the Local Bubble. The Local Bubble is a known zone of low gas density in which \cite{Local_Bubble} found that the expansion of its surface has driven the star formation near the Sun. These authors found that almost all star-forming complexes within a 200 pc radius from the Sun are situated along the surface boundary of the Local Bubble and have experienced an outward expansion mainly orthogonal to the surface.

Two young metal-poor groups of stars were identified by analysing their stellar parameters. They have an age estimate of younger than 2 Gyr, matching the sparsely populated tail of the Main Red Star sample metallicity distribution ([Fe/H] $\lesssim$ -1 dex) (refer to Figure \ref{fig:Box-Plot-Metallicity}, in Appendix \ref{Appendix:Extra statiscal plots}). These stars cluster in two distinct groups of abundance and stand out at a significance level exceeding 5$\sigma$ from the mean [Fe/H] in the [$\alpha$/Fe] versus [Fe/H] plane (refer to Figure \ref{fig:APOGEE-Young-Alpha-Fe}). Interestingly, they display orbital eccentricities of greater than 0.79. Such high values, for associated low [Fe/H], are known indicators of stars with halo kinematics \citep{Reddy-2006}. We confirmed this possibility using the probabilistic kinematic stellar component technique described in \citep{Bensby-et-al-2003,Bensby-2005}. The results, summarised in Appendix \ref{Appendix: Peculiar young stars}, namely in Table \ref{tab:Appendix-Kin-First-Table} and Table \ref{tab:Kin-Second-Table}, reveal that every star within these groups exhibits a probability of belonging to the halo of greater than 97\%.

Subsequently,  the dynamics of these groups was examined with the distribution of vertical angular momentum in relation to age, denoted as L$_{z}$ (refer to Figure \ref{fig: L$_{z}$-Age-Structure}). This distribution unveils bulk L$_{z}$ values, from which these stars emerge as notable outliers.\\
In each group, stars show vertical angular momentum close to the group mean < L$_{z}$>. Both group means deviate from the bulk mean by more than 3$\sigma$, indicating the emergence of two distinct kinematic groups. This notable difference in kinematical properties is likely attributed to the distinct gravitational potentials experienced by the stars in these widely separated regions (approximately 6 kpc apart). Indeed the first group is situated in the range (6.04 < X(kpc) < 6.34, -0.33 < Y(kpc) < -0.28,  0.53 < Z(kpc) < 0.62), while the second one occupies the range (7.24 < X(kpc) < 7.42, -6.89< Y(kpc) < -5.61,  0.88 < Z(kpc) < 1.07).

\begin{table*}[htb]
\centering
    \begin{tabular}{ccccccccc}
    \hline
    \textbf{\textit{Gaia} DR3}   & \textbf{T$_{\texttt{eff}}$} & \textbf{{[}Fe/H{]}} & \textbf{log(g)}  & \textbf{L} & \textbf{M} & \textbf{Age$_{Tracks}$} & \textbf{Age$_{Model}$} \\ \hline
     & K & dex & dex & L$_{\odot}$ & M$_{\odot}$  & Gyr & Gyr \\ \hline
    5413575344812320640 & $4936^{+100}_{-100}$   & $-1.27^{+0.1}_{-0.1}$  & $1.874^{+0.038}_{-0.038}$  & $ 362.56^{+103.13}_{-103.13}$               & $1.850^{+0.57}_{-0.57}$ & $1.04^{+1.47}_{-0.50}$ & $1.75^{+0.46}_{-0.46}$             \\ \hline
    6045465055252946560 & $4928^{+100}_{-100}$       & $-1.04^{+0.1}_{-0.1}$ & $2.443^{+0.034}_{-0.034}$   & $83.16^{+10.18}_{-10.18}$  & $1.582^{+0.263}_{-0.263}$  & $1.50^{+0.95}_{-0.50}$ & $1.48^{+0.38}_{-0.38}$ \\ \hline
    \end{tabular}
\caption{Stellar parameters of the two most massive young metal-poor stars. The T$_{\texttt{eff}}$, [Fe/H], and log(g) come from APOGEE DR17. The errors on log(g) originate from the predictions of the neural network reported in APOGEE DR17. The two last columns refer to the ages obtained with the BaSTI stellar tracks and the \texttt{CatBoostRegressor} model, respectively.}
\label{tab:Young-Robust-Peculiar-Stars-Masses}
\end{table*}

\begin{table*}[ht]
\centering
\begin{tabular}{ccccccc}
\hline
\textbf{\textit{Gaia} DR3} & \textbf{{[}Fe/H{]}} & \textbf{T$_{\texttt{eff}}$} & \textbf{log(g)} & \textbf{L} & \textbf{{[}$\alpha$/Fe{]}} & \textbf{Age} \\ \hline
                    & dex              & K                   & dex             & L$_{\odot}$        & dex             & Gyr             \\ \hline
4342806982510019456 & -0.54 $\pm$ 0.01 & 5015 $\pm$ 19.00 & 2.49 $\pm$ 0.04 & 60.35 $\pm$ 15.64  & 0.20 $\pm$ 0.04 & 5.77 $\pm$ 1.27 \\ \hline
1158019050967480064 & 0.15 $\pm$ 0.01  & 4709 $\pm$ 7.00  & 3.04 $\pm$ 0.02 & 8.78 $\pm$ 0.33    & 0.06 $\pm$ 0.02 & 8.28 $\pm$ 0.99 \\ \hline
1159422749358899072 & -0.24 $\pm$ 0.01 & 4923 $\pm$ 9.00  & 3.03 $\pm$ 0.02 & 13.97 $\pm$ 0.42   & 0.06 $\pm$ 0.02 & 4.70 $\pm$ 1.03 \\ \hline
4131550666634133888 & -0.91 $\pm$ 0.01 & 4643 $\pm$ 10.00 & 1.72 $\pm$ 0.03 & 203.66 $\pm$ 56.66 & 0.29 $\pm$ 0.03 & 8.60 $\pm$ 1.03 \\ \hline
4110973989419352704 & -0.04 $\pm$ 0.01 & 4416 $\pm$ 6.00  & 2.03 $\pm$ 0.02 & 153.15 $\pm$ 14.26 & 0.03 $\pm$ 0.02 & 6.11 $\pm$ 1.27 \\ \hline
... & ...  & ...   & ...  & ...  & ...  & ...  \\ \hline
\end{tabular}
\caption{Partial content of the age catalogue. Only the most used stellar parameters in stellar physics are displayed here. The [Fe/H], T$_{\texttt{eff}}$ and {[}$\alpha$/Fe{]} come from APOGEE DR17 and are known to be underestimated. The errors in log(g) are expected given the results of the neural network used by the authors of APOGEE DR17. The full age table with the complete set of stellar parameters can be found online.}
\label{tab:Catalogue-Final}
\end{table*}

The existence of these young metal-poor stars is consistent with recent evidence pointing to a metal-poor gas infall, as reported by \cite{Spitoni-2022}. This gas infall event is estimated to have taken place around 2.7 Gyr ago.\\
We performed several tests to ensure the robust age derivation of these young metal-poor stars and to eliminate the possibility of machine learning artefacts. As these young red giants are absent from the comprehensive asteroseismic parameter catalogue of \cite{Hon-2021}, and corresponding light curves are unavailable in the TASOC database \citep{TASOC-One-2021, TASOC-Two-2021}, their ages could not be reliably calculated using state-of-the-art stellar modelling codes. However, one advantage of having \textit{Gaia} luminosities is that the use of stellar tracks for mass determination is no longer required. Therefore, derivations of the stellar
mass for red giants can be independent of the systematic errors originating from the effective temperature scale and the stellar tracks.\\ Consequently, the procedure to determine the stellar mass implies the use of the Stefan-Boltzmann and surface gravity equations, leading to an algebraic formula (refer to Equation \ref{eq:Alg-Mass}) where L is the luminosity, g is the surface gravity, $\sigma$ is the Stefan-Boltzmann constant, G is the universal gravitational constant, and T is the effective temperature. This equation makes mass determination only dependent on the quality of the data, but not on stellar models. 
\begin{equation}
    M = \frac{L\cdot g}{4 \pi \sigma G \cdot T^{4}}
    \label{eq:Alg-Mass}
.\end{equation}
The associated uncertainties were propagated linearly. As the uncertainties on T$_{\texttt{eff}}$ and [Fe/H] are known to be underestimated from the ASPCAP abundance fitting procedure \citep{ASPCAP-2016}, it was decided to set them to 100K and 0.1 dex. This enabled us to obtain more reliable uncertainties on the computed stellar masses (refer to Table \ref{tab:Young-Peculiar-Stars-Masses}). Subsequently, we directly compared  the masses and associated [Fe/H] to the comprehensive grid of BaSTI  \citep{BASTI-ONE-2018, BASTI-TWO-2021} stellar tracks in order to derive the ages, using the publicly available online BaSTI stellar track table maker \footnote{http://basti-iac.oa-abruzzo.inaf.it/tracks.html}.
 Only one star per group has displayed a sufficiently high median mass estimate and a sufficiently low mass uncertainty to provide a robust young age estimate from the stellar tracks (refer to Table \ref{tab:Young-Robust-Peculiar-Stars-Masses}).\\ The fact that these stars have clustered chemical and kinematical values provides confidence in the observation that they share similar age values below 2.7 Gyr. However, definitive confirmation of the age for the entire two groups will necessitate acquiring lower and more robust uncertainties in T$_{\texttt{eff}}$ and [Fe/H] with the future public release of the APOGEE SDSS V survey \citep{APOGEE-SDSS-V} or acquiring their global asteroseismic parameters from dedicated surveys with TESS or the upcoming PLATO mission \citep{PLATO-2014}. 

The above tests provide evidence that at least two of these stars were recently made from the most recent metal-poor gas infall, as described in \cite{Spitoni-2022}. As the classical two-infall model cannot predict this type of young low-metallicity population, the discovery of these stars advocates for the three-infall chemical evolution model described in \cite{Spitoni-2022}.

%%%%%%%%%%%%%%%%%%%%%%%%%%%%%%%%%%%%%%%%%%%%%%%%%%%%%%%%%%%%%%%%%%%%%%%%%%%%%%%%%%%
\section{Conclusion}\label{sec:Conclusion}

We developed a machine learning model to provide a list of asteroseismically calibrated ages for the APOGEE DR17 catalogue, working with a sample of 6539 stars. One component of this sample comes from the TESS SCVZ catalogue \citep{MacKereth-2021} cross-matched with the APOGEE DR17 catalogue. The other component comes from the second APOKASC catalogue \citep{Second_APOKASC_Catalog:2018} updated with data from APOGEE DR17.\\
We introduce the main concepts underlying the construction and evaluation of a machine learning model. We justify the use of a \texttt{CatBoostRegressor} model by listing the advantages of tree-based models. In order to build the model, a feature-selection phase was first undertaken. We justify the selection of each feature. We then explain the several strategies used to optimise the performance of the model. We describe how we address the issue of the age-skewed target distribution by rescaling it with a logarithmic transformation. The imbalance in the data is managed by applying the random oversampling technique, which increases the representation of the minority class (ages older than 10 Gyr).\\ We identified a data shift between the APOKASC-2 and MCK samples. To address this shift,  the two datasets were combined. We discuss the unreliable nature of the lowest values of the [Ce/Fe] abundances and constrain the criteria to mitigate their impact.  By removing the unreliable [Ce/Fe], the performance of the model is improved, particularly in terms of standard deviation per age bin, which leads to a higher robustness in the resulting predictions.\\ The fully optimised model demonstrates performance characterised by a decreasing trend in the median fractional error and in the standard deviation per age bin as the age increases. The median fractional error reaches its lowest point at approximately 7\% for ages between 10 and 11 Gyr, and its highest point at 43\% for stars younger than 1 Gyr. The overall median fractional error reaches 20.8\%.\\ Moreover, we tested the integrity of the \texttt{CatBoostRegressor} performances on an independent data set of the K2 Galactic Archaeology Program. This test shows that there is no significant decay in the performances of the model, indicating that the model is generalisable.\\
The model yields an age map made of 125,445 red giants from the Main Red Star Sample within the APOGEE DR17 catalogue. The associated age catalogue is available online and Table \ref{tab:Catalogue-Final} depicts some of its columns.\\
The age map reveals features confirming the flaring of the disc young stars (Age $<$ 6 Gyr) towards the Galaxy's outer regions, as previously found in \cite{Ness-2018} and \cite{Anders-XGBoost}. Furthermore, features not previously found in \cite{Ness-2018} and \cite{Anders-XGBoost} are revealed. Notably, our map shows two groups of young metal-poor stars. Their chemical abundance and kinematical parameters appear to be clustered. In each group, the most massive star can be robustly dated using the BaSTI stellar tracks \citep{BASTI-ONE-2018,BASTI-TWO-2021} confirming an age of below 2.7 Gyr. This provides evidence supporting the most recent metal-poor gas infall proposed by \cite{Spitoni-2022} and therefore advocates for their three-infall chemical evolution model.

In the future, the performance of the CatBoost model is expected to improve with the release of larger datasets from the APOGEE consortium. Indeed the cross-match of APOGEE DR17 with the MCK catalogue leads to only 1025 stars among the 5574 stars benefiting from the most reliable asteroseismic data within the TESS-SCVZ. The first public release of SDSS V is expected to fill this gap. Moreover, the forthcoming PLATO mission is anticipated to make a substantial contribution to the augmentation of both the quantity and quality of data available for Galactic archaeology  \citep{Miglio-Plato-2017}. Indeed, it will bring higher precision in asteroseismic data associated with a 20 times larger field of view than the \textit{Kepler} mission (2132 deg$^2$ vs 105 deg$^2$) in two long-pointing surveys of two years each. Therefore, the portion of ages derived with stellar modelling codes will need to scale this increase in available data to benefit the machine learning training samples. Consequently, there are substantial grounds to anticipate an enhancement in the performance of machine learning models, which will play a crucial role in leading the field of Galactic archaeology to new discoveries.

\begin{acknowledgements}
I acknowledge the Fundação para Ciênça e Technologia for its financial support through the grant PD/BD/150426/2019. I would like to express my gratitude to the referee for her remarks and advice. Moreover, I thank Aldo Serenelli for his advice regarding the robust dating of the young metal-poor stars and the age errors in the Second APOKASC catalogue. I thank Tiago Campante for his advice regarding the choice of the training age-asteroseismic catalogues. I thank Vardan Adibekyan for his advice regarding the kinematics and chemical analysis. I thank Andrew Humphrey for his advice regarding the optimisation of the machine learning performances. I thank Elisa Delgado Meña for her advice regarding the Cerium abundance and the Barium stars. I thank Andreas Neitzel for his advice regarding the choice of the 3D dust maps and the criteria to refine the quality of the parallaxes. Finally, I thank them all for the general advice they gave me on the paper as a whole.
\end{acknowledgements}

\bibliographystyle{aa} % style aa.bst
\bibliography{export-bibtex} %

\clearpage
\onecolumn

%-----------------------------------------------------------
\begin{appendix} 

\section{Spearman coefficients matrix.}\label{Appendix:Spearman}

\begin{figure}[h]
       \centering
       \includegraphics[width=1.2\linewidth,angle=0]{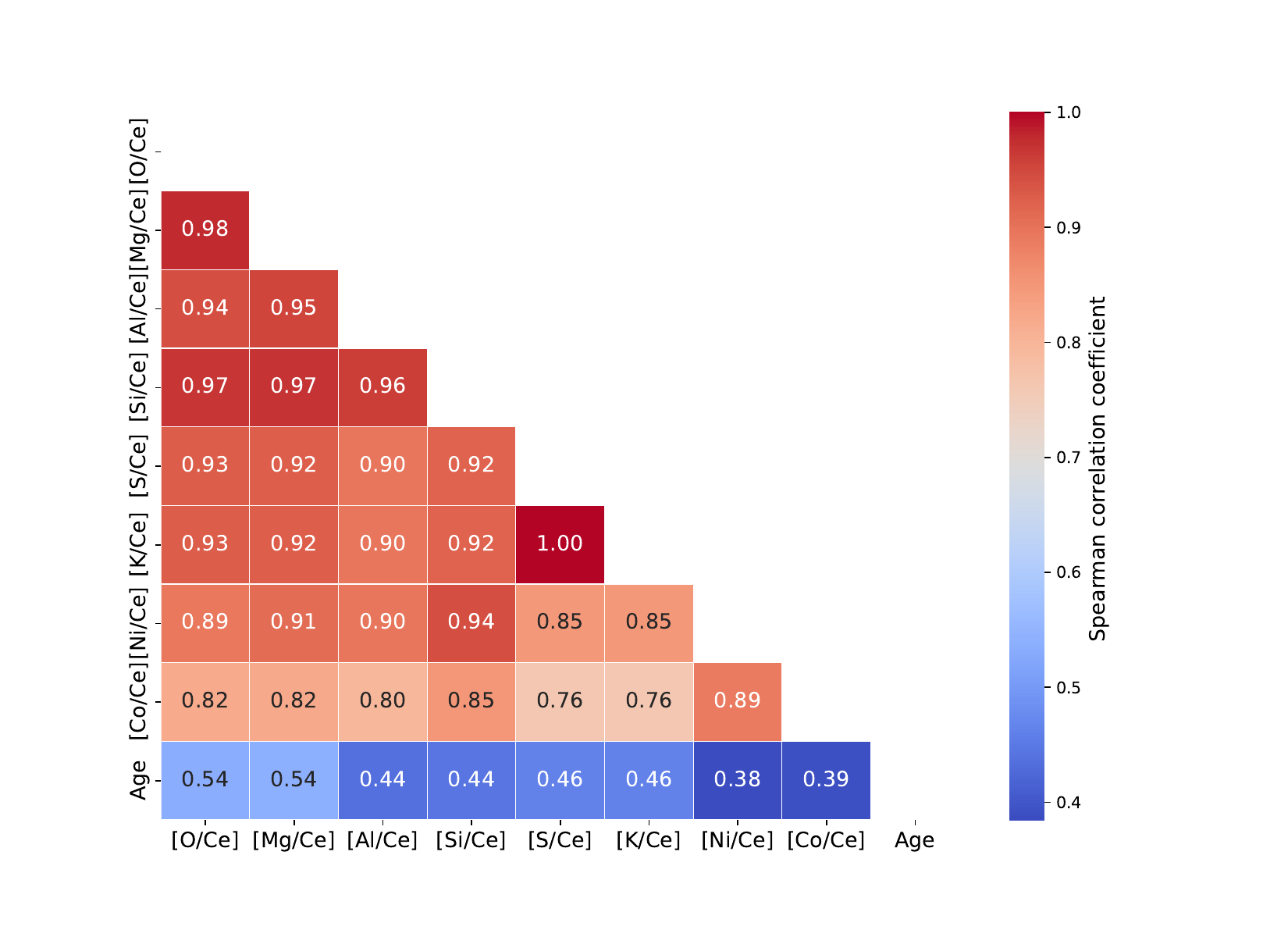}
       \caption{Map of the Spearman correlation coefficients for the [X/Ce] abundances.}
       \label{fig:Spearman-Matrix-Ce}
\end{figure}

%%%%%%%%%%%%%%%%%%%%%%%%%%%%%%%%%%%%%%%%%%%
\newpage
%%%%%%%%%%%%%%%%%%%%%%%%%%%%%%%%%%%%%%%%%%%

\section{Comparison between two scales of temperature}\label{Appendix:Temperature-Comparisons}

\begin{figure}[ht]
           \centering
           \includegraphics[width=0.6\linewidth,angle=0]{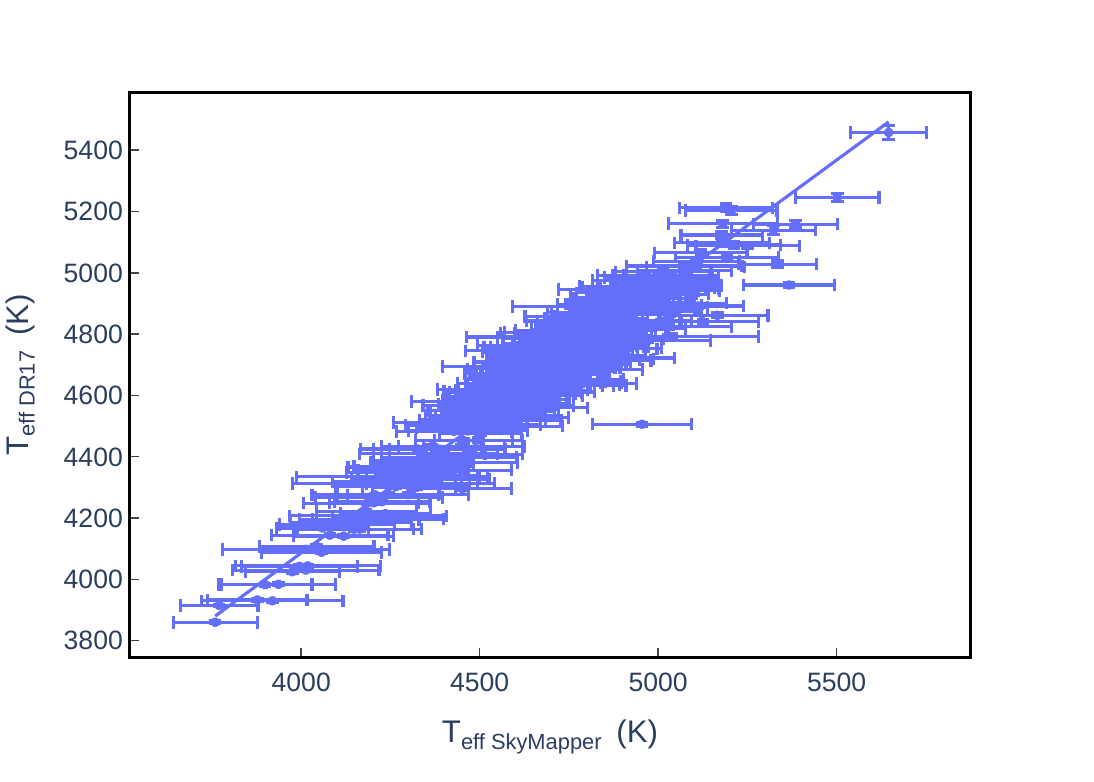}
           \caption{Regression plot between effective temperatures provided by APOGEE DR17 and SkyMapper on the MCK component of the training sample.}
           \label{fig:Teff-APODR17-Skymapper-MacKereth}
\end{figure}

The graphical representation of the ordinary least square regression between APOGEE DR17 and SkyMapper effective temperature values for the MCK component of the training sample is displayed in Figure \ref{fig:Teff-APODR17-Skymapper-MacKereth}. Regression quality parameters obtained for the fit can be found in Table \ref{tab:OLS-Table-Teff-APODR17-Skymapper}. The p-value of the F-statistic test reveals the linear relation is statistically significant. The regression slope and the R$^{2}$ are approximately close to one, and therefore one can state the two sets of temperatures form a coherent ensemble. 

\begin{longtable}{@{}ll@{}}
\toprule
Parameter                  & Value     \\
\midrule
R-squared                  & 0.947     \\
Adj. R-squared             & 0.947     \\
F-statistic                & 6351.     \\
Prob (F-statistic)         & 9.43e-228 \\
Log-Likelihood             & -1952.7   \\
AIC                        & 3909     \\
BIC                        & 3917     \\
slope                      & 0.8543    \\
std err                    & 0.011     \\
\bottomrule
\caption{Summary of some of the \textit{statsmodel} Python parameters provided for an ordinary least square regression.}
\label{tab:OLS-Table-Teff-APODR17-Skymapper}
\end{longtable}

%%%%%%%%%%%%%%%%%%%%%%%%%%%%%%%%%%%%%%%%%%
\newpage
%%%%%%%%%%%%%%%%%%%%%%%%%%%%%%%%%%%%%%%
\section{The data sets}\label{Appendix:Merged samples}

\begin{figure}[htb]
       \centering
       \includegraphics[width=1.0\linewidth]{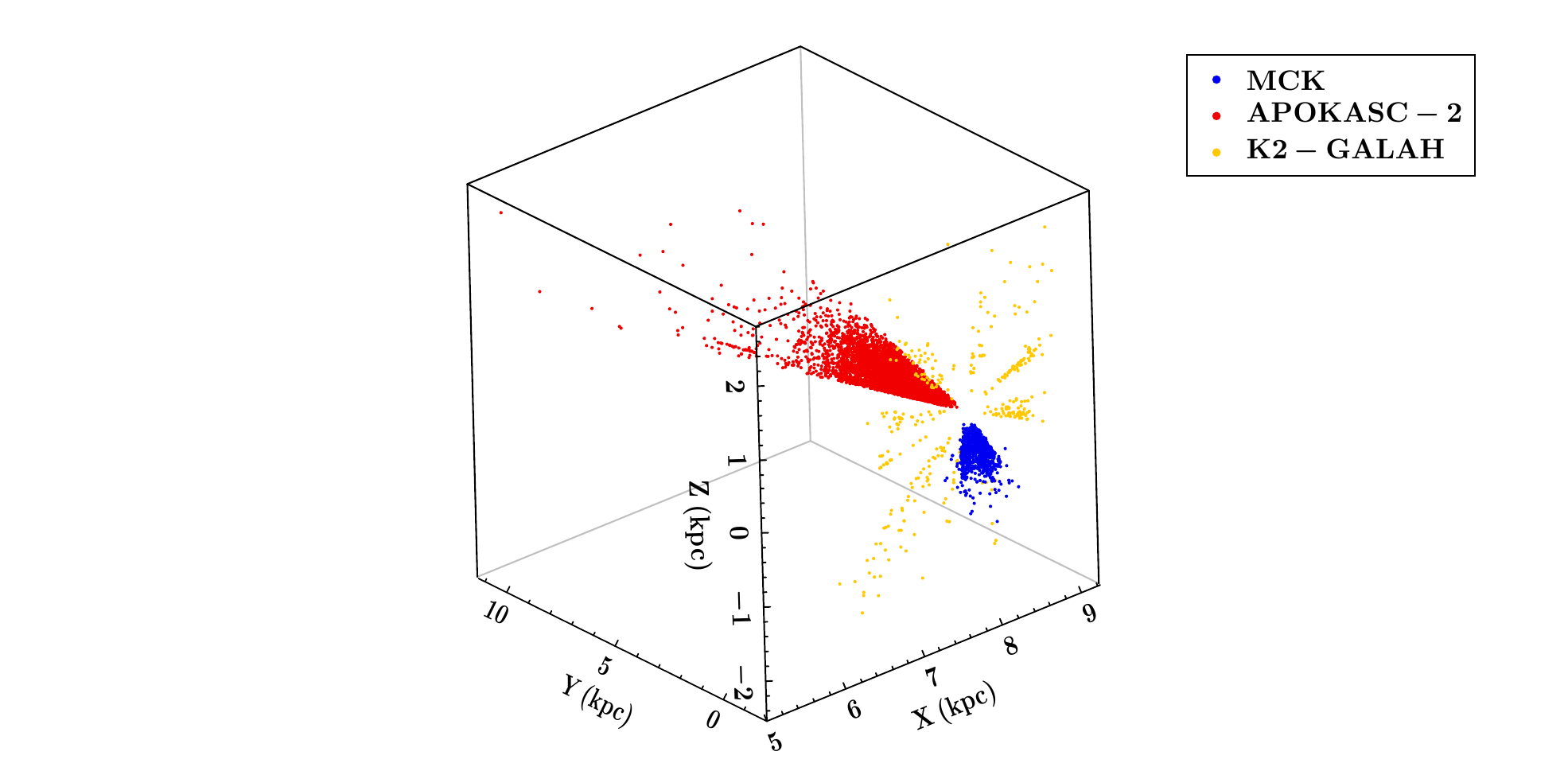}
       \caption{Visual representation of the spatial location of the APOKASC-2 and MCK components of the training--test sample, plus the K2-GALAH sample. The X-axis represents the distance from the Galactic centre.}
    \label{fig:Location-Merged-Sample}
\end{figure}

\section{Hyperparameters of CatBoost}\label{Appendix:CatBoost_Hyperparam}

\begin{table*}[htb]
\centering
\begin{tabular}{@{}c l @{}}
\toprule
\multicolumn{1}{l}{\textbf{CatBoost hyperparmeters}}  & \multicolumn{1}{c}{\textbf{Description}}     \\ \midrule
\textit{learning\_rate} & Controls the step size at each iteration of the gradient-boosting process.              \\ \midrule
\textit{depth}     & Specifies the maximum depth of each decision tree in the ensemble.   \\ \midrule
\textit{reg\_lambda} & Also known as L2 regularisation, it adds a penalty term to the loss function to prevent overfitting.  \\ \midrule
\textit{l2\_leaf\_reg} & It is another regularisation term that applies L2 regularisation specifically to the leaf weights of the trees. \\ \midrule
\textit{iterations} & Determines the number of boosting iterations or the number of decision trees to be built in the ensemble.\\ \midrule
\textit{random\_strength} & Controls the randomness of feature selection during tree construction.\\ \midrule
\textit{rsm} & Stands for row subsampling rate. Determines the portion of training data randomly sampled for each tree.\\ \midrule
\textit{subsample} & Similar to rsm but operates at the level of the entire dataset rather than individual trees.\\ \midrule
\textit{border\_count} & Determines the number of discrete values for numerical features. Allows for more accurate splits.\\ \midrule
\textit{bagging\_temperature}             & \begin{tabular}[c]{@{}l@{}} Controls the intensity of the internal bootstrap aggregation procedure.\end{tabular} \\ \bottomrule
\end{tabular}
\caption{List of the hyperparameters tuned for the grid optimisation of the data. The ones not mentioned kept their default values.}
\label{tab:Hyparameters-CatBoost}
\end{table*}

\newpage

\section{The young metal-poor stars}\label{Appendix: Peculiar young stars}

In this section, the probability threshold for a star to belong to a stellar kinematic component is 70\%. It is taken from \cite{Adibekyan-2012}.
The probabilities were computed using the probabilistic method described in \cite{Bensby-et-al-2003,Bensby-2005}.

\begin{table}[h]
\centering
\begin{tabular}{ccccccc}
\hline
\textbf{\textit{Gaia} DR3} & \textbf{U(LSR)} & \textbf{V(LSR)} & \textbf{W(LSR)} & \textbf{e} & \textbf{P} & \textbf{Group} \\ \hline
 & km/s & km/s & km/s & $\emptyset$ & $\emptyset$ & $\emptyset$ \\ \hline
6045456503982565120 & 51.182 & -205.478 & -8.106 & 0.953 & 0.990 & halo \\ 
6045459768157880192 & 58.786 & -207.177 & -5.415 & 0.946 & 0.993 & halo \\ 
6045464952173727104 & 51.538 & -200.899 & 1.956  & 0.917 & 0.984 & halo \\ 
6045465055252946560 & 49.520 & -224.141 & -9.548 & 0.994 & 0.999 & halo \\ 
6045477910100852992 & 59.078 & -195.153 & -4.376 & 0.893 & 0.975 & halo \\ 
6045464814730497152 & 54.368 & -228.714 & -4.290 & 0.931 & 0.999 & halo \\ 
6045487736986573056 & 58.375 & -194.460 & -3.221 & 0.895 & 0.972 & halo \\ 
6045487874425541760 & 54.732 & -197.180 & -5.316 & 0.906 & 0.978 & halo \\ 
\end{tabular}
\caption{Kinematic properties pertaining to the first group of young metal-poor stars. From left to right the parameters are the \textit{Gaia} DR3 identifier, the three velocity components in the local standard of rest, the orbital eccentricity, the probability of belonging to a given kinematic population, and the designation of the kinematic population to which the star is affiliated.}
\label{tab:Appendix-Kin-First-Table}
\end{table}

\begin{table}[h]
\centering
\begin{tabular}{cccccccc}
\hline
\textbf{\textit{Gaia} DR3} & \textbf{T$_{\texttt{eff}}$} & \textbf{{[}Fe/H{]}} & \textbf{log(g)}  & \textbf{L} & \textbf{M} & \textbf{Age$_{Model}$} \\ \hline
 & K & dex & dex & L$_{\odot}$ & M$_{\odot}$ & Gyr \\ \hline
6045465055252946560 & $4928^{+100}_{-100}$ & $-1.04^{+0.1}_{-0.1}$ & $2.443^{+0.034}_{-0.034}$ & $83.16^{+10.19}_{-10.19}$ & $1.584^{+0.264}_{-0.264}$ & $1.48^{+0.39}_{-0.39}$\\
6045456503982565120 & $5185^{+100}_{-100}$ & $-0.98^{+0.1}_{-0.1}$ & $3.364^{+0.041}_{-0.041}$ & $10.32^{+1.81}_{-1.81}$ & $1.335^{+0.285}_{-0.285}$ & $1.96^{+0.51}_{-0.51}$\\
6045464952173727104 & $4931^{+100}_{-100}$ & $-1.02^{+0.1}_{-0.1}$ & $2.485^{+0.038}_{-0.038}$ & $59.53^{+5.85}_{-5.85}$ & $1.244^{+0.192}_{-0.192}$ & $1.98^{+0.51}_{-0.51}$\\
6045459768157880192 & $5092^{+100}_{-100}$ & $-1.04^{+0.1}_{-0.1}$ & $2.94^{+0.035}_{-0.035}$ & $22.18^{+2.87}_{-2.87}$ & $1.162^{+0.199}_{-0.199}$ & $1.79^{+0.47}_{-0.47}$\\
6045464814730497152 & $4903^{+100}_{-100}$ & $-1.05^{+0.1}_{-0.1}$ & $2.327^{+0.037}_{-0.037}$ & $77.0^{+14.47}_{-14.47}$ & $1.146^{+0.254}_{-0.254}$ & $1.99^{+0.52}_{-0.52}$\\
6045487736986573056 & $5085^{+100}_{-100}$ & $-1.03^{+0.1}_{-0.1}$ & $2.953^{+0.037}_{-0.037}$ & $21.05^{+2.46}_{-2.46}$ & $1.144^{+0.188}_{-0.188}$ & $1.79^{+0.47}_{-0.47}$\\
6045487874425541760 & $5036^{+100}_{-100}$ & $-1.02^{+0.1}_{-0.1}$ & $2.769^{+0.035}_{-0.035}$ & $30.6^{+3.0}_{-3.0}$ & $1.131^{+0.169}_{-0.169}$ & $1.67^{+0.43}_{-0.43}$\\
6045477910100852992 & $5021^{+100}_{-100}$ & $-0.99^{+0.1}_{-0.1}$ & $2.69^{+0.034}_{-0.034}$ & $36.0^{+3.8}_{-3.8}$ & $1.122^{+0.172}_{-0.172}$ & $1.57^{+0.41}_{-0.41}$\\
\end{tabular}
\caption{Stellar parameters for the first group of young metal-poor stars. The errors on log(g) originate from the predictions of the neural network reported in APOGEE DR17. The ages are derived from the CatBoostRegressor model. The errors on age are computed based on the median fractional error obtained with the model for ages between one and two billion years, i.e. 26\%.}
\label{tab:Young-Peculiar-Stars-Masses}
\end{table}

\begin{table}[h]
\centering
\begin{tabular}{ccccccc}
\hline
\textbf{\textit{Gaia} DR3} & \textbf{U(LSR)} & \textbf{V(LSR)} & \textbf{W(LSR)} & \textbf{e} & \textbf{P} & \textbf{Group} \\ \hline
& km/s & km/s & km/s & $\emptyset$ & $\emptyset$ & $\emptyset$ \\ \hline
5413585519599048448 & 306.988 & -427.339 & 168.863 & 0.825 & 1.000 & halo \\ 
5413576860944908416 & 297.867 & -427.592 & 156.149 & 0.788 & 1.000 & halo \\ 
5413576689146193280 & 314.877 & -426.684 & 167.581 & 0.84  & 1.000 & halo \\ 
5413574073501926528 & 302.783 & -435.287 & 156.194 & 0.812 & 1.000 & halo \\ 
5413575344812320640 & 343.661 & -418.231 & 185.970 & 0.91  & 1.000 & halo \\ 
\end{tabular}
\caption{Kinematic properties pertaining to the second group of young metal-poor stars. The parameters are the same as in Table \ref{tab:Appendix-Kin-First-Table}.}
\label{tab:Kin-Second-Table}
\end{table}

\begin{table}[h]
\centering
\begin{tabular}{cccccccc}
\hline
\textbf{\textit{Gaia} DR3} & \textbf{T$_{\texttt{eff}}$} & \textbf{{[}Fe/H{]}} & \textbf{log(g)}  & \textbf{L} & \textbf{M} & \textbf{Age$_{Model}$} \\ \hline
 & K & dex & dex & L$_{\odot}$ & M$_{\odot}$ & Gyr \\ \hline
5413575344812320640 & $4936^{+100}_{-100}$ & $-1.28^{+0.1}_{-0.1}$ & $1.874^{+0.038}_{-0.038}$ & $362.56^{+103.13}_{-103.13}$ & $1.851^{+0.571}_{-0.571}$ & $1.75^{+0.46}_{-0.46}$\\
5413585519599048448 & $5047^{+100}_{-100}$ & $-1.3^{+0.1}_{-0.1}$ & $2.269^{+0.044}_{-0.044}$ & $129.51^{+35.66}_{-35.66}$ & $1.5^{+0.456}_{-0.456}$ & $1.66^{+0.43}_{-0.43}$\\
5413576860944908416 & $4881^{+100}_{-100}$ & $-1.27^{+0.1}_{-0.1}$ & $1.863^{+0.038}_{-0.038}$ & $258.3^{+51.37}_{-51.37}$ & $1.344^{+0.312}_{-0.312}$ & $1.98^{+0.51}_{-0.51}$\\
5413574073501926528 & $4922^{+100}_{-100}$ & $-1.33^{+0.1}_{-0.1}$ & $1.985^{+0.041}_{-0.041}$ & $187.57^{+50.39}_{-50.39}$ & $1.248^{+0.369}_{-0.369}$ & $1.95^{+0.51}_{-0.51}$\\
5413576689146193280 & $4952^{+100}_{-100}$ & $-1.29^{+0.1}_{-0.1}$ & $1.892^{+0.048}_{-0.048}$ & $222.61^{+59.81}_{-59.81}$ & $1.169^{+0.353}_{-0.353}$ & $1.54^{+0.4}_{-0.4}$\\
\end{tabular}
\caption{Stellar parameters for the second group of young metal-poor stars. For the description of the parameters, refer to Table \ref{tab:Young-Peculiar-Stars-Masses}}
\label{tab:Young-Second-Peculiar-Stars-Masses}
\end{table}

\section{Extra Age map plots }\label{Appendix:Age-map extra plots}

\begin{figure}[htb]
       \centering
       \includegraphics[width=0.9\linewidth]{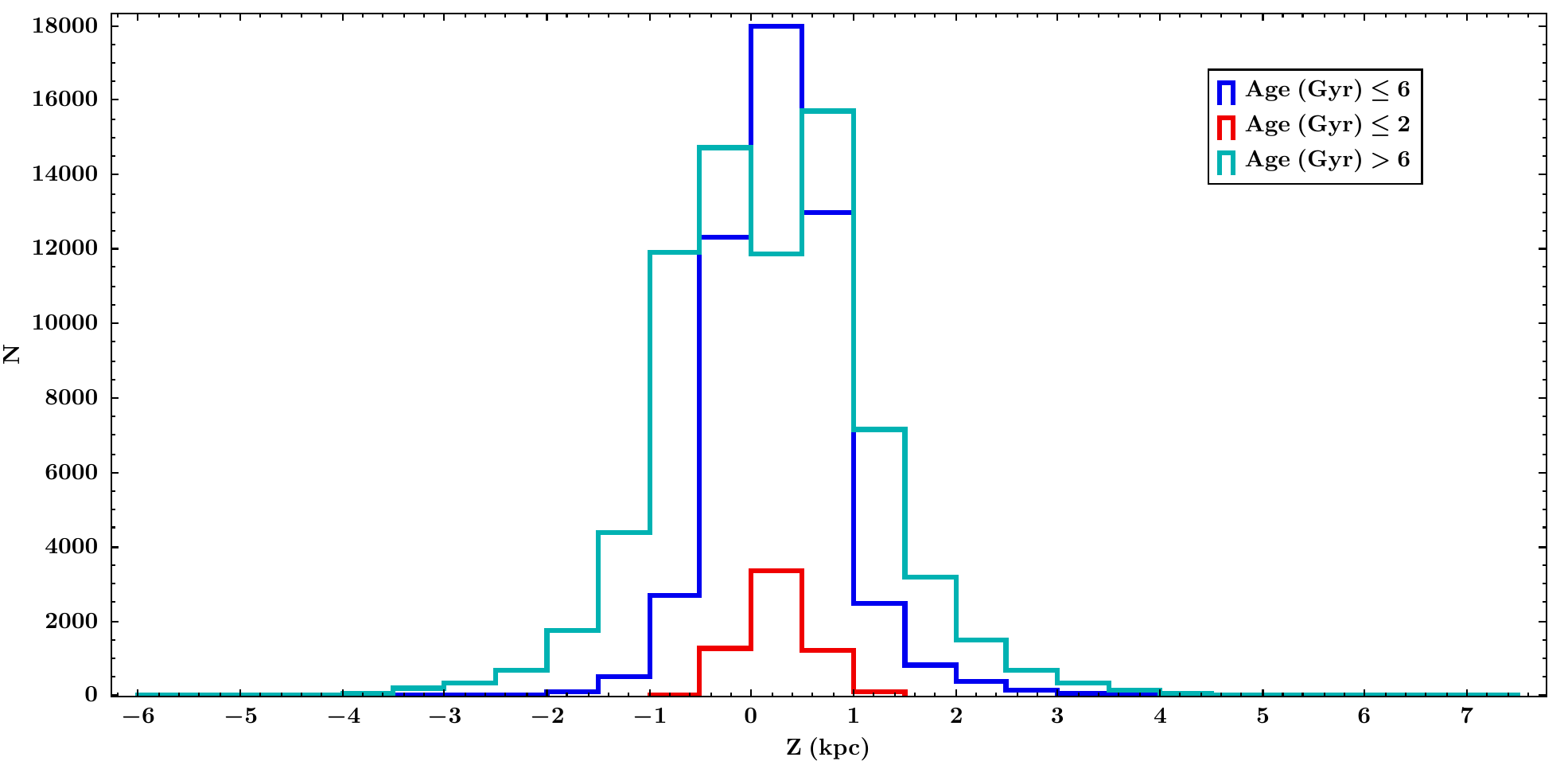}
       \caption{Plot of histograms regarding the Z feature across three age groups from the age map.}
    \label{fig:Histograms-Age-Z-map}
\end{figure}

One of the means to assess the quality of the model is to check for the expected general symmetry of ages in relation to the vertical distance from the Galactic star formation plane. Figure \ref{fig:Histograms-Age-Z-map} illustrates that the three chosen age groups exhibit a clear overlap. Since this study and previous ones \citep{Ness-2018, Anders-XGBoost} highlight that the flaring of the young Galactic disc is most noticeable for stars under 6 billion years old, this age was chosen as the threshold for the second group. Specifically, the three groups share a similar median (m $\sim$ 0.25 kpc). Progressing from the youngest to the oldest group, the skewness values are s=0.46, s=0.81, and s=0.15, indicating negligible skewness (|s|< 1). Notably, the skewness for all ages is s=0.18.

\section{Extra statistical plots}\label{Appendix:Extra statiscal plots}

\begin{figure}[ht]
       \centering
       \includegraphics[width=0.7\linewidth]{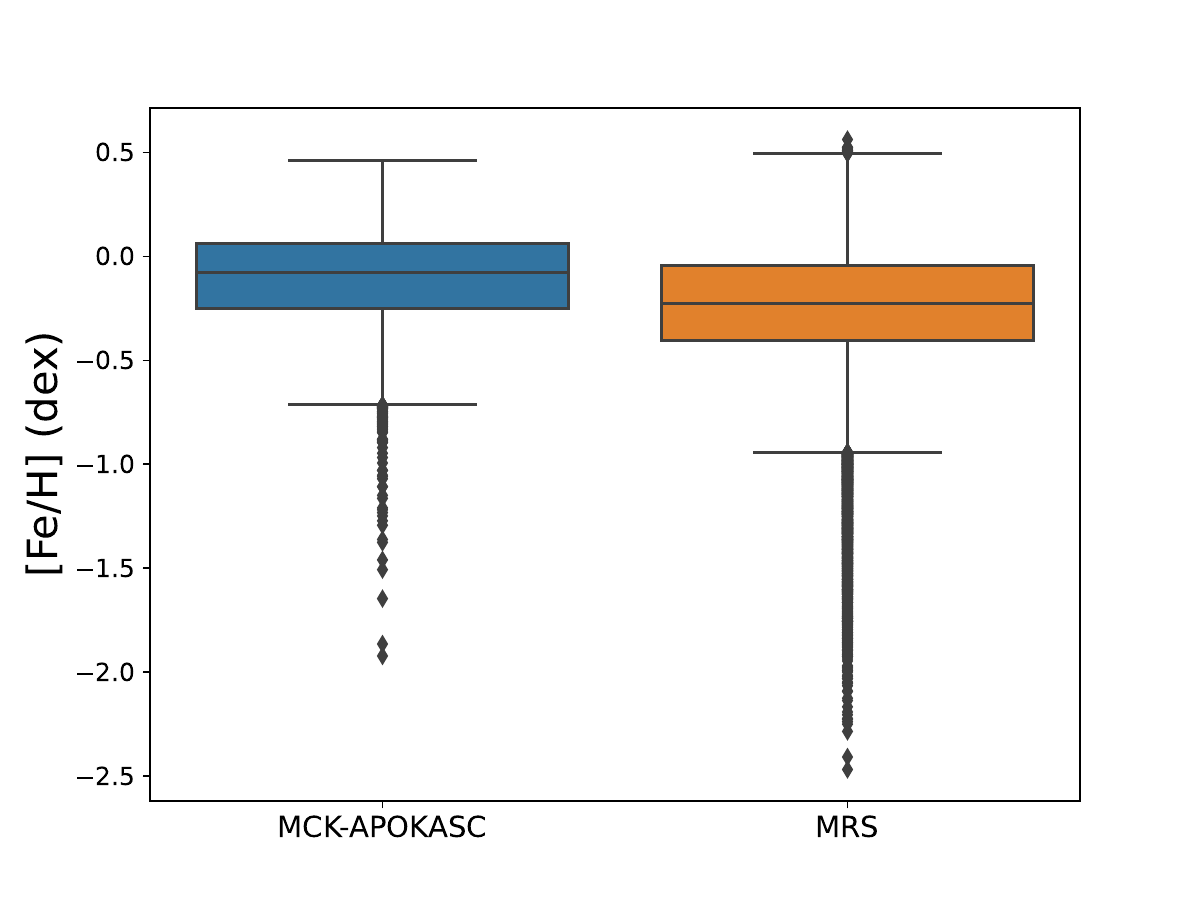}
       \caption{Box plots of the [Fe/H] distributions for the MCK-APOKASC and MRS-\textit{Gaia} samples.}
    \label{fig:Box-Plot-Metallicity}
\end{figure}

\begin{figure}[ht]
       \centering
       \includegraphics[width=0.75\linewidth]{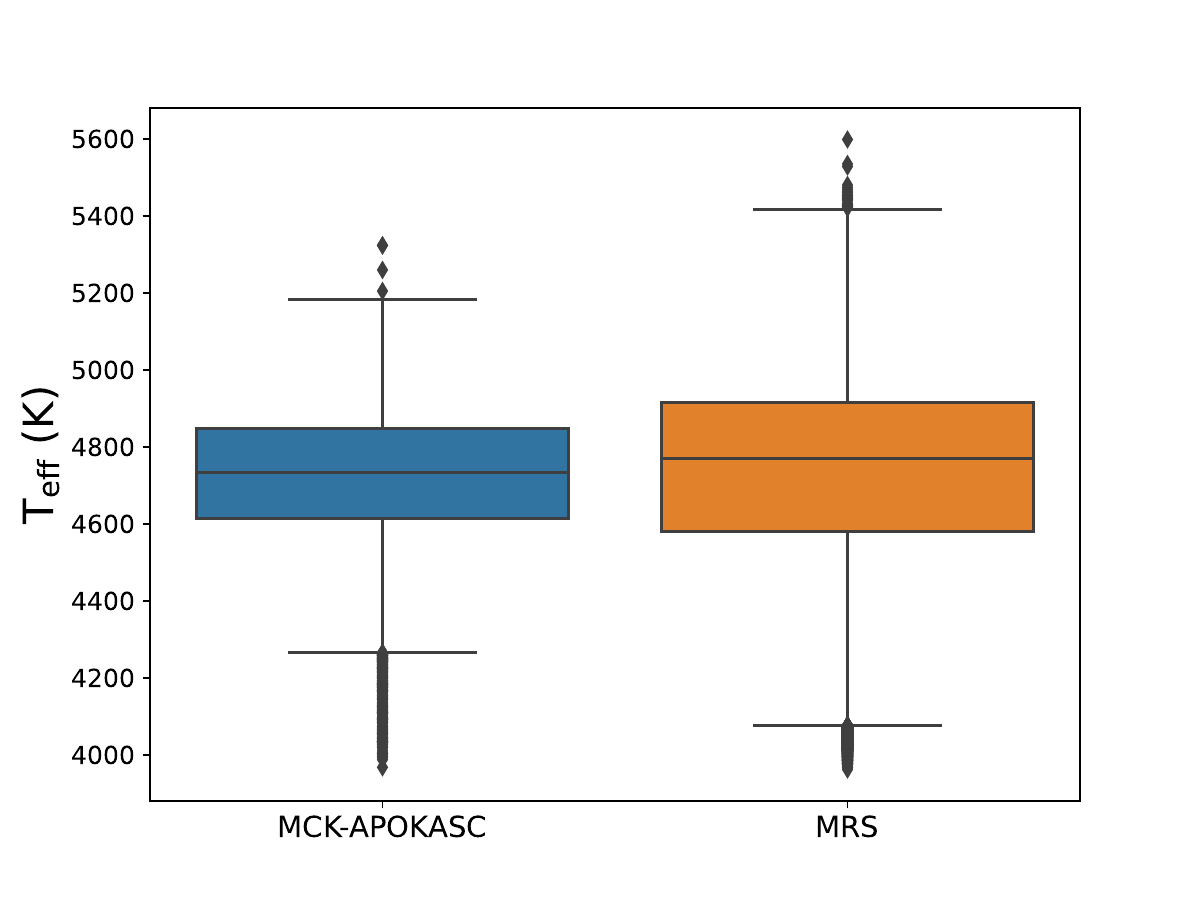}
       \caption{Box plots of the T$_{\texttt{eff}}$ distributions for the MCK-APOKASC and MRS-\textit{Gaia} samples.}
    \label{fig:Box-Plot-Teff}
\end{figure}

\begin{figure}[ht]
       \centering
       \includegraphics[width=0.7\linewidth]{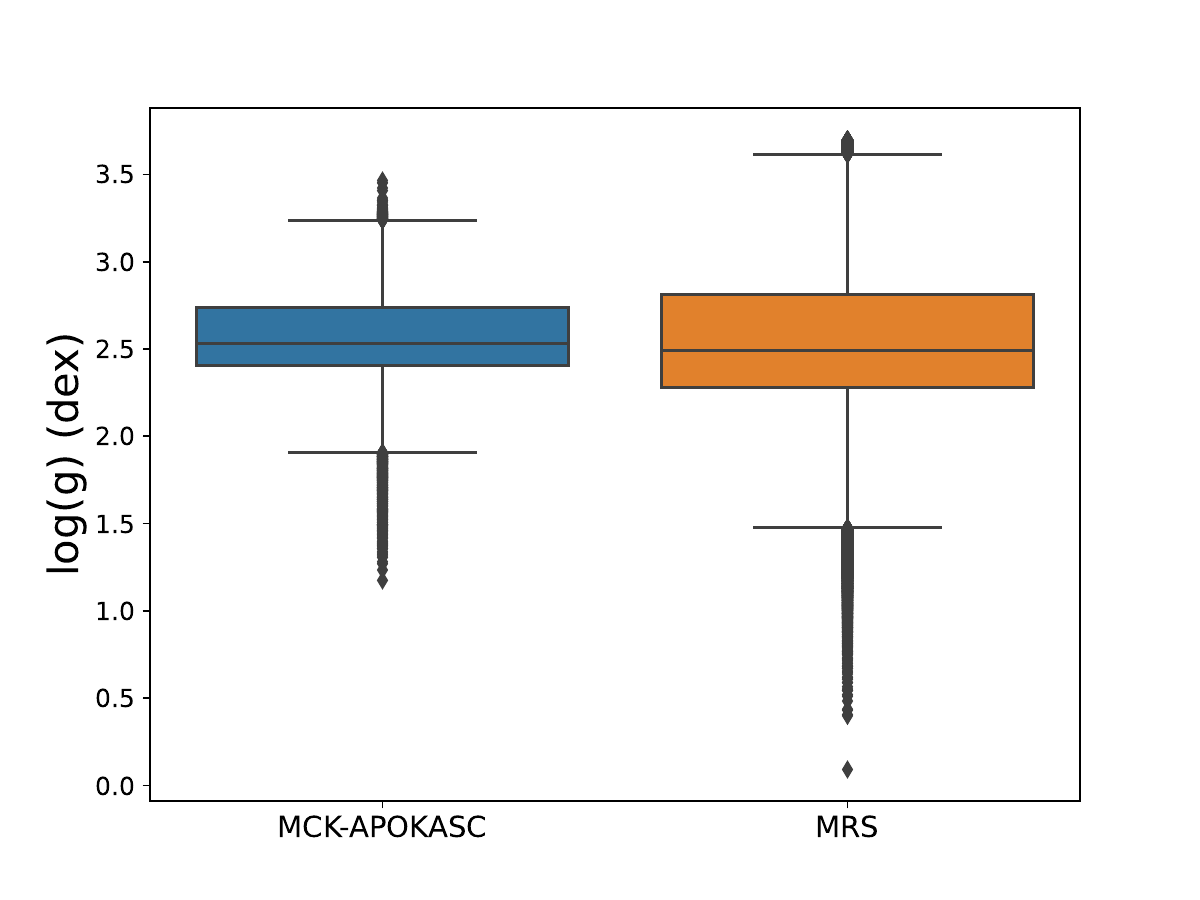}
       \caption{Box plots of the log(g) distributions for the MCK-APOKASC and MRS-\textit{Gaia} samples.}
    \label{fig:Box-Plot-Logg}
\end{figure}

\begin{figure}[ht]
       \centering
       \includegraphics[width=0.75\linewidth]{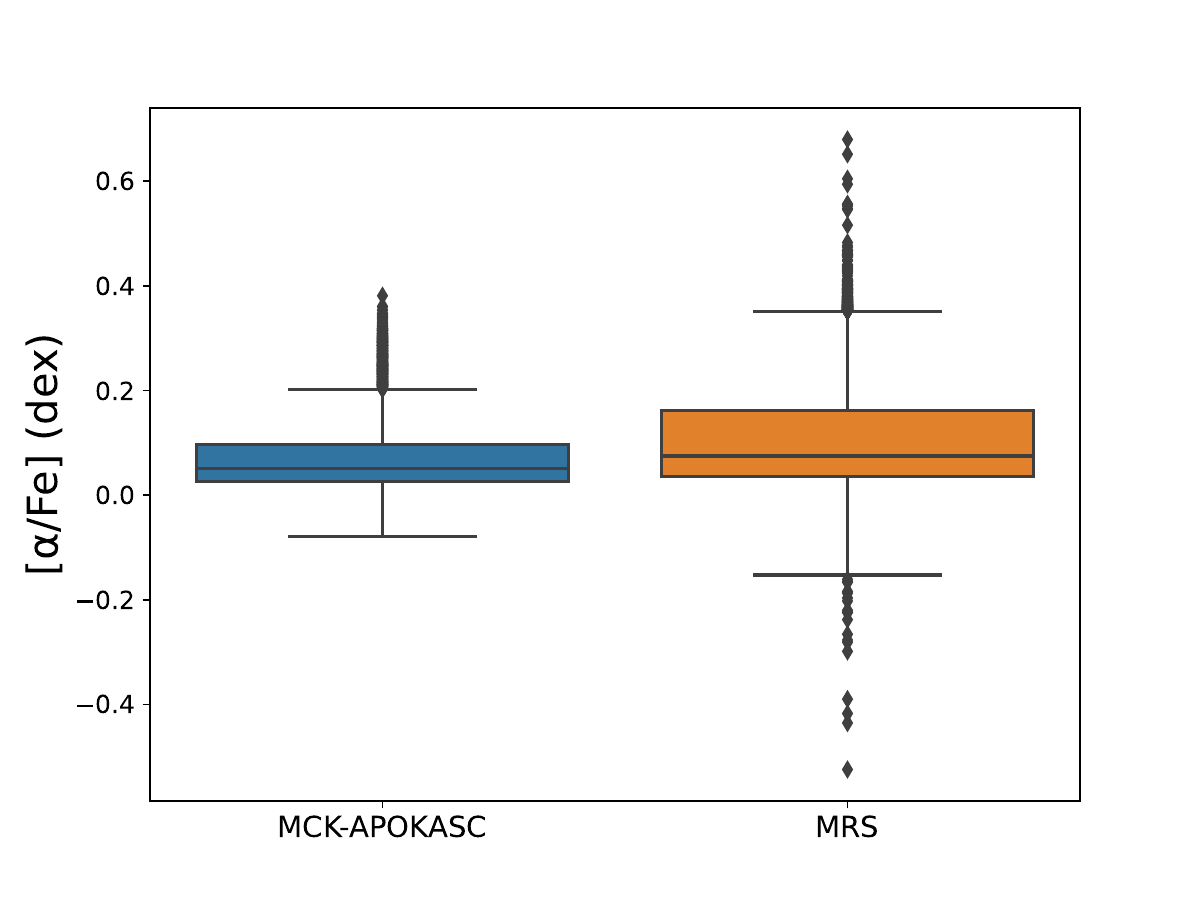}
       \caption{Box plots of the [$\alpha$/Fe] distributions for the MCK-APOKASC and MRS-\textit{Gaia} samples.}
    \label{fig:Box-Plot-alpha}
\end{figure}

\begin{figure}[ht]
       \centering
       \includegraphics[width=0.75\linewidth]{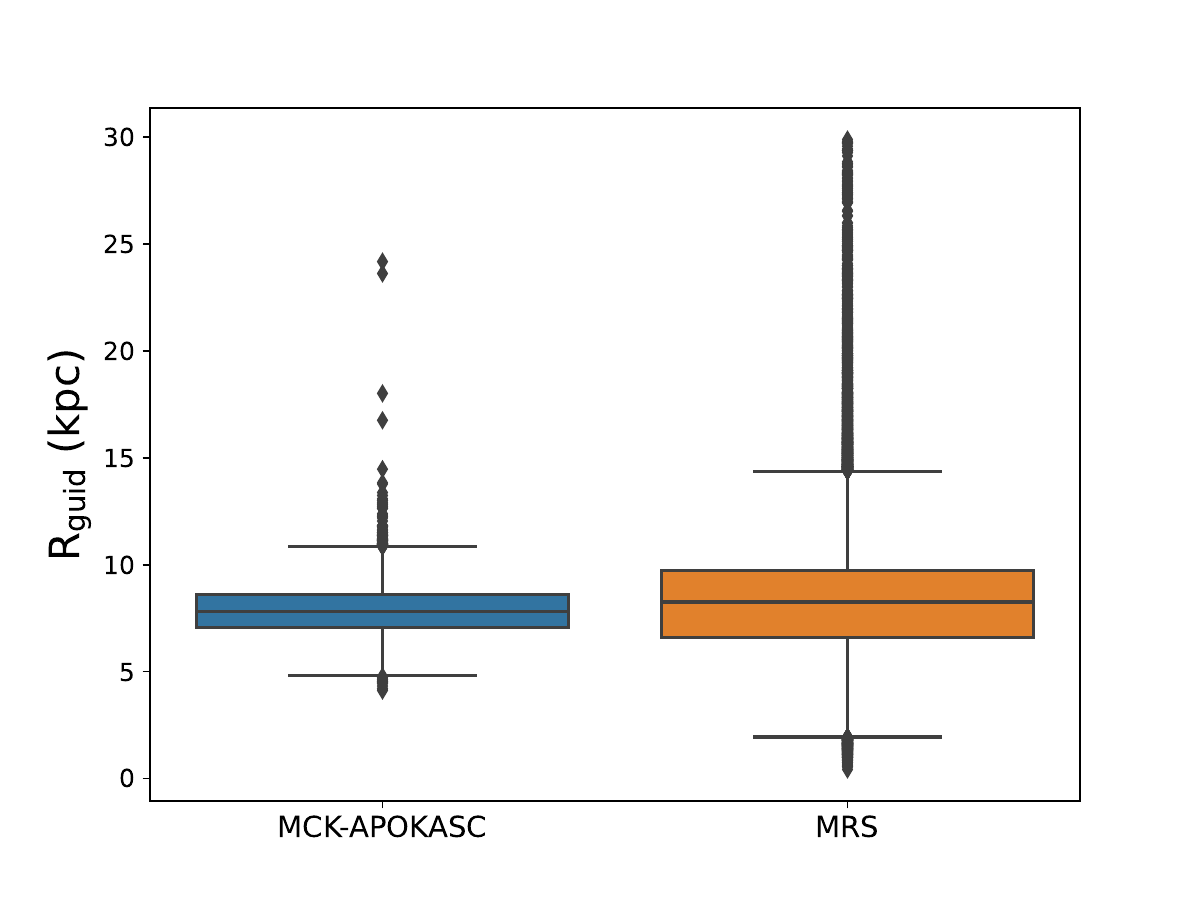}
       \caption{Box plots of the guiding radius distributions for the MCK-APOKASC and MRS-\textit{Gaia} samples.}
    \label{fig:Box-Plot-guiding-radius}
\end{figure}

\begin{figure}[ht]
       \centering
       \includegraphics[width=1\linewidth]{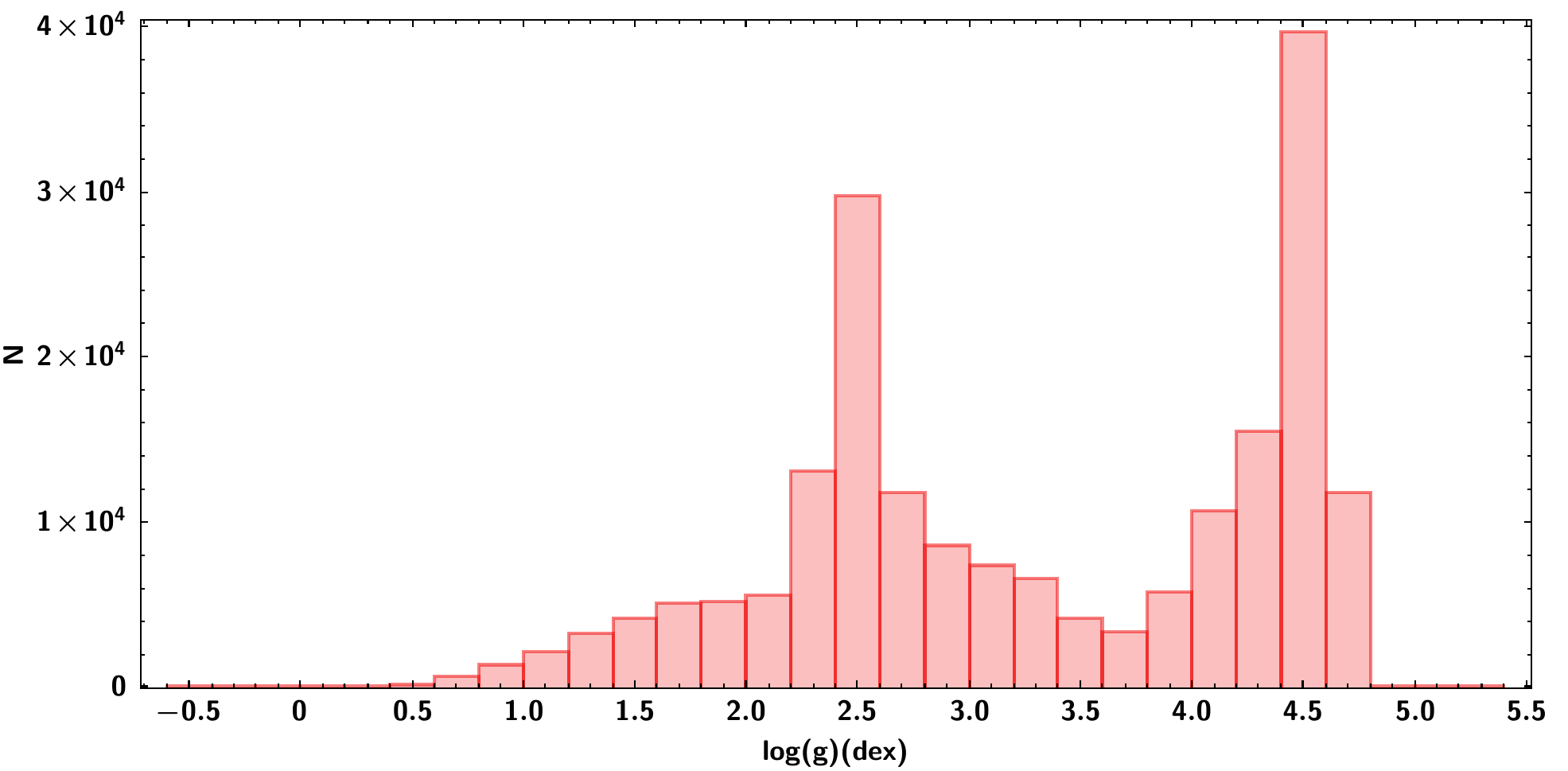}
       \caption{Histogram of log(g) for the MRS-\textit{Gaia} sample without the cut on log(g).}
    \label{fig:Histogram-log(g)-MRS-Gaia}
\end{figure}

\end{appendix}

\end{document}